\documentclass[aps,reprint,superscriptaddress,nofootinbib,nobibnotes]{revtex4-2}
\usepackage[italian]{varioref}
\usepackage{lipsum}
\usepackage{amssymb,amsmath,physics,booktabs,subfig,mathtools,xfrac}
\usepackage[]{graphicx}
\usepackage{ragged2e}
\usepackage{float}
\usepackage{caption}
\usepackage{gensymb,mathtools,amsfonts,amsthm,mathrsfs,slashed,relsize, cancel}
\usepackage[output-decimal-marker={.}]{siunitx}
\usepackage{xcolor,tcolorbox}
\usepackage{appendix}
\usepackage{dcolumn}
\usepackage{bm}
\usepackage{comment}
\usepackage{xcolor}
\usepackage[mathlines]{lineno}
\usepackage[colorlinks]{hyperref}
\modulolinenumbers[5]

\newcommand{\versor}[1]{\mathbf{\hat{#1}}}

\hypersetup{hidelinks}

\begin{document}
	
\preprint{AAPM/123-QED}
	
\title[]{A New Solution for the Observed Isotropic Cosmic Birefringence Angle and its Implications for the Anisotropic Counterpart through a Boltzmann Approach}
	
\author{Alessandro Greco}
\affiliation{Dipartimento di Fisica e Astronomia ``Galileo Galilei", Universit\`a degli Studi di Padova, via Marzolo 8, I-35131, Padova, Italy}
\affiliation{INFN, Sezione di Padova, via Marzolo 8, I-35131, Padova, Italy}
	
\author{Nicola Bartolo}
\affiliation{Dipartimento di Fisica e Astronomia ``Galileo Galilei", Universit\`a degli Studi di Padova, via Marzolo 8, I-35131, Padova, Italy}
\affiliation{INFN - Sezione di Padova, via Marzolo 8, I-35131, Padova, Italy}
\affiliation{INAF - Osservatorio Astronomico di Padova, vicolo dell'Osservatorio 5, I-35122 Padova, Italy}
	
\author{Alessandro Gruppuso}
\affiliation{INAF - Osservatorio di Astrofisica e Scienza dello Spazio di Bologna, via Gobetti 101, I-40129 Bologna, Italy}
\affiliation{INFN - Sezione di Bologna, viale Berti Pichat 6/2, I-40127 Bologna, Italy}
\affiliation{Dipartimento di Fisica e Scienze della Terra, Universit\`a degli Studi di Ferrara, via Saragat 1,I-44122, Ferrara, Italy}
	
\date{\today}

\begin{abstract}
Cosmic Birefringence (CB) is a phenomenon in which the polarization of the Cosmic Microwave Background (CMB) radiation is rotated as it travels through space due to the coupling between photons and an axion-like field. We look for a solution able to explain the result obtained from the \textit{Planck} Public Release 4 (PR4), which has provided a hint of detection of the CB angle, $\alpha=(0.30\pm0.11)^{\circ}$. In addition to the solutions, already present in the literature, which need a non-negligible evolution in time of the axion-like field during recombination, we find a new region of the parameter space that allows for a nearly constant time evolution of such a field in the same epoch. The latter reinforces the possibility to employ the commonly used relations connecting the observed CMB spectra with the unrotated ones, through trigonometric functions of the CB angle. However, if the homogeneous axion field sourcing isotropic birefringence is almost constant in time during the matter-dominated era, this does not automatically imply that the same holds also for the associated inhomogeneous perturbations. For this reason, in this paper we present a fully generalized Boltzmann treatment of this phenomenon, that is able, for the first time to our knowledge to deal with the time evolution of anisotropic cosmic birefringence (ACB). We employ this approach to provide predictions of ACB, in particular for the set of best-fit parameters found in the new solution of the isotropic case. If the latter is the correct model, we expect an ACB spectrum of the order of $(10^{-15}\divisionsymbol10^{-32})$ deg$^2$ for the auto-correlation, and  $(10^{-7}\divisionsymbol10^{-17})$ $\mu $K$\cdot\,$deg for the cross-correlations with the CMB $T$ and $E$ fields, depending on the angular scale.
\end{abstract}
	
\maketitle
	
\section{\label{sec:Intro}Introduction}
In the last decades, the investigation of parity-violating signatures in cosmology has become one of the most ambitious goals (see e.g. \cite{lue1999cosmological, komatsu2022new}). Many efforts have been made to constrain parity-breaking effects coming, e.g., from non-standard inflationary models, not only at the level of the CMB angular power spectra \cite{alexander2005birefringent, contaldi2008anomalous, alexander2009chern, sorbo2011parity, bartolo2015parity1, bartolo2015parity2, gerbino2016testing, qiao2019waveform, qiao2020polarized, campeti2023litebird}, but also by looking at higher-order correlation functions, such as bispectra and trispectra \cite{maldacena2011graviton, anber2012non, shiraishi2013parity, shiraishi2013probing, cook2013inflationary, shiraishi2015observed, shiraishi2016parity, meerburg2016cmb, bartolo2017parity, bartolo2019measuring, aghanim2020planck, liu2020probing, duivenvoorden2020cmb, bartolo2021tensor, niu2022parity, philcox2023cmb, philcox2023testing, philcox2023testing2}. Furthermore, besides CMB observables, recently the research on parity-breaking signals in large-scale structures \cite{dai2016antisymmetric, cahn2021test, hou2022measurement, philcox2022probing, cabass2023colliders, cabass2023parity, creque2023parity, coulton2023signatures, philcox2023galaxy, taylor2023unsupervised} and from astrophysical and cosmological gravitational waves at interferometers \cite{yunes2010testing, crowder2013measurement, kostelecky2016testing, alexander2018gravitational, yagi2018probing, shao2020combined, orlando2021measuring, wang2021gravitational, martinovic2021searching, niu2022constraining, okounkova2022constraining, califano2023parity, callister2023new} has known an increasing interest.

However, one of the most intriguing sources of cosmological parity violation seems to come from cosmic birefringence, which is nothing but the rotation of the linear polarization plane of CMB photons when free-streaming as a consequence of an electromagnetic Chern-Simons coupling with a pseudo-scalar field $\chi$ \cite{carroll1990limits},
\begin{equation}
\label{eqn:L_chern_simons}
\mathcal{L}=-\frac{1}{4}F_{\mu\nu}F^{\mu\nu}-\frac{\lambda}{4f}\chi F_{\mu\nu}\tilde{F}^{\mu\nu},
\end{equation}
where $\lambda/f$ is a parameter with the dimensions of the inverse of an energy, $\tilde{F}^{\mu\nu}\equiv\varepsilon^{\mu\nu\rho\sigma}F_{\rho\sigma}/(2\sqrt{-g})$ is the Hodge dual of the Maxwell tensor $F_{\mu\nu}$, and $\varepsilon^{\mu\nu\rho\sigma}$ is the Levi-Civita antisymmetric symbol. Indeed, this extension of the Maxwell theory induces a rotation of the observed Stokes parameters describing the linear CMB polarization (see e.g. \cite{lee2019dark, namikawa2021cmb, naokawa2023gravitational}),
\begin{equation}
\label{eqn:Stokes}
(Q\pm iU)=(Q\pm iU)_{\text{wcb}}e^{2i\alpha},
\end{equation}
where the label ``wcb'' denotes a quantity that one could obtain without cosmic birefringence, and the birefringence angle $\alpha$ is strictly related to the time evolution of the field $\chi$,
\begin{equation}
\label{eqn:alpha_0}
\alpha\equiv\frac{\lambda}{2f}\left(\chi_{\text{obs}}-\chi_{\text{emi}}\right),
\end{equation}
with $\chi_{\text{emi}}$ and $\chi_{\text{obs}}$ labeling the field's values at the moment of photons' emission and observation, respectively. An observational consequence of the rotation described in Eq.~\eqref{eqn:Stokes} is, e.g., the switching-on of a parity-breaking angular cross-correlation between the $E$ and $B$ modes of CMB polarization,
\begin{equation}
\label{eqn:EB_observed}
C_{\ell}^{EB}=\frac{1}{2}\sin(4\alpha)\left[C_{\ell, \text{wcb}}^{EE}-C_{\ell, \text{wcb}}^{BB}\right].
\end{equation}

Cosmic birefringence can be seen as a probe for the existence of such a field $\chi$, which could be a candidate for early and late dark energy \cite{poulin2019early, choi2021cosmic, fujita2021detection, gasparotto2022cosmic, murai2022isotropic, kamionkowski2022the, rezazadeh2022cascading, yin2023cosmic} or dark matter \cite{preskill1983cosmology, abbott1983cosmological, dine1983not, liu2017axion, nakagawa2021cosmic, obata2022implications, zhou2023cosmic}, in the form of an axion-like field \cite{arvanitaki2010string, hlozek2015search, marsh2016axion, poulin2018cosmological, kim2021cosmic, jain2021cmb, cao2022non, lin2022consistency, gonzalez2022stability, diegopalazuelos2023search, yin2023testing, gasparotto2023axiverse, ferreira2023axionic}. Other possible physical explanations for the birefringence are investigated e.g. in \cite{caloni2022probing, nakai2023explain, nilsson2023reexamining} and in the Refs. therein. The tantalizing idea of succeeding in unveiling the nature of the dark sector of the Universe by looking for cosmological parity-violating signatures has also brought with it the necessity to look for signatures that are able to discriminate among the different models able to induce the birefringence effect, and according to this purpose, a tomographic approach has been recently proposed \cite{sherwin2021cosmic, nakatsuka2022cosmic, lee2022probing, greco2023probing, namikawa2023cosmic}. A complete treatment of cosmic birefringence should consider the possibility that the field $\chi$, in general, may not be homogeneous, implying the presence of a non-zero anisotropic component in the birefringence angle \cite{li2008cosmological, caldwell2011cross, zhao2014fluctuations, zhai2020effects, capparelli2020cosmic, takahashi2021kilobyte, greco2022cosmic, cai2022impact, hagimoto2023measures, lee2023cosmological}: such anisotropies in the birefringence angle can provide by themselves a further and complementary observational test of models for birefringence.

An increasing number of observational constraints on both isotropic and anisotropic cosmological birefringence are present in the literature, as results of several CMB experiments: WMAP \cite{feng2006searching, jarosik2011seven, gluscevic2012first, eskilt2022improved}, POLARBEAR \cite{ade2015polarbear, polarbear2023constraints}, ACTPol \cite{namikawa2020atacama}, SPTpol \cite{bianchini2020searching}, BICEP/Keck \cite{ade2017bicep2, keck2022line}, and the \textit{Planck} satellite \cite{aghanim2016planck, contreras2017constraints, minami2020new, gruppuso2020planck, bortolami2022planck, eskilt2022frequency, eskilt2023constraint, zagatti2024planck}. In particular, the authors of \cite{diego2022cosmic}, exploiting the latest \textit{Planck} data release, have found a hint of detection of the isotropic birefringence angle $\alpha = (0.30\pm0.11)^{\circ}$ \cite{diego2022cosmic}. However, a more detailed analysis is required to be sure that such a rotation has effectively a cosmological origin, and it is not instead caused by, e.g., galactic dust or miscalibration angles \cite{miller2009cmb, keating2012self, minami2019simultaneous, clark2021origin, de2022determination, cukierman2022magnetic, vacher2022frequency, diego2022robustness, monelli2022impact, jost2022characterising, monelli2023impact, ritacco2023polarization}. Nevertheless, let us just mention that if the new physics hypothesis for the existence of a non-zero $EB$ cross-correlation would be confirmed, most probably this could only be explained by cosmic birefringence as shown by Eq.~\eqref{eqn:EB_observed} since any observed $EB$ correlation sourced by primordial chiral gravitational waves does not work due to the overproduction of the $B$ modes concerning the current constraints on the tensor-to-scalar ratio \cite{gerbino2016testing, fujita2022can}.

In this paper, we consider the field theory defined by the following action for the axion-like field\footnote{However, our results are not strongly dependent on the chosen potential for $\chi$ (see also App.~\ref{app:other_potential}).} together with the axion-photon coupling shown in Eq.~\eqref{eqn:L_chern_simons},
\begin{equation}
\begin{split}
\label{eqn:chi_action}
S_{\chi}=-\int\mathrm{d}^4x\,&\sqrt{-g}\left[\frac{1}{2}g^{\mu\nu}\partial_{\mu}\chi\partial_{\nu}\chi+\frac{1}{2}m^2_{\chi}\chi^2\right].
\end{split}
\end{equation}
We perform a chi-squared analysis to find the values for the axion mass $m_{\chi}$ and the Chern-Simons coupling $\lambda/f$ that best fit the previously mentioned \textit{Planck} result \cite{diego2022cosmic}, which nevertheless just refers to the regime of isotropic birefringence. For this reason, to consistently compare theory with observations, we initially restrict our analysis to the case of a homogeneous field, $\chi=\chi_0(\eta)$, where $\eta$ denotes the conformal time. 

One of the main findings of the present paper is the existence of a set of best-fit parameters for which the axion field is almost constant in time during the epoch of recombination, which has not been obtained in previous analyses. It is well known in the literature that the formula in Eq.~\eqref{eqn:Stokes} holds only in the sudden recombination approximation or when the axion time evolution is sufficiently slow during recombination. Otherwise, Eq.~\eqref{eqn:Stokes} could not be directly used for deriving an expression for the CMB spectra modified by cosmic birefringence. Indeed, it becomes necessary to solve the polarized Boltzmann equation for photons by taking into account the birefringence effect from the beginning, and Eq.~\eqref{eqn:Stokes} still holds, but photon-by-photon emission time \cite{loeb1996faraday, liu2006effect, finelli2009rotation, gubitosi2014including, galaverni2023redshift}, so that Eq.~\eqref{eqn:EB_observed} is recovered only in the regimes mentioned before. According to our results, the kind of evolution experienced by the axion seems to be compatible with the use of such an approximation, consistently with what recent data analysis seem to suggest, see e.g. in \cite{eskilt2023constraint}.

Moreover, our idea is to use the results of the fit to find the anisotropic birefringence signal associated with the set of best-fit parameters for the isotropic case. This approach has a twofold purpose: first of all, if the amount of anisotropic birefringence predicted by a theoretical model, whose parameters best fit the amount of isotropic birefringence, is found to be excluded by the constraints on anisotropic birefringence itself, this would mean that such a model is not a good theory for the axion field, making our approach a promising way for breaking degeneracies between different models. Second, as we are going to show in the next sections, the associated perturbations $\delta\chi(\eta,\mathbf{x})$ do not behave in the same way as the homogeneous part, giving us the motivation for generalizing the state-of-the-art of anisotropic cosmic birefringence and verify when it is possible to recover the treatment mainly used in the literature. Indeed, up to now, the anisotropic birefringence angle has been related to the value of the field at the epoch of recombination \cite{li2008cosmological},
\begin{equation}
\label{eqn:sudden_chi}
\delta\alpha(\versor{n})\equiv-\frac{\lambda}{2f}\delta\chi[\eta_{\text{rec}},(\eta_0-\eta_{\text{rec}})\versor{n}],
\end{equation}
where $\delta\chi\equiv\chi-\chi_0(\eta)$ is the inhomogeneous perturbation of the axion field, $-\versor{n}$ is the photons' coming direction, and $\eta_0$ ($\eta_{\text{rec}}$) is the conformal time today (at recombination). However, Eq.~\eqref{eqn:sudden_chi} holds only in the sudden recombination approximation or when $\delta\chi$ does not evolve significantly in time during the recombination epoch. To take into account that photons have been emitted according to a visibility function, partial generalizations of Eq.~\eqref{eqn:sudden_chi} have been adopted, either by convolving $\delta\chi[\eta,(\eta_0-\eta)\versor{n}]$ with the visibility function itself \cite{capparelli2020cosmic}, or by adopting a tomographic approach to separate the recombination contribution from the reionization one \cite{greco2023probing}. Nevertheless, for the first time, in this paper, we propose the most general treatment for anisotropic cosmic birefringence, which directly solves the modified Boltzmann equation, in analogy with what has been done in the literature for the isotropic counterpart. Our approach can completely characterize anisotropic birefringence as a second-order effect in perturbation theory, whose redshift evolution is now taken into account.

The structure of the paper is organized as follows. In Sec.~\ref{sec:Contours}, we present a chi-squared analysis for isotropic cosmic birefringence, and we show that an almost constant-in-time homogeneous field $\chi_{0}$ does not automatically imply that the same behavior is also shared with its inhomogeneous perturbations $\delta\chi$. In Sec.~\ref{sec:Boltzmann}, we describe our generalized Boltzmann treatment of anisotropic cosmic birefringence and we derive the analytical expression for the modified CMB angular power spectra. We conclude in Sec.~\ref{sec:Conclu} with a summary of our main findings and suggestions for future work. In App~\ref{app:other_potential} we have also extended our discussion to a different axion potential with respect to that considered in the main body of the paper.

\section{\label{sec:Contours}Constraining the Axion Parameters}
As described by Eq.~\eqref{eqn:alpha_0}, the value of the birefringence angle is related to the difference in the value of the field $\chi$ between the moment of the photon's observation and emission. We now focus on the case of isotropic birefringence, for which a homogeneous field $\chi=\chi_0(\eta)$ can induce a birefringence effect parameterized by an isotropic angle $\alpha=\alpha_0$. In particular, we want to find the set of parameters for the model defined in Eq.~\eqref{eqn:chi_action} that can explain $\alpha_0=(0.30\pm0.11)^{\circ}$ \cite{diego2022cosmic}. Since we are focusing now on the purely isotropic case, it is easy to derive the equation of motion for the axion $\chi_0$ by taking the functional derivative of Eq.~\eqref{eqn:chi_action} with respect to the scalar field\footnote{As shown in \cite{greco2023probing}, although the Chern-Simons term in Eq.~\eqref{eqn:L_chern_simons} is responsible for the birefringence mechanism, it gives no contribution to the equation of motion for $\chi$ at zero- and first-order in cosmological perturbation theory.}, and by working in the Friedmann-Lemaître-Robertson-Walker (FLRW) metric:
\begin{equation}
\label{eqn:chi_EOM}
\begin{split}
&\chi_0^{\prime\prime}+2\mathcal{H}\chi_0^{\prime}+a^2m^2_{\chi}\chi_0=0,
\end{split}
\end{equation}
where $a$ is the scale factor of the Universe and $\mathcal{H}\equiv a^{\prime}/a$ is the conformal Hubble parameter, and $\prime$ denotes the differentiation with respect to conformal time. The solution of Eq.~\eqref{eqn:chi_EOM} can be found once the initial conditions for the field are set. As already remarked e.g. in \cite{nakatsuka2022cosmic}, we can divide both sides of Eq.~\eqref{eqn:chi_EOM} by the initial value of the field $\chi_0^{\text{ini}}$, so that we can trade our mathematical problem with solving the same differential equation but for the field $\xi\equiv\chi_0(\eta)/\chi_0^{\text{ini}}$. In this way, one of the initial conditions in such a differential equation is automatically fixed as $\xi^{\text{ini}}=1$. Moreover, as already done in \cite{greco2023probing}, here we choose the axion initial velocity to be $\left(\mathrm{d}\chi_0/\mathrm{d}\eta\right)^{\text{ini}}=\left(\mathrm{d}\xi/\mathrm{d}\eta\right)^{\text{ini}}=0$, which is equivalent to impose the axion behaving as pure dark energy in its early evolution. Let us mention that, similarly to what has been done in \cite{nakatsuka2022cosmic, lee2022probing, greco2023probing}, to solve Eq.~\eqref{eqn:chi_EOM} we have assumed that the energy density of the axion is small so that we can take the field to be decoupled from the rest of the components in $\mathcal{H}$.

Now, we perform a chi-squared analysis to compare our theoretical prediction in the sudden recombination approximation,
\begin{equation}
\label{eqn:alpha_theory}
\alpha_0(\lambda\chi_0^{\text{ini}}/f,m_{\chi})=\frac{\lambda\chi_0^{\text{ini}}}{2f}\left[\xi(\eta_0)-\xi(\eta_{\text{rec}})\right]\big|_{m_{\chi}},
\end{equation}
with the \textit{Planck} result. The subscript $m_{\chi}$ on the right-hand side of Eq.~\eqref{eqn:alpha_theory} labels the fact that the dependence of $\alpha_0$ on the axion mass is ``hidden'' in the field variation, since different masses imply a different time evolution (see e.g. \cite{nakatsuka2022cosmic, lee2022probing, greco2023probing}). We have adopted a Bayesian approach and computed the posterior probability, identifying the best-fit parameters with those that maximize the posterior itself, by assuming a Gaussian likelihood and a uniform prior,
\begin{equation}
\label{eqn:posterior}
\begin{split}
\mathsf{P}&\left(\alpha_{\text{best}},\sigma|\lambda\chi_0^{\text{ini}}/f,m_{\chi},\right)\propto\\
&\quad\propto\frac{\exp\left\{-\frac{1}{2\sigma^2}\left[\alpha_{\text{best}}-\alpha_0\left(\lambda\chi_0^{\text{ini}}/f,m_{\chi}\right)\right]^2\right\}}{\sqrt{2\pi\sigma^2}},
\end{split}
\end{equation}
where $\alpha_{\text{best}}=0.30^{\circ}$ and $\sigma=0.11^{\circ}$. 

\begin{figure*}[htbp]
\centering
	
	
	
\subfloat[Contour plot of $\alpha_0$ for $\lambda\chi_0^{\text{ini}}/f\le0$.]{%
\includegraphics[width=0.5\textwidth]{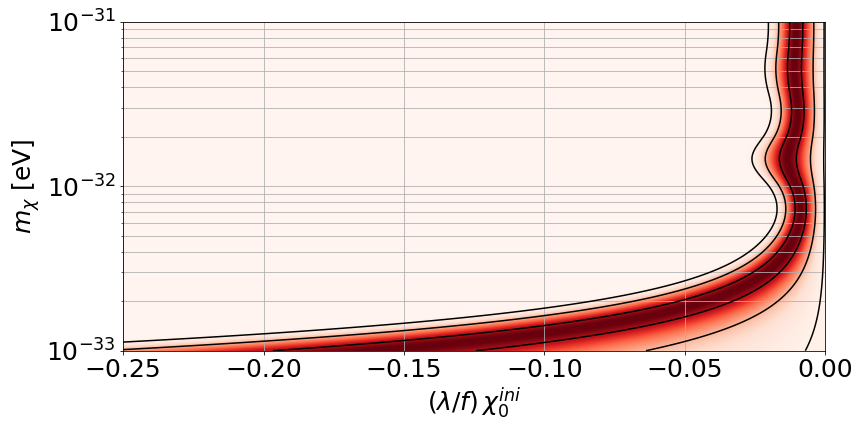}%
\label{fig:subplot3}%
}\hfill
\subfloat[Contour plot of $\alpha_0$ for $\lambda\chi_0^{\text{ini}}/f\ge0$.]{%
\includegraphics[width=0.5\textwidth]{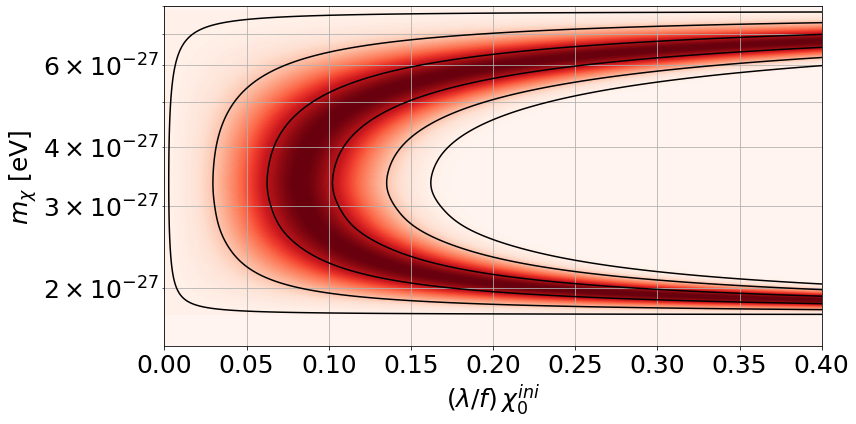}%
\label{fig:subplot4}%
}
	
\caption{Dependence of the isotropic birefringence angle on the axion mass $m_{\chi}$ and the coupling parameter $\lambda/f$. The numerical computation has been performed by running several simulations of the axion field dynamics, by setting the initial velocity of the axion field equal to zero, and by taking the fiducial values of the $\Lambda$CDM parameters provided in \cite{aghanim2020planck}. Figs.~\ref{fig:subplot3}-\ref{fig:subplot4} show the likelihood as a colormap in the parameter space, for positive and negative values of $\lambda\chi_0^{\text{ini}}/f$, respectively. The darkest red region between the innermost black curves explains the CMB birefringence angle at $1\sigma$ C.L. reported by \textit{Planck} PR4 \cite{diego2022cosmic}.}
\label{fig:alpha_0}
\end{figure*}

To evaluate $\alpha_0$ and the posterior probability $\mathsf{P}$ we have used a modified version of the Boltzmann code \texttt{CLASS} \cite{blas2011cosmic}, in which the dynamics of the axion field is implemented. Since, as mentioned before, the dependence of $\alpha_0$ on $m_{\chi}$ is not analytical, we have performed a large number of simulations running our version of \texttt{CLASS}, and then we have numerically interpolated the predicted theoretical birefringence angle over a grid of values for the axion mass and the product between the coupling parameter and the field's initial value. We have then obtained contour plots in the parameter space, by evaluating Eq.~\eqref{eqn:posterior} on such a grid, as shown in Figs.~\ref{fig:subplot3}-\ref{fig:subplot4}, where we have found two different regions, according to the sign of the dimensionless quantity $\lambda\chi_0^{\text{ini}}/f$.

\begin{figure*}
\centering
\includegraphics[width=\textwidth]{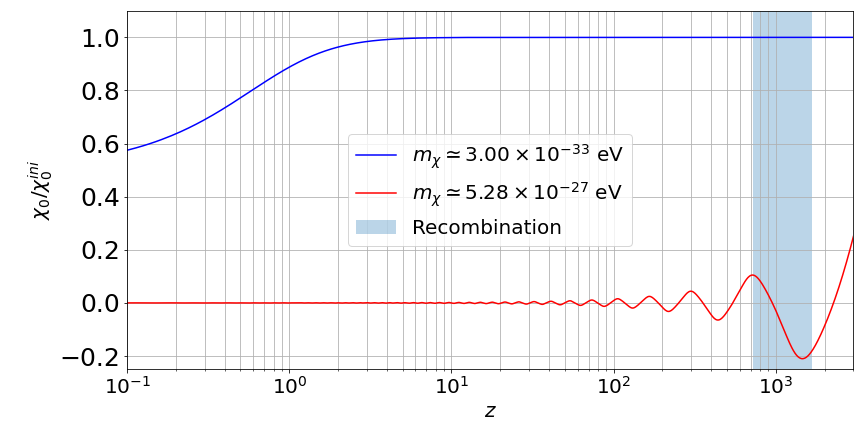}
\caption{Redshift evolution of the ratio between the homogeneous scalar field $\chi_0$ and its initial value $\chi_0^{\text{ini}}$ for the two best-fit masses resulting from our chi-squared analysis. The colored region is the range of redshifts corresponding to recombination evaluated using the \texttt{HyRec} algorithm \cite{Ali-Haimoud:2010tlj, Ali-Haimoud:2010hou, Lee:2020obi}. As in Fig.~\ref{fig:alpha_0}, the numerical computation has been performed by setting the initial velocity of the axion field equal to zero, and for the fiducial values of the $\Lambda$CDM parameters provided in \cite{aghanim2020planck}.}
\label{fig:chi_0_best}
\end{figure*}

As a result, the best-fit parameter $\lambda\chi_0^{\text{ini}}/f$ and the best-fit masses are estimated to be:
\begin{itemize}
\item $\lambda\chi_0^{\text{ini}}/f\simeq-0.02$ and $m_{\chi}\simeq\SI{3.00e-33}{\electronvolt}$;
\item $\lambda\chi_0^{\text{ini}}/f\simeq0.12$ and $m_{\chi}\simeq\SI{5.28e-27}{\electronvolt}$,
\end{itemize}	
according to the two different regions of the parameter space that we have found to be most consistent with the \textit{Planck} result. Different values for $m_{\chi}$ imply different evolution of $\chi$, as shown in Fig.~\ref{fig:chi_0_best}. Indeed, for $m_{\chi}\simeq\SI{3.00e-33}{\electronvolt}$ (which is associated with a $\lambda\chi_0^{\text{ini}}/f\le0$), the field seems to be constant during the epoch of recombination, whereas the opposite occurs for $m_{\chi}\simeq\SI{5.28e-27}{\electronvolt}$ (which is instead associated with $\lambda\chi_0^{\text{ini}}/f\ge0$).

The existence of two very different regions in the parameter space yielding a value of $\alpha_0$ consistent with $\textit{Planck}$ can be explained as follows: by looking at Eq.~\eqref{eqn:alpha_theory} it is clear that to get a small positive value close to $\alpha_{\text{best}}=0.30^{\circ}$ the product between $\lambda\chi_0^{\text{ini}}/f$ and the time variation of $\xi(\eta)$ must be positive as well, and this can happen when both these two quantities are positive or negative. By direct inspection of Fig.~\ref{fig:chi_0_best}, we can easily see that in the $m_{\chi}\sim\SI{e-33}{\electronvolt}$, the evolution of the axion field is really slow and $\xi(\eta_0)<\xi(\eta_{\text{rec}})$, so that a negative sign of $\lambda\chi_0^{\text{ini}}/f$ is needed to provide a positive angle. Instead, for  $m_{\chi}\sim\SI{e-27}{\electronvolt}$, the field evolution is faster than in the previous case, so that, because of the strong oscillatory behavior, it is possible to achieve $\xi(\eta_0)>\xi(\eta_{\text{rec}})$ in specific moments leaving the possibility to have a positive $\lambda\chi_0^{\text{ini}}/f$. However, let us note that in the latter situation, as can be seen by looking at Fig.~\ref{fig:subplot4}, the region of the parameter space consistent with $\alpha_{\text{best}}$ is smaller with respect to that of the former case, presented instead in Fig.~\ref{fig:subplot3}. The reason is that to get $\alpha_{\text{best}}$ starting from Eq.~\eqref{eqn:alpha_theory} exploiting a positive $\lambda\chi_0^{\text{ini}}/f$ and an oscillatory behavior, one needs to fine-tune the value of the axion mass, to reach exactly the desired amount of isotropic cosmic birefringence. On the contrary, in the case of a negative $\lambda\chi_0^{\text{ini}}/f$, it is possible to find a more stable solution with an almost constant field during the matter-dominated epoch. Now, if the axion field is not constant in time during recombination, an important consequence is that one should account for the birefringence effect for each photon emitted in that range of redshifts, i.e. the one associated with its own emission's redshift. This is something that has been already pointed out in the literature, and such an issue has been solved by including the birefringence effect directly in the polarized Boltzmann equation for CMB photons \cite{murai2022isotropic, loeb1996faraday, liu2006effect, finelli2009rotation, gubitosi2014including, galaverni2023redshift}. Before proceeding, let us just remember that our analysis has been adopted for the $V(\chi)\equiv m^2_{\chi}\chi^2/2$ potential, and so with a different potential, the situation could in principle change. However, we have found a similar result for an Early Dark Energy (EDE) potential (see App.~\ref{app:other_potential}). Let us highlight that in our analysis we have neglected the astrophysical constraints, but we are authorized to do that because the orders of magnitude at which they become relevant are well outside of the regions explored in our contour plots, as shown in \cite{fujita2021detection}. Moreover, let us note that in fact a solution associated with an ultra-light axion field appears also in \cite{fujita2021detection}, where the authors constrained models of cosmic birefringence by just relying on the absolute value of the isotropic angle. 
However, it is essential to predict the sign of the isotropic birefringence angle. Indeed, the analysis performed in \cite{fujita2021detection} is insensitive to the sign of $\chi_0(\eta_0)-\chi(\eta_{\text{rec}})$. In order to recover the amplitude \textit{and} the sign of the isotropic angle observed in \cite{diego2022cosmic}, with an ultra-light axion field it is necessary to set negative initial conditions for the homogeneous axion-field itself. Instead, by considering a positive initial condition, such a value of the isotropic angle can only be explained with a much larger mass, as shown above. As previously mentioned, we have identified two distinct regions in the parameter space that can reproduce the observed cosmic birefringence angle, rather than just two specific parameter values. Indeed, with our simple chi-squared analysis, our goal is to highlight the existence of these two regions, with the ‘best-fit solutions’ serving as representative examples of the behavior of the solutions within these regions.

\begin{figure*}[htbp]
\centering
\includegraphics[width=\textwidth]{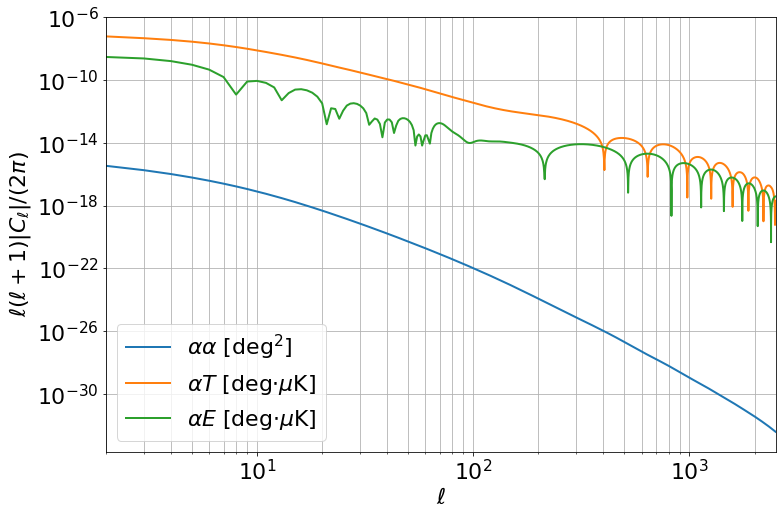}%
\caption{Angular power spectra involving anisotropic cosmic birefringence for the first one of the two sets of best-fit parameters resulting from the chi-squared analysis we performed for $\alpha_0$. The numerical computation has been performed by setting the axion mass equal to $\SI{3.0e-33}{\electronvolt}$, the parameter $\lambda/f$ equal to $\SI{1.6e-20}{\giga\electronvolt^{-1}}$, the initial value of the axion field $\chi_{0}^{\text{ini}}$ equal to $-m_{Pl}/2$ (with $m_{Pl}=M_{Pl}/\sqrt{8\pi}$ being the reduced Planck mass), so that, according to our parameter estimation, $\lambda/f\chi_{0}^{\text{ini}}\simeq-0.02$. Moreover, the initial velocity of the axion field has been set equal to zero, and we have used the fiducial values of the $\Lambda$CDM parameters provided in \cite{aghanim2020planck}.}
\label{fig:power_spectrum_aniso}
\end{figure*}

As mentioned in Sec.~\ref{sec:Intro}, one of our aims is to find the predicted amount of anisotropic cosmic birefringence for that set of parameters that predicts the level of the isotropic one found from \textit{Planck} data. First of all, we now expand the anisotropic angle defined in Eq.~\eqref{eqn:sudden_chi} over the celestial sphere as
\begin{equation}
\alpha_{\ell m}=\int\mathrm{d}^2\hat{n}\,Y_{\ell m}^*(\versor{n})\,\delta\alpha(\versor{n}),
\end{equation}
so that by performing a treatment similar to that adopted e.g. in \cite{caldwell2011cross, capparelli2020cosmic, zhai2020effects, greco2023probing}, we can compute the angular auto- and cross-correlations of anisotropic cosmic birefringence as
\begin{equation}
C_{\ell}^{\alpha M}=\frac{1}{2\ell+1}\sum_{mm'}\expval{\alpha^*_{\ell m} M_{\ell m'}},
\end{equation}
where $M$ can be $\alpha$, the CMB temperature $T$ or $E$ polarization modes. The results are shown in Fig.~\ref{fig:power_spectrum_aniso}: by looking at them we can notice that for such a set of parameters, the predicted amount of anisotropic birefringence is expected to be well below the current observational constraints reported e.g. from \textit{Planck} data \cite{bortolami2022planck}. This implies that (in this model) the anisotropic contribution to cosmic birefringence is predicted to be extremely subdominant with respect to the isotropic one.
Moreover, it is then clear that if a future observation detects an anisotropic signal larger than that predicted by our model, this will be evident proof that the model under consideration has to be ruled out. In this sense, we can regard anisotropic birefringence as a potentially further observable useful for consistency checks and sensitive to scientific falsifiability.

However, although we have discovered a region in the parameter space that seems to indicate that the approximation involving a constant isotropic birefringence angle is justified, this cannot be automatically extended to its anisotropic counterpart. The reason is simply that, according to Eq.~\eqref{eqn:sudden_chi} and Eq.~\eqref{eqn:alpha_theory}, the isotropic birefringence angle $\alpha_0$ and the anisotropic one $\delta\alpha$ are related to the dynamics of the homogeneous field $\chi_0$ and its associated perturbation $\delta\chi$, respectively, but these two fields \textit{obey different equations of motions}.  In the following, we are going to show that $\delta\chi$ can present a non-trivial time evolution even if $\chi_{0}$ is almost constant in time. As we know, the equation of motion for $\chi_0$ is Eq.~\eqref{eqn:chi_EOM}, whereas that for $\delta\chi$ can be obtained by simply varying the action in Eq.~\eqref{eqn:chi_action} with respect to $\delta\chi$ by working at linear order in perturbation theory. For instance, in the Newtonian conformal gauge, it can be easily evaluated to be \cite{caldwell2011cross, capparelli2020cosmic, greco2023probing},
\begin{equation}
\label{eqn:delta_chi}
\begin{split}
\delta\chi^{\prime\prime}+2\mathcal{H}\delta\chi^{\prime}&+\left(k^2+a^2m^2_{\chi}\right)\delta\chi=\\
&\qquad\quad=\chi_{0}^{\prime}(3\Phi^{\prime}+\Psi^{\prime})-2a^2m^2_{\chi}\chi_0\Psi,
\end{split}
\end{equation}
where we also moved to the Fourier space. Now, let us try to understand how $\delta\chi$ evolves in time during the matter-dominated epoch, i.e. when almost all the CMB photons have been emitted. Since the two Newtonian potentials are constant in time and $a\simeq\Omega_{m0}H_0^2\eta^2/2$ during that epoch, the equation above reduces to
\begin{equation}
\label{eqn:DifferentialRoot}
\delta\chi^{\prime\prime}+\frac{4}{\eta}\delta\chi^{\prime}+\left(k^2+A^2\eta^4\right)\delta\chi=-2A^2\chi_0\Psi(k)\eta^4,
\end{equation}
where we have defined $A\equiv m_{\chi}\Omega_{m0}H_0^2/2$. Hence, we have found a second-order linear ordinary differential equation that does not admit a solution that can be expressed in terms of elementary functions. However, our purpose here is not to find an analytical solution, but just to show that a constant $\chi_{0}$ does not imply a constant in time $\delta\chi$. For this reason, let us make a further reasonable simplification: according to our parameter estimation, the background field $\chi_{0}$ is almost constant in time during matter-domination when its mass is about $m_{\chi}\simeq\SI{3e-33}{\electronvolt}$, as can be seen by looking at Fig.~\ref{fig:chi_0_best}, so that we can safely disregard the $A^2\eta^4\delta\chi\ll(4/\eta)\delta\chi^{\prime}$ contribution\footnote{Indeed, by recalling our definitions, we can roughly estimate the ratio between $A^2\eta^4\delta\chi$ and $(4/\eta)\delta\chi^{\prime}$ to be $\sim(m_{\chi}/H)^2$, and we checked that in the $\Lambda$CDM model the Hubble parameter is such that $m_{\chi}\ll H$ for almost the entire duration of the matter-dominated epoch, if $m_{\chi}\simeq\SI{3e-33}{\electronvolt}$.} on the left-hand side of Eq.~\eqref{eqn:DifferentialRoot}. After doing that, the solution of the differential equation can be easily found to be
\begin{widetext}
\begin{equation}
\begin{split}
\delta\chi(\eta,k)&=\frac{2A^2\chi_{0}\Psi(k)}{k^6}\left(28k^2\eta^2-k^4\eta^4-280\right)+\sqrt{\frac{2}{\pi k\eta^4}}\Bigg[\left(\frac{\mathcal{C}_1}{k\eta}-\mathcal{C}_2\right)\sin(k\eta)-\left(\mathcal{C}_1+\frac{\mathcal{C}_2}{k\eta}\right)\cos(k\eta)\Bigg],
\end{split}
\end{equation}
\end{widetext}
where $\mathcal{C}_1$ and $\mathcal{C}_2$ are integration constants. The approximation we made is valid as long as the axion mass $m_{\chi}$ is small, and in the limit $m_{\chi}\to0$, the homogeneous axion $\chi$ field becomes completely constant (see e.g. \cite{greco2023probing}). On the contrary, we can see that, even in such a regime, $\delta\chi$ still encodes a non-trivial time dependence.

This proves our statement about the fact that, although our parameter estimation for isotropic birefringence has selected a region of the parameter space for which $\chi_{0}$ does not evolve significantly during matter-domination, we cannot simply extend this result also to $\delta\chi$. Therefore, our goal is now to propose a new treatment that can be seen as the most complete generalization of the current state of the art concerning cosmic birefringence, in which the redshift evolution of both the pseudoscalar field inducing the rotation and its inhomogeneous perturbations are taken into account.

\section{\label{sec:Boltzmann}Generalized Boltzmann Equation for Cosmic Birefringence}
Let us start with the standard polarized Boltzmann equation for CMB photons \cite{kosowsky1995cosmic,hu1997cmb,zaldarriaga1997all},
\begin{equation}
\label{eqn:Boltzmann_standard}
\begin{split}
\left[\frac{\partial}{\partial\eta}-i\mathbf{k}\cdot\versor{n}-\frac{\mathrm{d}\tau}{\mathrm{d}\eta}\right]&\,_{\pm}\Delta_{P}(\eta,\mathbf{k},\versor{n})=\,_{\pm}\mathcal{S}_P(\eta, \mathbf{k},\versor{n}),
\end{split}
\end{equation}
where the optical depth is defined as
\begin{equation}
\tau(\eta)\equiv\sigma_T\int_{\eta}^{\eta_0}\mathrm{d}\tilde{\eta}\,n_e(\tilde{\eta})a(\tilde{\eta}),
\end{equation}
$n_e$ being the free electrons' number density and $\sigma_T$ being the Thomson cross-section, respectively. $_{\pm}\mathcal{S}_{P}$ instead is the polarization's source function, encoding the contributions due to Thomson scattering \cite{liu2002polarization},
\begin{equation}
\label{eqn:pol_source}
_{\pm}\mathcal{S}_{P}=-\frac{\mathrm{d}\tau}{\mathrm{d}\eta}\sum_{\lambda=-2}^{2}\sqrt{\frac{6\pi}{5}}\,_{\pm2}Y_{2,\lambda}(\versor{n})\Pi_{\lambda}(\eta,\mathbf{k}),
\end{equation}
where $\Pi_{\lambda}$ is the polarization source. The quantity $_{\pm}\Delta_{P}$ is the Fourier transform of the linear combination of Stokes parameters $Q$ and $U$,
\begin{equation}
\label{eqn:def_transfer}
\left[Q\pm iU\right](\eta, \mathbf{x}, \versor{n})=\int\frac{\mathrm{d}^3k}{(2\pi)^3}\,_{\pm}\Delta_{P}(\eta,\mathbf{k},\versor{n})\,e^{i\mathbf{k}\cdot\mathbf{x}}.
\end{equation} 
To include cosmic birefringence in Eq.~\eqref{eqn:Boltzmann_standard}, we consider Eq.~\eqref{eqn:Stokes} as valid at each redshift. By differentiating it with respect to conformal time we get
\begin{equation}
\begin{split}
\frac{\mathrm{d}}{\mathrm{d}\eta}\left[Q\pm iU\right]&(\eta, \mathbf{x}, \versor{n})=\\
&=\left[\frac{\mathrm{d}\mathbf{x}}{\mathrm{d}\eta}\cdot\grad+\frac{\partial}{\partial\eta}\right]\left[Q\pm iU\right](\eta,\mathbf{x},\versor{n}),\\
&\\
&
\end{split}
\end{equation}
which can be rewritten as
\begin{equation}
\label{eqn:Fourier}
\begin{split}
\frac{\mathrm{d}}{\mathrm{d}\eta}&\left[Q\pm iU\right](\eta, \mathbf{x}, \versor{n})=\\
&=\int\frac{\mathrm{d}^3k}{(2\pi)^3}\left[\frac{\partial}{\partial\eta}-i\mathbf{k}\cdot\versor{n}\right]\,_{\pm}\Delta_{P}(\eta,\mathbf{k},\versor{n})\,e^{i\mathbf{k}\cdot\mathbf{x}}.\\
&
\end{split}
\end{equation}
We can define now a birefringence angle for any photon emitted, and Eq.~\eqref{eqn:alpha_0} generalizes to
\begin{equation}
\begin{split}
\alpha(\eta,\mathbf{x})&\equiv\alpha_0(\eta)+\delta\alpha(\eta,\mathbf{x}),\\
\end{split}
\end{equation}
where $\alpha_0$ and $\delta\alpha$ are the isotropic and anisotropic angles:
\begin{align}
\alpha_0(\eta)&\equiv\frac{\lambda}{2f}\left[\chi_0(\eta_0)-\chi_0(\eta)\right],\\
\delta\alpha(\eta,\mathbf{x})&\equiv\frac{\lambda}{2f}\left[\delta\chi(\eta_0,\mathbf{x}_0)-\delta\chi(\eta,\mathbf{x})\right].
\end{align}

Since our goal here is to find how cosmic birefringence impacts the Boltzmann equation, let us assume just for now that CMB polarization is only affected by the presence of the axion, so that we will include the contribution from the source function $_{\pm}\mathcal{S}_P$ only later. By starting from Eq.~\eqref{eqn:Stokes}, we can easily compute the total conformal time derivative of the Stokes parameters as
\begin{equation}
\begin{split}
&\frac{\mathrm{d}}{\mathrm{d}\eta}\left[Q\pm iU\right](\eta,\mathbf{x},\versor{n})=\\
&\quad=\mp2i\left[Q\pm iU\right](\eta,\mathbf{x},\versor{n})\frac{\mathrm{d}}{\mathrm{d}\eta}\left[\alpha_0(\eta)+\delta\alpha(\eta,\mathbf{x})\right].
\end{split}
\end{equation}
We have already Fourier-transformed the left-hand side of the equation above in Eq.~\eqref{eqn:Fourier}, so by doing the same for the right-hand side we obtain 

\begin{widetext}
\begin{equation}
\label{eqn:long_Fourier}
\begin{split}
\int\frac{\mathrm{d}^3k}{(2\pi)^3}\left[\frac{\partial}{\partial\eta}-i\mathbf{k}\cdot\versor{n}\right]\,_{\pm}\Delta_{P}(\eta,\mathbf{k},\versor{n})e^{i\mathbf{k}\cdot\mathbf{x}}&=\mp2i\Bigg\{\frac{\mathrm{d}\alpha_0}{\mathrm{d}\eta}\int\frac{\mathrm{d}^3k}{(2\pi)^3}\,_{\pm}\Delta_{P}(\eta,\mathbf{k},\versor{n})e^{i\mathbf{k}\cdot\mathbf{x}}\\
&\,\,\,\,+\int\frac{\mathrm{d}^3k_1\,\mathrm{d}^3k_2}{(2\pi)^6}\,_{\pm}\Delta_{P}(\eta,\mathbf{k}_1,\versor{n})\left[\frac{\partial}{\partial\eta}-i\mathbf{k}_2\cdot\versor{n}\right]\delta\alpha(\eta,\mathbf{k}_2)\,e^{i(\mathbf{k}_1+\mathbf{k}_2)\cdot\mathbf{x}}\Bigg\}.\\
&
\end{split}
\end{equation}
\end{widetext}

It is now the moment in which we add the contribution from the Thomson scattering. However, before doing that, let us expand the transfer function for CMB polarization and its source function at second-order in perturbation theory \cite{bartolo2006cosmic,bartolo2007cmb,bartolo2007cosmic, naruko2013second, pettinari2013intrinsic, saito2014geodesic, su2014formulating}, for a reason which will be clarified very soon:
\begin{align}
\,_{\pm}\Delta_{P}(\eta,\mathbf{k},\versor{n})&=\,_{\pm}\Delta^{(1)}_{P}(\eta,\mathbf{k},\versor{n})+\,_{\pm}\Delta^{(2)}_{P}(\eta,\mathbf{k},\versor{n}),\\
\,_{\pm}\mathcal{S}_{P}(\eta,\mathbf{k},\versor{n})&=\,_{\pm}\mathcal{S}_{P}^{(1)}(\eta,\mathbf{k},\versor{n})+\,_{\pm}\mathcal{S}_{P}^{(2)}(\eta,\mathbf{k},\versor{n}),\\
\Pi_{m}(\eta,\mathbf{k})&=\Pi_{m}^{(1)}(\eta,\mathbf{k})+\Pi_m^{(2)}(\eta,\mathbf{k}).
\end{align} 
Indeed, if we take the inverse Fourier transform of Eq.~\eqref{eqn:long_Fourier} and we plug this result in Eq.~\eqref{eqn:Boltzmann_standard}, we find two generalized Boltzmann equations: the former is valid at first-order in perturbation theory,
\begin{equation}
\label{eqn:first_order}
\begin{split}
\left[\frac{\partial}{\partial\eta}-i\mathbf{k}\cdot\versor{n}-\frac{\mathrm{d}\tau}{\mathrm{d}\eta}\pm 2i\frac{\mathrm{d}\alpha_0}{\mathrm{d}\eta}\right]&\,_{\pm}\Delta^{(1)}_{P}(\eta,\mathbf{k},\versor{n})=\\
&\quad=\,_{\pm}\mathcal{S}_{P}^{(1)}(\eta,\mathbf{k},\versor{n}),\\
&
\end{split}
\end{equation}
whereas the latter at second-order, encoding an extra-term due to anisotropic cosmic birefringence,
\begin{widetext}
\begin{equation}
\label{eqn:second_order}
\begin{split}
&\left[\frac{\partial}{\partial\eta}-i\mathbf{k}\cdot\versor{n}-\frac{\mathrm{d}\tau}{\mathrm{d}\eta}\pm 2i\frac{\mathrm{d}\alpha_0}{\mathrm{d}\eta}\right]\,_{\pm}\Delta^{(2)}_{P}(\eta,\mathbf{k},\versor{n})=\,_{\pm}\mathcal{S}_{P}^{(2)}(\eta,\mathbf{k},\versor{n})\\
&\hspace{150pt}\mp2i\int\frac{\mathrm{d}^3k_1\,\mathrm{d}^3k_2}{(2\pi)^3}\delta^{(3)}(\mathbf{k}-\mathbf{k}_1-\mathbf{k}_2)\,_{\pm}\Delta^{(1)}_{P}(\eta,\mathbf{k}_1,\versor{n})\left[\frac{\partial}{\partial\eta}-i\mathbf{k}_2\cdot\versor{n}\right]\delta\alpha(\eta,\mathbf{k}_2).
\end{split}
\end{equation}	
\end{widetext}
Now it is clear why we have adopted a perturbative expansion of the relevant quantities. As can be seen by looking at Eqs.~\eqref{eqn:first_order}-\eqref{eqn:second_order}, isotropic cosmic birefringence affects CMB polarization at any order in perturbation theory, whereas anisotropic cosmic birefringence does it starting from the second-order. This is obvious, since the inhomogeneous fluctuation of the axion field is, in fact, an extra cosmological perturbation. To solve the two differential equations, we firstly integrate along the line-of-sight for a generic final time $\eta$ both sides of Eq.~\eqref{eqn:first_order},
\begin{equation}
\begin{split}
\,_{\pm}\Delta_{P}^{(1)}(\eta,\mathbf{k},\versor{n})=\int_{0}^{\eta}&\mathrm{d}\tilde{\eta}\,_{\pm}\mathcal{S}_{P}^{(1)}(\tilde{\eta},\mathbf{k},\versor{n})e^{i\mathbf{k}\cdot\versor{n}(\eta-\tilde{\eta})}\\
&e^{-\left[\tau(\tilde{\eta})-\tau(\eta)\right]}e^{\pm2i\left[\alpha_0(\tilde{\eta})-\alpha_0(\eta)\right]}.
\end{split}
\end{equation}
This procedure is standard in cosmological perturbation theory, and it is performed because we need such a quantity not only to find $_{\pm}\Delta_{P}^{(1)}(\eta_0,\mathbf{k},\versor{n})$ (by simply replacing $\eta$ with $\eta_0$), but also $_{\pm}\Delta_{P}^{(2)}(\eta_0,\mathbf{k},\versor{n})$, since, as can been seen by looking at Eq.~\eqref{eqn:second_order}, it depends on the first-order transfer function. Therefore, we find
\begin{equation}
\label{eqn:Boltzmann_iso}
\begin{split}
\,_{\pm}\Delta_{P}^{(1)}(\eta_0,\mathbf{k},\versor{n})=\int_{0}^{\eta_0}\mathrm{d}\eta\,_{\pm}\mathcal{S}_{P}^{(1)}&(\eta,\mathbf{k},\versor{n})e^{i\mathbf{k}\cdot\versor{n}(\eta_0-\eta)}\\
&\quad e^{-\tau(\eta)}e^{\pm2i\alpha_0(\eta)},
\end{split}
\end{equation}
where we have neglected the value of the optical depth today, and we have used that the isotropic birefringence angle for a photon emitted today is identically zero, according to the definition given in Eq.~\eqref{eqn:alpha_0}. Let us notice that Eq.~\eqref{eqn:Boltzmann_iso} is exactly the same formula used in \cite{murai2022isotropic, loeb1996faraday, liu2006effect, finelli2009rotation, gubitosi2014including, galaverni2023redshift}. Similarly, the second-order transfer function, after integrating by parts, reads
\begin{widetext}
\begin{equation}
\label{eqn:Boltzmann_aniso}
\,_{\pm}\Delta_{P}^{(2)}(\eta_0,\mathbf{k},\versor{n})=\int_{0}^{\eta_0}\mathrm{d}\eta\,e^{i\mathbf{k}\cdot\versor{n}(\eta_0-\eta)}e^{-\tau(\eta)}e^{\pm2i\alpha_0(\eta)}\left[\,_{\pm}\mathcal{S}_{P}^{(2)}(\eta,\mathbf{k},\versor{n})\pm2i\int\frac{\mathrm{d}^3q}{(2\pi)^3}\,\delta\alpha(\eta,\mathbf{k}-\mathbf{q})\,_{\pm}\mathcal{S}_{P}^{(1)}(\eta,\mathbf{q},\versor{n})\right].
\end{equation}
\end{widetext}
Eq.~\eqref{eqn:Boltzmann_iso} and Eq.~\eqref{eqn:Boltzmann_aniso} are the core of our generalized treatment of CMB polarization. Armed with these expressions, we can investigate how they are related to the more common approaches used in the literature, and what is the impact on the CMB power spectra.

\subsection{Recovering the Sudden Recombination Approximation}
Let us now show how it is possible to recover the formulas used up to now in the literature for cosmic birefringence in specific regimes. Indeed, as we are going to prove, our treatment is completely general and reduces to the standard one in the sudden recombination approximation, i.e. by assuming that all the CMB photons have been emitted at the same time during the recombination epoch. 

In order to see this, let us firstly substitute Eq.~\eqref{eqn:pol_source} in Eq.~\eqref{eqn:Boltzmann_aniso}, so that, according to Eq.~\eqref{eqn:def_transfer}, we can easily relate $_{\pm}\Delta_{P}^{(2)}$ to the second-order linear combination of Stokes parameters $Q$ and $U$ observed today ($\eta =\eta_0$) on Earth ($\mathbf{x}=\mathbf{x}_0=\mathbf{0}$) as
\begin{widetext}
\begin{equation}	
\label{eqn:eraSbagliata}
\begin{split}
\left[Q^{(2)}\pm iU^{(2)}\right](\eta_0,\mathbf{x}_0,\versor{n})&=\sqrt{\frac{6\pi}{5}}\sum_{\lambda=-2}^{2}\,_{\pm2}Y_{2,\lambda}(\versor{n})\int_{0}^{\eta_0}\mathrm{d}\eta\,g(\eta)e^{\pm2i\alpha_0(\eta)}\Bigg\{\int\frac{\mathrm{d}^3k}{(2\pi)^3}\,e^{i\mathbf{k}\cdot\versor{n}(\eta_0-\eta)}\Pi_{\lambda}^{(2)}(\eta,\mathbf{k})\\
&\qquad\qquad\qquad\qquad\qquad\qquad\qquad\quad\pm2i\int\frac{\mathrm{d}^3k}{(2\pi)^3}\,e^{i\mathbf{k}\cdot\versor{n}(\eta_0-\eta)}\left[\delta\alpha(\eta)\ast\Pi_{\lambda}^{(1)}(\eta)\right](\mathbf{k})\Bigg\},
\end{split}
\end{equation}
\end{widetext}
where we defined the definitions of the convolution product,
\begin{equation}
\begin{split}
&\left[\delta\alpha(\eta)\ast\Pi_{\lambda}^{(2)}(\eta)\right](\mathbf{k})\equiv\\
&\qquad\qquad\quad\equiv\int\frac{\mathrm{d}^3q}{(2\pi)^3}\,\delta\alpha(\eta,\mathbf{k}-\mathbf{q})\Pi_{\lambda}^{(2)}(\eta,\mathbf{q}),
\end{split}
\end{equation}
and the photons' visibility function,
\begin{equation}
\label{eqn:visibility}
g(\eta)\equiv-\left(\frac{\mathrm{d}\tau}{\mathrm{d}\eta}\right)e^{-\tau(\eta)}.
\end{equation} 
Let us stress that the term on the right-hand side of Eq.~\eqref{eqn:eraSbagliata} proportional to $\Pi_{\lambda}^{(2)}(\eta,\mathbf{k})$ has never been considered in the literature as far as the birefringence effect is concerned. 

We can now simplify Eq.~\eqref{eqn:eraSbagliata}, by exploiting the convolution theorem, which allows us to deal with the Fourier transform of the convolution in the last line so that we can write 
\begin{widetext}
\begin{equation}	
\label{eqn:convolution}
\begin{split}
\left[Q^{(2)}\pm iU^{(2)}\right](\eta_0,\mathbf{x}_0,\versor{n})&=\sqrt{\frac{6\pi}{5}}\sum_{\lambda=-2}^{2}\,_{\pm2}Y_{2,\lambda}(\versor{n})\int_{0}^{\eta_0}\mathrm{d}\eta\,g(\eta)e^{\pm2i\alpha_0(\eta)}\Bigg\{\Pi_{\lambda}^{(2)}[\eta,(\eta_0-\eta)\versor{n}]\\
&\qquad\qquad\qquad\qquad\qquad\qquad\qquad\quad\pm2i\delta\alpha[\eta,(\eta_0-\eta)\versor{n}]\Pi_{\lambda}^{(1)}[\eta,(\eta_0-\eta)\versor{n}]\Bigg\},
\end{split}
\end{equation}
\end{widetext}

Notice that up to now we have made no approximations, but if we assume the sudden recombination regime, i.e. we trade the photons' visibility function for a Dirac delta peaked at the recombination, then the time integral would be trivially computed leading to the following results:
\begin{widetext}
\begin{align}
\label{eqn:first-order}
\left[Q^{(1)}\pm iU^{(1)}\right](\versor{n})&=e^{\pm2i\alpha_0}\left[Q^{(1)}\pm iU^{(1)}\right]_{\text{wcb}}(\versor{n}),\\
\label{eqn:second-order}
\left[Q^{(2)}\pm iU^{(2)}\right](\versor{n})&=e^{\pm2i\alpha_0}\Bigg\{\left[Q^{(2)}\pm iU^{(2)}\right]_{\text{wcb}}(\versor{n})\pm2i\delta\alpha(\versor{n})\left[Q^{(1)}\pm iU^{(1)}\right]_{\text{wcb}}(\versor{n})\Bigg\},
\end{align}
\end{widetext}
where the Stokes parameters on the right-hand side are those that one could obtain without any birefringence effects, whereas the first-order expression for $(Q\pm iU)$ has been obtained by simply mimicking the procedure we adopted for the second-order one. As a consequence of the sudden recombination approximation, here $\alpha_0$ and $\delta\alpha(\versor{n})$ are the same defined in Eq.~\eqref{eqn:alpha_theory} and Eq.~\eqref{eqn:sudden_chi}, respectively, because $\delta\chi$ for $\eta=\eta_0$ gives rise to an unobservable monopole contribution.  If we sum together Eqs.~\eqref{eqn:first-order}-\eqref{eqn:second-order} we get
\begin{widetext}
\begin{equation}
\label{eqn:SRA}
\begin{split}
\left[Q\pm iU\right](\versor{n})&=e^{\pm2i\alpha_0}\Bigg\{\sum_{x=1}^{2}\left[Q^{(x)}\pm iU^{(x)}\right]_{\text{wcb}}(\versor{n})\pm2i\delta\alpha(\versor{n})\left[Q^{(1)}\pm iU^{(1)}\right]_{\text{wcb}}(\versor{n})+\mathcal{O}(\delta^3)\Bigg\},
\end{split}
\end{equation}
\end{widetext}
where $\mathcal{O}(\delta^3)$ denotes terms at third-order in perturbation theory. Indeed, we notice that Eq.~\eqref{eqn:SRA} matches exactly Eq.~\eqref{eqn:Stokes} expanded at second-order in perturbation theory once the full birefringence angle is decomposed into its isotropic and anisotropic parts.

Therefore, this proves that our generalized expression reduces to the standard ones \cite{greco2023probing, li2008cosmological, caldwell2011cross, zhao2014fluctuations, zhai2020effects, capparelli2020cosmic, takahashi2021kilobyte, greco2022cosmic, cai2022impact} in the sudden recombination approximation. Similarly, if in Eq.~\eqref{eqn:convolution} we substitute the photon visibility function with a series of Dirac deltas associated with the peaks of the original $g(\eta)$, we could take into account also the contribution from the reionization epoch:
\begin{equation}
\label{eqn:tomographic_approximation}
g(\eta)\simeq g_{\text{rec}}\delta(\eta-\eta_{\text{rec}})+g_{\text{rei}}\delta(\eta-\eta_{\text{rei}}),
\end{equation} 
with $g_{\text{rec}}\gg g_{\text{rei}}$. In such a case we would recover exactly the results of a tomographic approach \cite{sherwin2021cosmic,nakatsuka2022cosmic,lee2022probing,greco2023probing}, i.e.
\begin{equation}
	\begin{split}
		&\left[Q\pm i U\right](\versor{n})=\\
		&\,=\sum_{c\,=\,\text{rec, rei}}e^{\pm2i\left[\alpha_0(\eta_c)+\delta\alpha(\eta_c,\versor{n})\right]}\left[Q_c\pm iU_c\right]_{\text{wcb}}(\versor{n}).
	\end{split}
\end{equation}

\subsection{Harmonic Expansion of CMB Polarization}
Once we have proved that our treatment is mathematically consistent with the current state-of-the-art about cosmic birefringence, it is time to find the expression of the CMB angular power spectra. To do that, it is convenient to come back to $_{\pm}\Delta_{P}$ and adopt a more efficient notation so that we have to do the same procedure for all the perturbation orders just once. Therefore, let us now define the two following quantities:
\begin{align}
\label{eqn:mathcal_T1}
_{\pm}\mathcal{T}_{\lambda}^{(1)}(\eta,\mathbf{k})&\equiv e^{\pm2i\alpha_0(\eta)}\Pi_{\lambda}^{(1)}(\eta,\mathbf{k}),
\end{align}
and similarly
\begin{align}
\label{eqn:mathcal_T2}
_{\pm}\mathcal{T}_{\lambda}^{(2)}(\eta,\mathbf{k})&\equiv e^{\pm2i\alpha_0(\eta)}\Bigg\{\Pi_{\lambda}^{(2)}(\eta,\mathbf{k})\nonumber\\
&\qquad\qquad\pm2i\left[\delta\alpha(\eta)	\ast\Pi_{\lambda}^{(1)}(\eta)\right](\mathbf{k})\Bigg\}.
\end{align}
Indeed, if we replace Eq.~\eqref{eqn:pol_source} in Eqs.~\eqref{eqn:Boltzmann_iso}-\eqref{eqn:Boltzmann_aniso}, we can then write a compact expression valid for any perturbative order $x=1,2$:
\begin{widetext}
\begin{equation}
\label{eqn:my_transfer}
\,_{\pm}\Delta_{P}^{(x)}(\eta_0,\mathbf{k},\versor{n})=\sqrt{\frac{6\pi}{5}}\int_{0}^{\eta_0}\mathrm{d}\eta\,e^{i\mathbf{k}\cdot\versor{n}(\eta_0-\eta)}g(\eta)\sum_{\lambda=-2}^{2}\,_{\pm2}Y_{2,\lambda}(\versor{n})\,_{\pm}\mathcal{T}_{\lambda}^{(x)}(\eta,\mathbf{k}).
\end{equation}
\end{widetext}
Eq.~\eqref{eqn:my_transfer} is the main result of this section, and we have put it in such a specific form because now the mathematical computation becomes less challenging since it has the same form of the standard transfer function of CMB polarization: for instance, it looks like exactly Eq.~(14) of \cite{liu2002polarization}. Indeed, we can appreciate that cosmic birefringence affects CMB polarization as a modification of the transfer function $_{\pm}\mathcal{T}_{\lambda}^{(x)}(\eta, \mathbf{k})$, and this occurs because cosmic birefringence is a propagation effect. As previously shown, Eq.~\eqref{eqn:my_transfer} yields the standard formalism of cosmic birefringence when assuming the sudden recombination approximation. However, it is easy to show that the same happens also when the birefringence angle is independent of the photons' emission time, which is the case occurring when the axion field is constant in time during the matter-dominated epoch.
 
To test our generalized treatment of cosmic birefringence, let us compute the CMB angular power spectra. First of all, let us notice that the dependence of $_{\pm}\Delta_P^{\,(x)}$ on $\versor{n}$ is encoded in $_{\pm2}Y_{2,\lambda}(\versor{n})$ but also in the complex exponential. Then, for our purposes it is convenient to move to the multipole space, by evaluating the following harmonic transform:
\begin{equation}
\label{eqn:harmonic}
\begin{split}
&P_{\pm2,\ell m}^{(x)}(\eta_0,\mathbf{x}_0)\equiv\\
&\qquad\equiv\int\frac{\mathrm{d}^2\hat{n}}{4\pi}\,_{\pm2}Y_{\ell m}^*(\versor{n})\int\frac{\mathrm{d}^3k}{(2\pi)^3}\,_{\pm}\Delta_{P}^{(x)}(\eta_0,\mathbf{k},\versor{n}).
\end{split}
\end{equation}
We now adopt the plane wave-expansion \cite{mehrem2011plane} for the complex exponential involving $\mathbf{k}$,
\begin{equation}
\label{eqn:plane_wave}
\frac{e^{i\mathbf{k}\cdot\versor{n}(\eta_0-\eta)}}{4\pi}=\sum_{LM}i^Lj_L[k(\eta_0-\eta)]Y_{LM}^*(\versor{k})Y_{LM}(\versor{n}),
\end{equation}
where $j_L$ is the $L$-th spherical Bessel function. By substituting Eq.~\eqref{eqn:plane_wave} in Eq.~\eqref{eqn:my_transfer}, we can easily see that now the dependence on $\versor{n}$ is encoded in the product of two spin-weighted spherical harmonics, which can be rewritten as a single one using the composition of angular momenta~\cite{varshalovich1988quantum},
\begin{widetext}
\begin{equation}
\label{eqn:composition}
\begin{split}
\,_{s_1}Y_{\ell_1 ,m_1}(\versor{n})\,_{s_2}Y_{\ell_2,m_2}(\versor{n})=\sum_{\ell_3m_3s_3}\sqrt{\frac{(2\ell_1+1)(2\ell_2+1)(2\ell_3+1)}{4\pi}}\begin{pmatrix}
\ell_1 & \ell_2 & \ell_3 \\
-s_1 & -s_2 & -s_3
\end{pmatrix}\begin{pmatrix}
\ell_1 & \ell_2 & \ell_3 \\
m_1 & m_2 & m_3
\end{pmatrix}\,_{s_3}Y_{\ell_3m_3}^*(\versor{n}),
\end{split}
\end{equation}
\end{widetext}
where the ``matrix'' is a  Wigner 3-j symbol, which satisfies the following selection rules:
\begin{align}
\label{eqn:flip}
&|\ell_1-\ell_2|\le\ell_3\le|\ell_1+\ell_2|,\\
&\nonumber\\
&m_1+m_2+m_3=0=s_1+s_2+s_3.
\end{align}
 However, to better digest such a long computation, it is better to adopt a standard trick in CMB calculations: indeed, instead of directly evaluating Eq.~\eqref{eqn:harmonic}. we apply a sort of ``fake'' rotation of the $\versor{k}$ unit vector, that is we firstly compute
 \begin{equation}
\label{eqn:rotated_my_transfer}
\begin{split}
\,_{\pm}\Delta_{P}^{(x)}(\eta_0,\mathbf{k},\versor{n})=\mathsf{R}_{k\versor{z}\mapsto\mathbf{k}}\left[\,_{\pm}\Delta_{P}^{(x)}(\eta_0,k\versor{z},\versor{n})\right].\\
\end{split}
 \end{equation}
In other words, we choose to work in the coordinate system where $\mathbf{k}\parallel\versor{z}$, and then, before performing the angular integration, we apply a rotation $\mathsf{R}_{k\versor{z}\mapsto\mathbf{k}}$ which brings $\mathbf{k}$ in a generic direction. Nevertheless, rotating the reference system implies that also $\versor{n}$ rotates and we have to take this into account.
 
We substitute Eq.~\eqref{eqn:plane_wave} and Eq.~\eqref{eqn:composition} within Eq.~\eqref{eqn:my_transfer} and we plug all together in Eq.~\eqref{eqn:rotated_my_transfer} , so that we obtain
\begin{widetext}
\begin{equation}
\label{eqn:long_Delta}
\begin{split}
\,_{\pm}\Delta_{P}^{(x)}(\eta_0,\mathbf{k},\versor{n})&=\sqrt{\frac{3}{2}}\int_{0}^{\eta_0}\mathrm{d}\eta\,g(\eta)\sum_{LM}i^Lj_L[k(\eta_0-\eta)]\sum_{L'M'}\sqrt{(2L+1)(2L'+1)}\begin{pmatrix}
	L & 2 & L' \\
	0 & \mp2 & \pm2
\end{pmatrix}\\
&\qquad\qquad\qquad\qquad\qquad\sum_{\lambda=-2}^{2}\begin{pmatrix}
	L & 2 & L' \\
	M & \lambda & M'
\end{pmatrix}\mathsf{R}_{k\versor{z}\mapsto\mathbf{k}}\left[Y_{LM}^*(\versor{z})\,_{\pm}\mathcal{T}_{\lambda}^{(x)}(\eta,k\versor{z})\,_{\mp2}Y_{L'M'}^*(\versor{n})\right].
\end{split}
\end{equation}
\end{widetext}
Thanks to our choice of $\versor{k}=\versor{z}$, the associated spherical harmonics is simply given as \cite{landau2013quantum}
\begin{equation}
Y_{LM}^*(\versor{z})=\delta_{M,0}\sqrt{\frac{2L+1}{4\pi}},
\end{equation}
so that in the last line of Eq.~\eqref{eqn:long_Delta} we have
\begin{equation}
\label{eqn:multi_rotation}
\begin{split}
&\mathsf{R}_{k\versor{z}\mapsto\mathbf{k}}\left[Y_{LM}^*(\versor{z})\,_{\pm}\mathcal{T}_{\lambda}^{(x)}(\eta,k\versor{z})\,_{\mp2}Y_{L'M'}^*(\versor{n})\right]=\\
&=\delta_{M,0}\sqrt{\frac{2L+1}{4\pi}}\,_{\pm}\mathcal{T}_{\lambda}^{(x)}(\eta,\mathbf{k})\,\mathsf{R}_{k\versor{z}\mapsto\mathbf{k}}\left[\,_{\mp2}Y_{L'M'}^*(\versor{n})\right]
\end{split}
\end{equation}
where we have exploited that the rotation operator is unitary, and so applying it to a product of quantities is equivalent to multiplying the rotated quantities themselves. The action of the rotation operator on the spin-weighted spherical harmonics is given as \cite{landau2013quantum}
\begin{equation}
\label{eqn:D_matrix}
\begin{split}
&\mathsf{R}_{k\versor{z}\mapsto\mathbf{k}}\left[\,_{\mp2}Y_{L'M'}^*(\versor{n})\right]=\\
&\quad=\sum_{m'=-L'}^{L'}D_{m'M'}^{(L')}\left[\mathsf{R}^{-1}_{k\versor{z}\mapsto\mathbf{k}}\right]\,_{\mp2}Y_{L'm'}^*(\versor{n})\\
&\quad=\sum_{m'=-L'}^{L'}\sqrt{\frac{4\pi}{2L'+1}}\,_{M'}Y^*_{L',m'}(\versor{k})\,_{\pm2}Y_{L',m'}(\versor{n}),
\end{split}
\end{equation}
where the $D_{m'M'}^{(L)}$'s are elements of the Wigner D-matrix. We now substitute the results of Eqs.~\eqref{eqn:multi_rotation}-\eqref{eqn:D_matrix} in Eq.~\eqref{eqn:long_Delta}. By exploiting the orthonormality of spin-weighted spherical harmonics \cite{goldberg1967spin},
\begin{equation}
\int\mathrm{d}^2\hat{n}\,_{s}Y_{\ell_1m_1}^*(\versor{n})\,_{s}Y_{\ell_2m_2}(\versor{n})=\delta_{\ell_1\ell_2}\delta_{m_1m_2},
\end{equation}
we can finally evaluate the right-hand side of Eq.~\eqref{eqn:harmonic},
\begin{widetext}
\begin{equation}
\label{eqn:P_lm}
\begin{split}
P_{\pm2,\ell m}^{(x)}(\eta_0,\mathbf{x}_0)=\sqrt{\frac{3}{2}}\sum_{L=|\ell-2|}^{\ell+2}&i^L(2L+1)\begin{pmatrix}
	L & 2 & \ell \\
	0 & \mp2 & \pm2
\end{pmatrix}\sum_{\lambda=-2}^{2}\begin{pmatrix}
	L & 2 & \ell \\
	0 & \lambda & -\lambda
\end{pmatrix}\\
&\int\frac{\mathrm{d}^3k}{(2\pi)^3}\,_{-\lambda}Y^*_{\ell m}(\versor{k})\int_{0}^{\eta_0}\mathrm{d}\eta\,g(\eta)\,_{\pm}\mathcal{T}_{\lambda}^{(x)}(\eta,\mathbf{k})j_L[k(\eta_0-\eta)].
\end{split}
\end{equation}
\end{widetext}
We now move to the standard decomposition in $E$ and $B$ modes, that is
\begin{equation}
P_{\pm2,\ell m}^{(x)}(\eta_0,\mathbf{x}_0)\equiv-\left[E_{\ell m}^{(x)}\pm iB_{\ell m}^{(x)}\right](\eta_0,\mathbf{x}_0).
\end{equation}
Therefore, we are now in the position to give the most general expression for the harmonic coefficients of the CMB polarization. By recalling all the procedure that we have made, it can be easily understood that the results of Eq.~\eqref{eqn:E_modes} and Eq.~\eqref{eqn:B_modes} are valid for any kind of cosmological perturbations (scalar, vector or tensor) up to the second-order in perturbation theory ($x=1,2$), and for any kind of initial conditions:
\begin{widetext}
\begin{equation}
\label{eqn:E_modes}
\begin{split}
E_{\ell m}^{(x)}(\eta_0,\mathbf{x}_0)=\sqrt{\frac{3}{8}}\sum_{L=|\ell-2|}^{\ell+2}&i^{L+2}(2L+1)\begin{pmatrix}
	L & 2 & \ell \\
	0 & -2 & 2
\end{pmatrix}\sum_{\lambda=-2}^{2}\begin{pmatrix}
	L & 2 & \ell \\
	0 & \lambda & -\lambda
\end{pmatrix}\int\frac{\mathrm{d}^3k}{(2\pi)^3}\,_{-\lambda}Y^*_{\ell m}(\versor{k})\\
&\int_{0}^{\eta_0}\mathrm{d}\eta\,g(\eta)\left[\,_{+}\mathcal{T}_{\lambda}^{(x)}(\eta,\mathbf{k})+(-1)^{\ell + L}\,_{-}\mathcal{T}_{\lambda}^{(x)}(\eta,\mathbf{k})\right]j_L[k(\eta_0-\eta)],
\end{split}
\end{equation}
\begin{equation}
\label{eqn:B_modes}
\begin{split}
B_{\ell m}^{(x)}(\eta_0,\mathbf{x}_0)=\sqrt{\frac{3}{8}}\sum_{L=|\ell-2|}^{\ell+2}&i^{L+1}(2L+1)\begin{pmatrix}
L & 2 & \ell \\
0 & -2 & 2
\end{pmatrix}\sum_{\lambda=-2}^{2}\begin{pmatrix}
L & 2 & \ell \\
0 & \lambda & -\lambda
\end{pmatrix}\int\frac{\mathrm{d}^3k}{(2\pi)^3}\,_{-\lambda}Y^*_{\ell m}(\versor{k})\\
&\int_{0}^{\eta_0}\mathrm{d}\eta\,g(\eta)\left[\,_{+}\mathcal{T}_{\lambda}^{(x)}(\eta,\mathbf{k})-(-1)^{\ell + L}\,_{-}\mathcal{T}_{\lambda}^{(x)}(\eta,\mathbf{k})\right]j_L[k(\eta_0-\eta)].
\end{split}
\end{equation}
\end{widetext}
The inflationary information is encoded in the $\mathbf{k}$ dependence of $\,_{\pm}\mathcal{T}_{\lambda}^{(x)}$, which is left general so that we have not assumed primordial Gaussianity, statistical isotropy or adiabaticity of the initial conditions a priori. Here we have not considered the contributions from weak lensing and spectral distortions, that could be relevant in general, but this is beyond the purpose of our paper.

\subsection{CMB Power Spectra}
To better appreciate the power of the formulas for the $E$ and $B$ modes of CMB polarization given by Eqs.~\eqref{eqn:E_modes}-\eqref{eqn:B_modes}, we can now compute the CMB power spectra, that are defined as
\begin{equation}
\label{eqn:Cl}
\begin{split}
C_{\ell}^{MN}\equiv\frac{1}{2\ell+1}\sum_{mm'}&\Big[\expval{M^{(1)*}_{\ell m}N^{(1)}_{\ell m}}\\
&\qquad\quad+\expval{M^{(2)*}_{\ell m}N^{(2)}_{\ell m}}+\dots\Big],\\
&
\end{split}
\end{equation}
where $M,N=T,E,B$. The dots in Eq.~\eqref{eqn:Cl} denote terms that we are not taking into account, such as correlations involving higher-order terms, e.g. $\expval{M^{(3)*}_{\ell m}N^{(1)}_{\ell m}}$, and eventual non-Gaussian contributions.

Now, to evaluate the $C_{\ell}$'s we can use Eqs.~\eqref{eqn:E_modes} and Eq.~\eqref{eqn:B_modes}, which are completely general, but for the sake of simplicity, we are now going to give a simplified estimation of the power spectra. First of all, let us consider just the contributions coming from the $\lambda=0$ case.  Indeed, if we recall the definition of Eq.~\eqref{eqn:pol_source} and Eqs.~\eqref{eqn:mathcal_T1}-\eqref{eqn:mathcal_T2}, setting $\lambda=0$ means selecting just scalar perturbations (see e.g. \cite{zaldarriaga1997all}). By the way, let us notice that when going at second order in perturbation theory, the axis-symmetry of the radiation field around the mode axis is broken by coupling to other modes \cite{liu2002polarization, bartolo2006cosmic}.

However, as said, here we want just to provide an estimation, and this is why we are now going to provide the harmonic coefficients of CMB polarization with $\lambda=0$. At first order we get
\begin{widetext}
\begin{equation}
\label{eqn:Elm1}
\begin{split}
E_{\ell m}^{(1)}(\eta_0,\mathbf{x}_0)\big|_{\lambda=0}=\sqrt{\frac{3}{2}}\sum_{L=|\ell-2|}^{\ell+2}i^{L+2}(2L+1)&\begin{pmatrix}
L & 2 & \ell \\
0 & -2 & 2
\end{pmatrix}\begin{pmatrix}
L & 2 & \ell \\
0 & 0 & 0
\end{pmatrix}\int\frac{\mathrm{d}^3k}{(2\pi)^3}\,Y^*_{\ell m}(\versor{k})\\
&\int_{0}^{\eta_0}\mathrm{d}\eta\,g(\eta)\cos[2\alpha_0(\eta)]\Pi_{0}^{(1)}(\eta,\mathbf{k})j_L[k(\eta_0-\eta)],
\end{split}
\end{equation}
\begin{equation}
\label{eqn:Blm1}
\begin{split}
B_{\ell m}^{(1)}(\eta_0,\mathbf{x}_0)\big|_{\lambda=0}=\sqrt{\frac{3}{2}}\sum_{L=|\ell-2|}^{\ell+2}i^{L+2}(2L+1)&\begin{pmatrix}
L & 2 & \ell \\
0 & -2 & 2
\end{pmatrix}\begin{pmatrix}
L & 2 & \ell \\
0 & 0 & 0
\end{pmatrix}\int\frac{\mathrm{d}^3k}{(2\pi)^3}\,Y^*_{\ell m}(\versor{k})\\
&\int_{0}^{\eta_0}\mathrm{d}\eta\,g(\eta)\sin[2\alpha_0(\eta)]\Pi_{0}^{(1)}(\eta,\mathbf{k})j_L[k(\eta_0-\eta)].
\end{split}
\end{equation}
whereas at second-order in perturbations we have
\begin{equation}
\label{eqn:Elm2}
\begin{split}
E_{\ell m}^{(2)}&(\eta_0,\mathbf{x}_0)\big|_{\lambda=0}=\sqrt{\frac{3}{2}}\sum_{L=|\ell-2|}^{\ell+2}i^{L+2}(2L+1)\begin{pmatrix}
L & 2 & \ell \\
0 & -2 & 2
\end{pmatrix}\begin{pmatrix}
L & 2 & \ell \\
0 & 0 & 0
\end{pmatrix}\int\frac{\mathrm{d}^3k}{(2\pi)^3}\,Y^*_{\ell m}(\versor{k})\\
&\int_{0}^{\eta_0}\mathrm{d}\eta\,g(\eta)\Bigg\{\cos[2\alpha_0(\eta)]\Pi_{0}^{(2)}(\eta,\mathbf{k})-2\sin[2\alpha_0(\eta)]\left[\delta\alpha(\eta)\ast\Pi_{0}^{(1)}(\eta)\right](\mathbf{k})\Bigg\}j_L[k(\eta_0-\eta)],
\end{split}
\end{equation}
\begin{equation}
\label{eqn:Blm2}
\begin{split}
B_{\ell m}^{(2)}&(\eta_0,\mathbf{x}_0)\big|_{\lambda=0}=\sqrt{\frac{3}{2}}\sum_{L=|\ell-2|}^{\ell+2}i^{L+2}(2L+1)\begin{pmatrix}
L & 2 & \ell \\
0 & -2 & 2
\end{pmatrix}\begin{pmatrix}
L & 2 & \ell \\
0 & 0 & 0
\end{pmatrix}\int\frac{\mathrm{d}^3k}{(2\pi)^3}\,Y^*_{\ell m}(\versor{k})\\
&\int_{0}^{\eta_0}\mathrm{d}\eta\,g(\eta)\Bigg\{\sin[2\alpha_0(\eta)]\Pi_{0}^{(2)}(\eta,\mathbf{k})+2\cos[2\alpha_0(\eta)]\left[\delta\alpha(\eta)\ast\Pi_{0}^{(1)}(\eta)\right](\mathbf{k})\Bigg\}j_L[k(\eta_0-\eta)].
\end{split}
\end{equation}
\end{widetext}
By inspecting the first-order expressions, we can see that, if the isotropic birefringence angle equals zero, we have no $B$-modes, and this is something completely expected since $B$ modes are primordially sourced just by tensor perturbations. This is fully consistent with what is already present in the literature \cite{murai2022isotropic, liu2006effect, finelli2009rotation, gubitosi2014including, galaverni2023redshift}, that our treatment has been able to generalize. 

\begin{figure*}
\centering
\includegraphics[width=\textwidth]{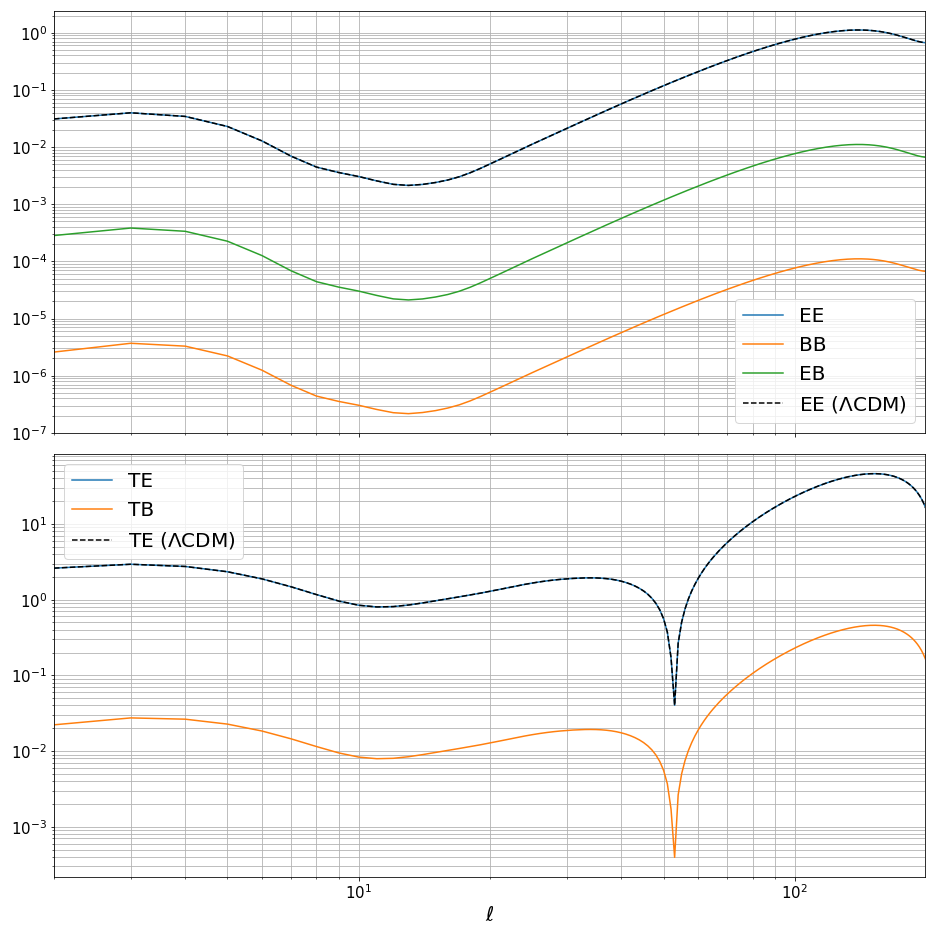}
\caption{Absolute value of the unlensed angular power spectra of CMB, $\ell(\ell+1)|C_{\ell}|/(2\pi)$, in units of $\mu K^2$ affected by isotropic and anisotropic cosmic birefringence compared with the $\Lambda$CDM predictions without any birefringence effects (black dashed lines). The numerical computation has been performed by setting the axion mass equal to $\SI{3.0e-33}{\electronvolt}$, the parameter $\lambda/f$ equal to $\SI{1.6e-20}{\giga\electronvolt^{-1}}$, the initial value of the axion field $\chi_{0}^{\text{ini}}$ equal to $-m_{Pl}/2$ (so that, according to our parameter estimation, $\lambda/f\chi_{0}^{\text{ini}}\simeq-0.02$), and its initial velocity equal to zero, and for the fiducial values of the $\Lambda$CDM parameters provided in \cite{aghanim2020planck}. The tensor-to-scalar ratio has been set equal to zero.}
\label{fig:cmb_spectra}
\end{figure*}

As previously said, one of the main goals of our paper is to estimate the impact on CMB observables of the predicted anisotropic birefringence associated with the set of best-fit parameters most consistent with the isotropic birefringence angle measured from \textit{Planck} data. To do that, we can numerically evaluate the CMB power spectra using Eqs.~\eqref{eqn:Elm1}-\eqref{eqn:Blm2}, and this is precisely what we have done in Fig.~\ref{fig:cmb_spectra}. 
However, dealing with second-order terms in perturbation theory could be extremely challenging and time-consuming: this is why to obtain the plot shown in Fig.~\ref{fig:cmb_spectra} we have applied the tomographic approximation of Eq.~\eqref{eqn:tomographic_approximation} to a modified version of Boltzmann code \texttt{SONG} \cite{pettinari2013intrinsic} so that we have partially bypassed the integration over the conformal time. Indeed, we have used a modified version of \texttt{SONG} for computing the first term in the time-integrals at the right-hand side of Eqs.~\eqref{eqn:Elm2}-\eqref{eqn:Blm2}, which represent the second-order polarization term just affected by isotropic birefringence, whereas we have exploited our modified version of \texttt{CLASS} to evaluate the remaining first- and second-order terms. In particular, we have taken advantage of the tomographic approximation to deal with the convolution product $\delta\alpha\ast\Pi^{(1)}_0$. Of course, a more rigorous treatment should involve a full integration like that present in our general theoretical formulas, which requires a much higher computational cost, but this is something beyond the purpose of our paper. 

By direct inspection of Fig.~\ref{fig:cmb_spectra}, we can recover a completely expected behavior of the birefringent CMB correlation functions. Indeed, it seems that the spectra involving the $B$-modes can be obtained using the standard rescaling of those involving the $E$ ones, i.e.  
\begin{align}
C_{\ell}^{BB}&\sim C_{\ell}^{EE}\tan^2(2\times0.30^{\circ}),\\ 
C_{\ell}^{EB}&\sim C_{\ell}^{EE}\tan(2\times0.30^{\circ}),\\
C_{\ell}^{TB}&\sim C_{\ell}^{TE}\tan(2\times0.30^{\circ}).
\end{align}
Such a behavior is in complete agreement with our previous results: first of all, since we have set the tensor-to-scalar ratio equal to zero, it is clear that the only source of unlensed $B$ modes is cosmic birefringence. Secondly, we have evaluated the CMB spectra for the first set of best-fit axion parameters we found in Sec.~\ref{sec:Contours}, which predicts the contribution coming from anisotropic birefringence to be subdominant with respect to the isotropic one. Thirdly, the physical source of isotropic cosmic birefringence, i.e. the homogeneous axion-like field $\chi_{0}$, experiences a phase in which it is almost constant in time during matter-domination for such a set of parameters. This translates into an isotropic angle which is independent of the photons' emission time, and this is why the birefringence effect manifests itself as just a multiplicative factor in Fig.~\ref{fig:cmb_spectra}. Let us notice that such features are not only theoretical predictions but seem to be perfectly consistent with the data analysis of both the isotropic and anisotropic angle, done e.g. in \cite{eskilt2023constraint, bortolami2022planck}, respectively, which suggest a negligible redshift evolution of the birefringence angle and an anisotropic contribution compatible with the null hypothesis.

\section{\label{sec:Conclu}Conclusions}
In this paper, we have tested the possibility of explaining the \textit{Planck}'s hint of detection for isotropic cosmic birefringence with the physics of an axion-like field $\chi$ interacting with the electromagnetic one through a Chern-Simons coupling. In particular, we have considered a pseudoscalar field described by a standard quadratic potential, and we have found that there exist two different regions of the parameter space that are consistent with $\alpha_0=(0.30\pm0.11)^{\circ}$: one characterized by a negative product between the initial value of the field and the coupling parameter ($\lambda\chi_0^{\text{ini}}/f\simeq-0.02$, $m_{\chi}\simeq\SI{3.00e-33}{\electronvolt}$), the other by a positive one ($\lambda\chi_0^{\text{ini}}/f\simeq0.12$, $m_{\chi}\simeq\SI{5.28e-27}{\electronvolt}$). Interestingly enough, the first of these two solutions can explain an isotropic birefringence angle $\sim0.30^{\circ}$ for a very light mass of the axion-like field (of the order $\SI{3e-33}{\electronvolt}$), whose time evolution is almost constant during the matter-dominated epoch of the Universe. This behavior is perfectly consistent with the results of \cite{eskilt2023constraint}. which seem to prefer a CMB $EB$ spectrum, coming from such dynamics with respect to the case in which the homogeneous field evolves in time during the recombination epoch.

In the second part of the paper we then moved to investigating the implications of such results for the anisotropic component of cosmic birefringence. For this reason, we also provided the angular power spectrum of $\delta\alpha(\versor{n})$ and its cross-correlation with CMB temperature and the $E$ polarization mode. We found that the amount of anisotropic birefringence sourced by the perturbations of the axion-like field $\delta\chi$ for the previously (first) mentioned set of best-fit parameters is predicted to be well below the latest observational constraints, see also Fig.~\ref{fig:power_spectrum_aniso}. Indeed, for the model defined in Eq.~\ref{eqn:chi_action} we predict that the amount of anisotropic birefringence is of the order of $(10^{-15}\divisionsymbol10^{-32})$ deg$^2$ for the auto-correlation, and  $(10^{-7}\divisionsymbol10^{-17})$ $\mu $K$\cdot\,$deg for the cross-correlations with the CMB $T$ and $E$ fields, according to the angular scale. This is compatible with what has been found e.g. in \cite{bortolami2022planck}, i.e. a signal consistent with the null hypothesis. Similar results are found for a different potential, see App.~\ref{app:other_potential}.

However, as shown at the end of Sec.~\ref{sec:Contours}, a set of parameters predicting an almost constant-in-time $\chi_{0}(\eta)$, does not automatically imply that the same behavior is also followed by the inhomogeneous perturbations of the axion-like field $\delta\chi(\eta,\mathbf{x})$. Hence, in Sec.~\ref{sec:Boltzmann} we have presented a generalized Boltzmann approach, which can completely take into account the dependence of isotropic and anisotropic birefringence on the photons' emission time. To our knowledge, this is the first time in the literature in which anisotropic cosmic birefringence is formally characterized, i.e. as a second-order effect arising because of the convolution between the first-order axion perturbations and the standard first-order CMB polarization. As proof of the validity of our treatment, we have checked that such formalism recovers the frequently used equations when assuming sudden recombination or when considering the axion-like field as almost constant during the matter-dominated epoch.

Since dealing with second-order quantities is numerically challenging, we have evaluated our generalized expressions for $E_{\ell m}$ and $B_{\ell m}$ by invoking the ``tomographic approximation'', already exploited for similar purposes also in \cite{greco2023probing}, so that we have been able to compute the angular power spectra of CMB polarization and its cross-correlation with temperature anisotropies, by using the fiducial values of the $\Lambda$CDM model together with the (first) set of best-fit parameters found in Sec.~\ref{sec:Contours}. As shown in Fig.~\ref{fig:cmb_spectra}, such spectra are consistent with a purely redshift-independent isotropic birefringence effect, and, according to our previous discussions, this is not surprising. Indeed, as just mentioned, for such a set of parameters the homogeneous axion-like field is predicted to be almost constant in time, and the anisotropic component of cosmic birefringence to be extremely subdominant. Despite the smallness of such anisotropic signal, this result is nevertheless really important, since it shows how the models investigated can provide promising falsifiability checks for future observations.

\begin{acknowledgments}
A. Greco would like to thank E. Komatsu, J. Hou, G. Perna, and S. Gasparotto for useful discussions and comments. N. Bartolo and A. Gruppuso acknowledge support from the COSMOS network (www.cosmosnet.it) through the ASI (Italian Space Agency) Grants 2016-24-H.0, 2016-24-H.1-2018 and 2020-9-HH.0. This work is supported in part by the MUR PRIN2022 Project ``BROWSEPOL: Beyond standaRd mOdel With coSmic microwavE background POLarization”-2022EJNZ53 financed by the European Union - Next Generation EU. 
\end{acknowledgments}
\hspace{20pt}
\appendix

\section{\label{app:other_potential} The EDE Potential}
\begin{figure*}[htbp]
\centering
	
\subfloat[Contour plot of $\alpha_0$.]{
\includegraphics[width=0.475\textwidth]{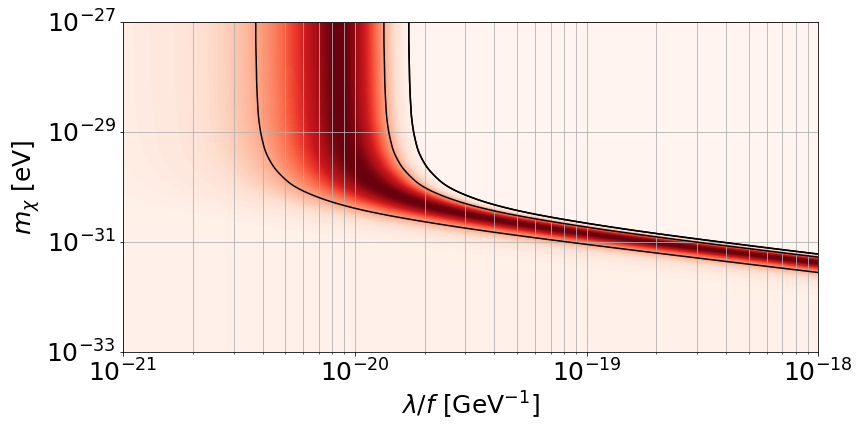}
\label{fig:new_subplot2}
}\hfill
\subfloat[Redshift evolution of the ratio between $\chi_0$ and its initial value for the best-fit mass resulting from our chi-squared analysis.]{
\includegraphics[width=0.475\textwidth]{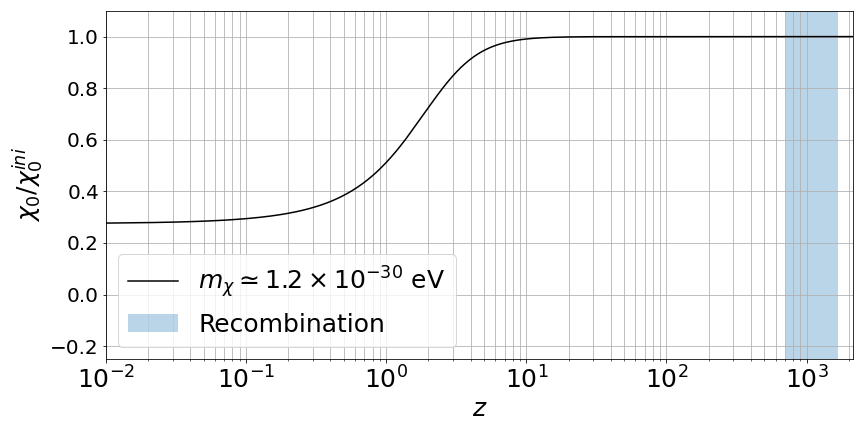}
\label{fig:new_subplot3}%
}

\vspace{\baselineskip}

\subfloat[Angular power spectra involving anisotropic cosmic birefringence for the set of best-fit parameters resulting from our chi-squared analysis we performed in this appendix for $\alpha_0$, i.e. $m_{\chi}\simeq\SI{1.2e-30}{\electronvolt}$ and $\lambda/f\simeq\SI{1.2e-20}{\giga\electronvolt^{-1}}$.]{\includegraphics[width=1.0\textwidth]{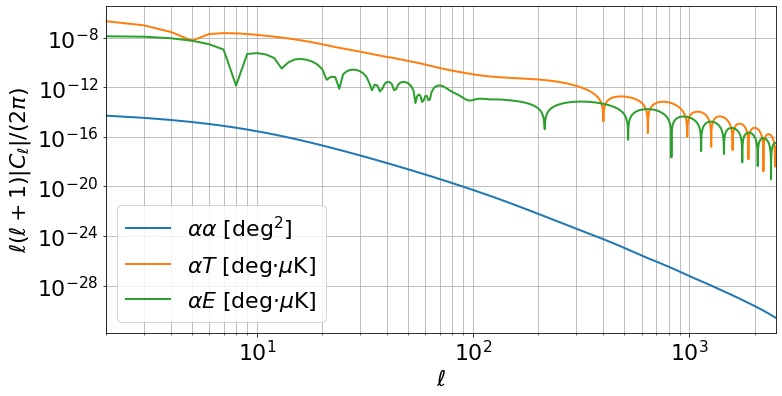}
\label{fig:new_subplot4}}


	

\caption{Cosmic birefringence from the model described in Eq.~\eqref{eqn:new_action}. The numerical computation has been performed by setting the initial value of the axion field $\chi_{0}^{\text{ini}}$ equal to $-m_{Pl}/2$, the initial velocity of the axion field equal to zero, and for the fiducial values of the $\Lambda$CDM parameters provided in \cite{aghanim2020planck}.}
\label{fig:new_alpha_0}
\end{figure*}

In this appendix we are going to show the results of a chi-squared analysis, very similar to what we performed in Sec.~\ref{sec:Contours}, but with a different theory for the pseudoscalar often used to characterize the axion as early dark energy (see e.g. \cite{capparelli2020cosmic, eskilt2023constraint})
\begin{equation}
\begin{split}
\label{eqn:new_action}
S_{\chi}=-\int\mathrm{d}^4x\,&\sqrt{-g}\Bigg\{\frac{1}{2}g^{\mu\nu}\partial_{\mu}\chi\partial_{\nu}\chi\\
&+m_{\chi}^2M_{Pl}^2\left[1-\cos\left(\frac{\chi}{M_{Pl}}\right)\right]^3\Bigg\}.
\end{split}
\end{equation}
The equation of motion for the homogeneous axion-like field $\chi=\chi_{0}(\eta)$ following by Eq.~\eqref{eqn:new_action} reads
\begin{equation}
\label{eqn:new_EOM}
\begin{split}
&\chi_0^{\prime\prime}+2\mathcal{H}\chi_0^{\prime}=\\&=-3a^2m^2_{\chi}M_{Pl}\sin\left(\frac{\chi_0}{M_{Pl}}\right)\left[1-\cos\left(\frac{\chi}{M_{Pl}}\right)\right]^2,\\
&
\end{split}
\end{equation}
which, contrary to Eq.~\eqref{eqn:chi_EOM}, is not linear in $\chi_{0}$ and so it cannot be rescaled by a factor $\chi_0^{\text{ini}}$ as we did in Sec.~\ref{sec:Contours}. Therefore, in this case the parameter space for the isotropic birefringence angle is represented as a grid of values for $m_{\chi}$ and $\lambda/f$, whereas we have fixed the initial value of $\chi_0^{\text{ini}}=-m_{Pl}/2$: the results are shown in Fig.~\ref{fig:new_alpha_0}.

As in the case of a quadratic potential, we have found that there is a region of the parameter space consistent with an almost constant in time, homogeneous field $\chi_{0}$ during recombination. Moreover, this can be achieved by setting the initial value of the field equal to a negative amount, which was indeed one of the possibilities already examined in Sec.~\ref{sec:Contours}. Hence, also with an EDE potential, it is possible to find a set of parameters in agreement with the \textit{Planck} result, which yields an almost constant in-time evolution of the axion field at recombination.

\bibliography{ThirdPaper.bib}

\begin{thebibliography}{159}%
\makeatletter
\providecommand \@ifxundefined [1]{%
 \@ifx{#1\undefined}
}%
\providecommand \@ifnum [1]{%
 \ifnum #1\expandafter \@firstoftwo
 \else \expandafter \@secondoftwo
 \fi
}%
\providecommand \@ifx [1]{%
 \ifx #1\expandafter \@firstoftwo
 \else \expandafter \@secondoftwo
 \fi
}%
\providecommand \natexlab [1]{#1}%
\providecommand \enquote  [1]{``#1''}%
\providecommand \bibnamefont  [1]{#1}%
\providecommand \bibfnamefont [1]{#1}%
\providecommand \citenamefont [1]{#1}%
\providecommand \href@noop [0]{\@secondoftwo}%
\providecommand \href [0]{\begingroup \@sanitize@url \@href}%
\providecommand \@href[1]{\@@startlink{#1}\@@href}%
\providecommand \@@href[1]{\endgroup#1\@@endlink}%
\providecommand \@sanitize@url [0]{\catcode `\\12\catcode `\$12\catcode
  `\&12\catcode `\#12\catcode `\^12\catcode `\_12\catcode `\%12\relax}%
\providecommand \@@startlink[1]{}%
\providecommand \@@endlink[0]{}%
\providecommand \url  [0]{\begingroup\@sanitize@url \@url }%
\providecommand \@url [1]{\endgroup\@href {#1}{\urlprefix }}%
\providecommand \urlprefix  [0]{URL }%
\providecommand \Eprint [0]{\href }%
\providecommand \doibase [0]{https://doi.org/}%
\providecommand \selectlanguage [0]{\@gobble}%
\providecommand \bibinfo  [0]{\@secondoftwo}%
\providecommand \bibfield  [0]{\@secondoftwo}%
\providecommand \translation [1]{[#1]}%
\providecommand \BibitemOpen [0]{}%
\providecommand \bibitemStop [0]{}%
\providecommand \bibitemNoStop [0]{.\EOS\space}%
\providecommand \EOS [0]{\spacefactor3000\relax}%
\providecommand \BibitemShut  [1]{\csname bibitem#1\endcsname}%
\let\auto@bib@innerbib\@empty
\bibitem [{\citenamefont {Lue}\ \emph {et~al.}(1999)\citenamefont {Lue},
  \citenamefont {Wang},\ and\ \citenamefont
  {Kamionkowski}}]{lue1999cosmological}%
  \BibitemOpen
  \bibfield  {author} {\bibinfo {author} {\bibfnamefont {A.}~\bibnamefont
  {Lue}}, \bibinfo {author} {\bibfnamefont {L.}~\bibnamefont {Wang}},\ and\
  \bibinfo {author} {\bibfnamefont {M.}~\bibnamefont {Kamionkowski}},\
  }\href@noop {} {\bibfield  {journal} {\bibinfo  {journal} {Physical Review
  Letters}\ }\textbf {\bibinfo {volume} {83}},\ \bibinfo {pages} {1506}
  (\bibinfo {year} {1999})}\BibitemShut {NoStop}%
\bibitem [{\citenamefont {Komatsu}(2022)}]{komatsu2022new}%
  \BibitemOpen
  \bibfield  {author} {\bibinfo {author} {\bibfnamefont {E.}~\bibnamefont
  {Komatsu}},\ }\href@noop {} {\bibfield  {journal} {\bibinfo  {journal}
  {Nature Reviews Physics}\ ,\ \bibinfo {pages} {1}} (\bibinfo {year}
  {2022})}\BibitemShut {NoStop}%
\bibitem [{\citenamefont {Alexander}\ and\ \citenamefont
  {Martin}(2005)}]{alexander2005birefringent}%
  \BibitemOpen
  \bibfield  {author} {\bibinfo {author} {\bibfnamefont {S.}~\bibnamefont
  {Alexander}}\ and\ \bibinfo {author} {\bibfnamefont {J.}~\bibnamefont
  {Martin}},\ }\href@noop {} {\bibfield  {journal} {\bibinfo  {journal}
  {Physical Review D}\ }\textbf {\bibinfo {volume} {71}},\ \bibinfo {pages}
  {063526} (\bibinfo {year} {2005})}\BibitemShut {NoStop}%
\bibitem [{\citenamefont {Contaldi}\ \emph {et~al.}(2008)\citenamefont
  {Contaldi}, \citenamefont {Magueijo},\ and\ \citenamefont
  {Smolin}}]{contaldi2008anomalous}%
  \BibitemOpen
  \bibfield  {author} {\bibinfo {author} {\bibfnamefont {C.~R.}\ \bibnamefont
  {Contaldi}}, \bibinfo {author} {\bibfnamefont {J.}~\bibnamefont {Magueijo}},\
  and\ \bibinfo {author} {\bibfnamefont {L.}~\bibnamefont {Smolin}},\
  }\href@noop {} {\bibfield  {journal} {\bibinfo  {journal} {Physical review
  letters}\ }\textbf {\bibinfo {volume} {101}},\ \bibinfo {pages} {141101}
  (\bibinfo {year} {2008})}\BibitemShut {NoStop}%
\bibitem [{\citenamefont {Alexander}\ and\ \citenamefont
  {Yunes}(2009)}]{alexander2009chern}%
  \BibitemOpen
  \bibfield  {author} {\bibinfo {author} {\bibfnamefont {S.}~\bibnamefont
  {Alexander}}\ and\ \bibinfo {author} {\bibfnamefont {N.}~\bibnamefont
  {Yunes}},\ }\href@noop {} {\bibfield  {journal} {\bibinfo  {journal} {Physics
  Reports}\ }\textbf {\bibinfo {volume} {480}},\ \bibinfo {pages} {1} (\bibinfo
  {year} {2009})}\BibitemShut {NoStop}%
\bibitem [{\citenamefont {Sorbo}(2011)}]{sorbo2011parity}%
  \BibitemOpen
  \bibfield  {author} {\bibinfo {author} {\bibfnamefont {L.}~\bibnamefont
  {Sorbo}},\ }\href@noop {} {\bibfield  {journal} {\bibinfo  {journal} {Journal
  of Cosmology and Astroparticle Physics}\ }\textbf {\bibinfo {volume}
  {2011}}\bibinfo  {number} { (06)},\ \bibinfo {pages} {003}}\BibitemShut
  {NoStop}%
\bibitem [{\citenamefont {Bartolo}\ \emph
  {et~al.}(2015{\natexlab{a}})\citenamefont {Bartolo}, \citenamefont
  {Matarrese}, \citenamefont {Peloso},\ and\ \citenamefont
  {Shiraishi}}]{bartolo2015parity1}%
  \BibitemOpen
\bibfield  {number} {  }\bibfield  {author} {\bibinfo {author} {\bibfnamefont
  {N.}~\bibnamefont {Bartolo}}, \bibinfo {author} {\bibfnamefont
  {S.}~\bibnamefont {Matarrese}}, \bibinfo {author} {\bibfnamefont
  {M.}~\bibnamefont {Peloso}},\ and\ \bibinfo {author} {\bibfnamefont
  {M.}~\bibnamefont {Shiraishi}},\ }\href@noop {} {\bibfield  {journal}
  {\bibinfo  {journal} {Journal of Cosmology and Astroparticle Physics}\
  }\textbf {\bibinfo {volume} {2015}}\bibinfo  {number} { (01)},\ \bibinfo
  {pages} {027}}\BibitemShut {NoStop}%
\bibitem [{\citenamefont {Bartolo}\ \emph
  {et~al.}(2015{\natexlab{b}})\citenamefont {Bartolo}, \citenamefont
  {Matarrese}, \citenamefont {Peloso},\ and\ \citenamefont
  {Shiraishi}}]{bartolo2015parity2}%
  \BibitemOpen
\bibfield  {number} {  }\bibfield  {author} {\bibinfo {author} {\bibfnamefont
  {N.}~\bibnamefont {Bartolo}}, \bibinfo {author} {\bibfnamefont
  {S.}~\bibnamefont {Matarrese}}, \bibinfo {author} {\bibfnamefont
  {M.}~\bibnamefont {Peloso}},\ and\ \bibinfo {author} {\bibfnamefont
  {M.}~\bibnamefont {Shiraishi}},\ }\href@noop {} {\bibfield  {journal}
  {\bibinfo  {journal} {Journal of Cosmology and Astroparticle Physics}\
  }\textbf {\bibinfo {volume} {2015}}\bibinfo  {number} { (07)},\ \bibinfo
  {pages} {039}}\BibitemShut {NoStop}%
\bibitem [{\citenamefont {Gerbino}\ \emph {et~al.}(2016)\citenamefont
  {Gerbino}, \citenamefont {Gruppuso}, \citenamefont {Natoli}, \citenamefont
  {Shiraishi},\ and\ \citenamefont {Melchiorri}}]{gerbino2016testing}%
  \BibitemOpen
\bibfield  {number} {  }\bibfield  {author} {\bibinfo {author} {\bibfnamefont
  {M.}~\bibnamefont {Gerbino}}, \bibinfo {author} {\bibfnamefont
  {A.}~\bibnamefont {Gruppuso}}, \bibinfo {author} {\bibfnamefont
  {P.}~\bibnamefont {Natoli}}, \bibinfo {author} {\bibfnamefont
  {M.}~\bibnamefont {Shiraishi}},\ and\ \bibinfo {author} {\bibfnamefont
  {A.}~\bibnamefont {Melchiorri}},\ }\href@noop {} {\bibfield  {journal}
  {\bibinfo  {journal} {Journal of Cosmology and Astroparticle Physics}\
  }\textbf {\bibinfo {volume} {2016}}\bibinfo  {number} { (07)},\ \bibinfo
  {pages} {044}}\BibitemShut {NoStop}%
\bibitem [{\citenamefont {Qiao}\ \emph {et~al.}(2019)\citenamefont {Qiao},
  \citenamefont {Zhu}, \citenamefont {Zhao},\ and\ \citenamefont
  {Wang}}]{qiao2019waveform}%
  \BibitemOpen
\bibfield  {number} {  }\bibfield  {author} {\bibinfo {author} {\bibfnamefont
  {J.}~\bibnamefont {Qiao}}, \bibinfo {author} {\bibfnamefont {T.}~\bibnamefont
  {Zhu}}, \bibinfo {author} {\bibfnamefont {W.}~\bibnamefont {Zhao}},\ and\
  \bibinfo {author} {\bibfnamefont {A.}~\bibnamefont {Wang}},\ }\href@noop {}
  {\bibfield  {journal} {\bibinfo  {journal} {Physical Review D}\ }\textbf
  {\bibinfo {volume} {100}},\ \bibinfo {pages} {124058} (\bibinfo {year}
  {2019})}\BibitemShut {NoStop}%
\bibitem [{\citenamefont {Qiao}\ \emph {et~al.}(2020)\citenamefont {Qiao},
  \citenamefont {Zhu}, \citenamefont {Zhao},\ and\ \citenamefont
  {Wang}}]{qiao2020polarized}%
  \BibitemOpen
  \bibfield  {author} {\bibinfo {author} {\bibfnamefont {J.}~\bibnamefont
  {Qiao}}, \bibinfo {author} {\bibfnamefont {T.}~\bibnamefont {Zhu}}, \bibinfo
  {author} {\bibfnamefont {W.}~\bibnamefont {Zhao}},\ and\ \bibinfo {author}
  {\bibfnamefont {A.}~\bibnamefont {Wang}},\ }\href@noop {} {\bibfield
  {journal} {\bibinfo  {journal} {Physical Review D}\ }\textbf {\bibinfo
  {volume} {101}},\ \bibinfo {pages} {043528} (\bibinfo {year}
  {2020})}\BibitemShut {NoStop}%
\bibitem [{\citenamefont {{The LiteBIRD Collaboration}}\ \emph
  {et~al.}(2023)\citenamefont {{The LiteBIRD Collaboration}}, \citenamefont
  {Campeti}, \citenamefont {Komatsu}, \citenamefont {Baccigalupi},
  \citenamefont {Ballardini}, \citenamefont {Bartolo}, \citenamefont {Carones},
  \citenamefont {Errard}, \citenamefont {Finelli}, \citenamefont {Flauger},
  \citenamefont {Galli}, \citenamefont {Galloni}, \citenamefont {Giardiello},
  \citenamefont {Hazumi}, \citenamefont {Henrot-Versillé}, \citenamefont
  {Hergt}, \citenamefont {Kohri}, \citenamefont {Leloup}, \citenamefont
  {Lesgourgues}, \citenamefont {Macias-Perez}, \citenamefont
  {Martínez-González}, \citenamefont {Matarrese}, \citenamefont {Matsumura},
  \citenamefont {Montier}, \citenamefont {Namikawa}, \citenamefont {Paoletti},
  \citenamefont {Poletti}, \citenamefont {Remazeilles}, \citenamefont
  {Shiraishi}, \citenamefont {van Tent}, \citenamefont {Tristram},
  \citenamefont {Vacher}, \citenamefont {Vittorio}, \citenamefont
  {Weymann-Despres}, \citenamefont {Anand}, \citenamefont {Aumont},
  \citenamefont {Aurlien}, \citenamefont {Banday}, \citenamefont {Barreiro},
  \citenamefont {Basyrov}, \citenamefont {Bersanelli}, \citenamefont {Blinov},
  \citenamefont {Bortolami}, \citenamefont {Brinckmann}, \citenamefont
  {Calabrese}, \citenamefont {Carralot}, \citenamefont {Casas}, \citenamefont
  {Clermont}, \citenamefont {Columbro}, \citenamefont {Conenna}, \citenamefont
  {Coppolecchia}, \citenamefont {Cuttaia}, \citenamefont {D'Alessandro},
  \citenamefont {de~Bernardis}, \citenamefont {Petris}, \citenamefont {Torre},
  \citenamefont {Giorgi}, \citenamefont {Diego-Palazuelos}, \citenamefont
  {Eriksen}, \citenamefont {Franceschet}, \citenamefont {Fuskeland},
  \citenamefont {Galloway}, \citenamefont {Georges}, \citenamefont {Gerbino},
  \citenamefont {Gervasi}, \citenamefont {Ghigna}, \citenamefont {Gimeno-Amo},
  \citenamefont {Gjerløw}, \citenamefont {Gruppuso}, \citenamefont
  {Gudmundsson}, \citenamefont {Krachmalnicoff}, \citenamefont {Lamagna},
  \citenamefont {Lattanzi}, \citenamefont {Lembo}, \citenamefont {Lonappan},
  \citenamefont {Masi}, \citenamefont {Massa}, \citenamefont {Micheli},
  \citenamefont {Moggi}, \citenamefont {Monelli}, \citenamefont {Morgante},
  \citenamefont {Mot}, \citenamefont {Mousset}, \citenamefont {Nagata},
  \citenamefont {Natoli}, \citenamefont {Novelli}, \citenamefont {Obata},
  \citenamefont {Pagano}, \citenamefont {Paiella}, \citenamefont {Pavlidou},
  \citenamefont {Piacentini}, \citenamefont {Pinchera}, \citenamefont {Pisano},
  \citenamefont {Puglisi}, \citenamefont {Raffuzzi}, \citenamefont {Ritacco},
  \citenamefont {Rizzieri}, \citenamefont {Ruiz-Granda}, \citenamefont
  {Savini}, \citenamefont {Scott}, \citenamefont {Signorelli}, \citenamefont
  {Stever}, \citenamefont {Stutzer}, \citenamefont {Sullivan}, \citenamefont
  {Tartari}, \citenamefont {Tassis}, \citenamefont {Terenzi}, \citenamefont
  {Thompson}, \citenamefont {Vielva}, \citenamefont {Wehus},\ and\
  \citenamefont {Zhou}}]{campeti2023litebird}%
  \BibitemOpen
  \bibfield  {author} {\bibinfo {author} {\bibnamefont {{The LiteBIRD
  Collaboration}}}, \bibinfo {author} {\bibfnamefont {P.}~\bibnamefont
  {Campeti}}, \bibinfo {author} {\bibfnamefont {E.}~\bibnamefont {Komatsu}},
  \bibinfo {author} {\bibfnamefont {C.}~\bibnamefont {Baccigalupi}}, \bibinfo
  {author} {\bibfnamefont {M.}~\bibnamefont {Ballardini}}, \bibinfo {author}
  {\bibfnamefont {N.}~\bibnamefont {Bartolo}}, \bibinfo {author} {\bibfnamefont
  {A.}~\bibnamefont {Carones}}, \bibinfo {author} {\bibfnamefont
  {J.}~\bibnamefont {Errard}}, \bibinfo {author} {\bibfnamefont
  {F.}~\bibnamefont {Finelli}}, \bibinfo {author} {\bibfnamefont
  {R.}~\bibnamefont {Flauger}}, \bibinfo {author} {\bibfnamefont
  {S.}~\bibnamefont {Galli}}, \bibinfo {author} {\bibfnamefont
  {G.}~\bibnamefont {Galloni}}, \bibinfo {author} {\bibfnamefont
  {S.}~\bibnamefont {Giardiello}}, \bibinfo {author} {\bibfnamefont
  {M.}~\bibnamefont {Hazumi}}, \bibinfo {author} {\bibfnamefont
  {S.}~\bibnamefont {Henrot-Versillé}}, \bibinfo {author} {\bibfnamefont
  {L.~T.}\ \bibnamefont {Hergt}}, \bibinfo {author} {\bibfnamefont
  {K.}~\bibnamefont {Kohri}}, \bibinfo {author} {\bibfnamefont
  {C.}~\bibnamefont {Leloup}}, \bibinfo {author} {\bibfnamefont
  {J.}~\bibnamefont {Lesgourgues}}, \bibinfo {author} {\bibfnamefont
  {J.}~\bibnamefont {Macias-Perez}}, \bibinfo {author} {\bibfnamefont
  {E.}~\bibnamefont {Martínez-González}}, \bibinfo {author} {\bibfnamefont
  {S.}~\bibnamefont {Matarrese}}, \bibinfo {author} {\bibfnamefont
  {T.}~\bibnamefont {Matsumura}}, \bibinfo {author} {\bibfnamefont
  {L.}~\bibnamefont {Montier}}, \bibinfo {author} {\bibfnamefont
  {T.}~\bibnamefont {Namikawa}}, \bibinfo {author} {\bibfnamefont
  {D.}~\bibnamefont {Paoletti}}, \bibinfo {author} {\bibfnamefont
  {D.}~\bibnamefont {Poletti}}, \bibinfo {author} {\bibfnamefont
  {M.}~\bibnamefont {Remazeilles}}, \bibinfo {author} {\bibfnamefont
  {M.}~\bibnamefont {Shiraishi}}, \bibinfo {author} {\bibfnamefont
  {B.}~\bibnamefont {van Tent}}, \bibinfo {author} {\bibfnamefont
  {M.}~\bibnamefont {Tristram}}, \bibinfo {author} {\bibfnamefont
  {L.}~\bibnamefont {Vacher}}, \bibinfo {author} {\bibfnamefont
  {N.}~\bibnamefont {Vittorio}}, \bibinfo {author} {\bibfnamefont
  {G.}~\bibnamefont {Weymann-Despres}}, \bibinfo {author} {\bibfnamefont
  {A.}~\bibnamefont {Anand}}, \bibinfo {author} {\bibfnamefont
  {J.}~\bibnamefont {Aumont}}, \bibinfo {author} {\bibfnamefont
  {R.}~\bibnamefont {Aurlien}}, \bibinfo {author} {\bibfnamefont {A.~J.}\
  \bibnamefont {Banday}}, \bibinfo {author} {\bibfnamefont {R.~B.}\
  \bibnamefont {Barreiro}}, \bibinfo {author} {\bibfnamefont {A.}~\bibnamefont
  {Basyrov}}, \bibinfo {author} {\bibfnamefont {M.}~\bibnamefont {Bersanelli}},
  \bibinfo {author} {\bibfnamefont {D.}~\bibnamefont {Blinov}}, \bibinfo
  {author} {\bibfnamefont {M.}~\bibnamefont {Bortolami}}, \bibinfo {author}
  {\bibfnamefont {T.}~\bibnamefont {Brinckmann}}, \bibinfo {author}
  {\bibfnamefont {E.}~\bibnamefont {Calabrese}}, \bibinfo {author}
  {\bibfnamefont {F.}~\bibnamefont {Carralot}}, \bibinfo {author}
  {\bibfnamefont {F.~J.}\ \bibnamefont {Casas}}, \bibinfo {author}
  {\bibfnamefont {L.}~\bibnamefont {Clermont}}, \bibinfo {author}
  {\bibfnamefont {F.}~\bibnamefont {Columbro}}, \bibinfo {author}
  {\bibfnamefont {G.}~\bibnamefont {Conenna}}, \bibinfo {author} {\bibfnamefont
  {A.}~\bibnamefont {Coppolecchia}}, \bibinfo {author} {\bibfnamefont
  {F.}~\bibnamefont {Cuttaia}}, \bibinfo {author} {\bibfnamefont
  {G.}~\bibnamefont {D'Alessandro}}, \bibinfo {author} {\bibfnamefont
  {P.}~\bibnamefont {de~Bernardis}}, \bibinfo {author} {\bibfnamefont {M.~D.}\
  \bibnamefont {Petris}}, \bibinfo {author} {\bibfnamefont {S.~D.}\
  \bibnamefont {Torre}}, \bibinfo {author} {\bibfnamefont {E.~D.}\ \bibnamefont
  {Giorgi}}, \bibinfo {author} {\bibfnamefont {P.}~\bibnamefont
  {Diego-Palazuelos}}, \bibinfo {author} {\bibfnamefont {H.~K.}\ \bibnamefont
  {Eriksen}}, \bibinfo {author} {\bibfnamefont {C.}~\bibnamefont
  {Franceschet}}, \bibinfo {author} {\bibfnamefont {U.}~\bibnamefont
  {Fuskeland}}, \bibinfo {author} {\bibfnamefont {M.}~\bibnamefont {Galloway}},
  \bibinfo {author} {\bibfnamefont {M.}~\bibnamefont {Georges}}, \bibinfo
  {author} {\bibfnamefont {M.}~\bibnamefont {Gerbino}}, \bibinfo {author}
  {\bibfnamefont {M.}~\bibnamefont {Gervasi}}, \bibinfo {author} {\bibfnamefont
  {T.}~\bibnamefont {Ghigna}}, \bibinfo {author} {\bibfnamefont
  {C.}~\bibnamefont {Gimeno-Amo}}, \bibinfo {author} {\bibfnamefont
  {E.}~\bibnamefont {Gjerløw}}, \bibinfo {author} {\bibfnamefont
  {A.}~\bibnamefont {Gruppuso}}, \bibinfo {author} {\bibfnamefont
  {J.}~\bibnamefont {Gudmundsson}}, \bibinfo {author} {\bibfnamefont
  {N.}~\bibnamefont {Krachmalnicoff}}, \bibinfo {author} {\bibfnamefont
  {L.}~\bibnamefont {Lamagna}}, \bibinfo {author} {\bibfnamefont
  {M.}~\bibnamefont {Lattanzi}}, \bibinfo {author} {\bibfnamefont
  {M.}~\bibnamefont {Lembo}}, \bibinfo {author} {\bibfnamefont {A.~I.}\
  \bibnamefont {Lonappan}}, \bibinfo {author} {\bibfnamefont {S.}~\bibnamefont
  {Masi}}, \bibinfo {author} {\bibfnamefont {M.}~\bibnamefont {Massa}},
  \bibinfo {author} {\bibfnamefont {S.}~\bibnamefont {Micheli}}, \bibinfo
  {author} {\bibfnamefont {A.}~\bibnamefont {Moggi}}, \bibinfo {author}
  {\bibfnamefont {M.}~\bibnamefont {Monelli}}, \bibinfo {author} {\bibfnamefont
  {G.}~\bibnamefont {Morgante}}, \bibinfo {author} {\bibfnamefont
  {B.}~\bibnamefont {Mot}}, \bibinfo {author} {\bibfnamefont {L.}~\bibnamefont
  {Mousset}}, \bibinfo {author} {\bibfnamefont {R.}~\bibnamefont {Nagata}},
  \bibinfo {author} {\bibfnamefont {P.}~\bibnamefont {Natoli}}, \bibinfo
  {author} {\bibfnamefont {A.}~\bibnamefont {Novelli}}, \bibinfo {author}
  {\bibfnamefont {I.}~\bibnamefont {Obata}}, \bibinfo {author} {\bibfnamefont
  {L.}~\bibnamefont {Pagano}}, \bibinfo {author} {\bibfnamefont
  {A.}~\bibnamefont {Paiella}}, \bibinfo {author} {\bibfnamefont
  {V.}~\bibnamefont {Pavlidou}}, \bibinfo {author} {\bibfnamefont
  {F.}~\bibnamefont {Piacentini}}, \bibinfo {author} {\bibfnamefont
  {M.}~\bibnamefont {Pinchera}}, \bibinfo {author} {\bibfnamefont
  {G.}~\bibnamefont {Pisano}}, \bibinfo {author} {\bibfnamefont
  {G.}~\bibnamefont {Puglisi}}, \bibinfo {author} {\bibfnamefont
  {N.}~\bibnamefont {Raffuzzi}}, \bibinfo {author} {\bibfnamefont
  {A.}~\bibnamefont {Ritacco}}, \bibinfo {author} {\bibfnamefont
  {A.}~\bibnamefont {Rizzieri}}, \bibinfo {author} {\bibfnamefont
  {M.}~\bibnamefont {Ruiz-Granda}}, \bibinfo {author} {\bibfnamefont
  {G.}~\bibnamefont {Savini}}, \bibinfo {author} {\bibfnamefont
  {D.}~\bibnamefont {Scott}}, \bibinfo {author} {\bibfnamefont
  {G.}~\bibnamefont {Signorelli}}, \bibinfo {author} {\bibfnamefont {S.~L.}\
  \bibnamefont {Stever}}, \bibinfo {author} {\bibfnamefont {N.}~\bibnamefont
  {Stutzer}}, \bibinfo {author} {\bibfnamefont {R.~M.}\ \bibnamefont
  {Sullivan}}, \bibinfo {author} {\bibfnamefont {A.}~\bibnamefont {Tartari}},
  \bibinfo {author} {\bibfnamefont {K.}~\bibnamefont {Tassis}}, \bibinfo
  {author} {\bibfnamefont {L.}~\bibnamefont {Terenzi}}, \bibinfo {author}
  {\bibfnamefont {K.~L.}\ \bibnamefont {Thompson}}, \bibinfo {author}
  {\bibfnamefont {P.}~\bibnamefont {Vielva}}, \bibinfo {author} {\bibfnamefont
  {I.~K.}\ \bibnamefont {Wehus}},\ and\ \bibinfo {author} {\bibfnamefont
  {Y.}~\bibnamefont {Zhou}},\ }\href@noop {} {} (\bibinfo {year} {2023}),\
  \Eprint {https://arxiv.org/abs/2312.00717} {arXiv:2312.00717 [astro-ph.CO]}
  \BibitemShut {NoStop}%
\bibitem [{\citenamefont {Maldacena}\ and\ \citenamefont
  {Pimentel}(2011)}]{maldacena2011graviton}%
  \BibitemOpen
  \bibfield  {author} {\bibinfo {author} {\bibfnamefont {J.~M.}\ \bibnamefont
  {Maldacena}}\ and\ \bibinfo {author} {\bibfnamefont {G.~L.}\ \bibnamefont
  {Pimentel}},\ }\href@noop {} {\bibfield  {journal} {\bibinfo  {journal}
  {Journal of High Energy Physics}\ }\textbf {\bibinfo {volume} {2011}},\
  \bibinfo {pages} {1} (\bibinfo {year} {2011})}\BibitemShut {NoStop}%
\bibitem [{\citenamefont {Anber}\ and\ \citenamefont
  {Sorbo}(2012)}]{anber2012non}%
  \BibitemOpen
  \bibfield  {author} {\bibinfo {author} {\bibfnamefont {M.~M.}\ \bibnamefont
  {Anber}}\ and\ \bibinfo {author} {\bibfnamefont {L.}~\bibnamefont {Sorbo}},\
  }\href@noop {} {\bibfield  {journal} {\bibinfo  {journal} {Physical Review
  D}\ }\textbf {\bibinfo {volume} {85}},\ \bibinfo {pages} {123537} (\bibinfo
  {year} {2012})}\BibitemShut {NoStop}%
\bibitem [{\citenamefont {Shiraishi}\ \emph {et~al.}(2013)\citenamefont
  {Shiraishi}, \citenamefont {Ricciardone},\ and\ \citenamefont
  {Saga}}]{shiraishi2013parity}%
  \BibitemOpen
  \bibfield  {author} {\bibinfo {author} {\bibfnamefont {M.}~\bibnamefont
  {Shiraishi}}, \bibinfo {author} {\bibfnamefont {A.}~\bibnamefont
  {Ricciardone}},\ and\ \bibinfo {author} {\bibfnamefont {S.}~\bibnamefont
  {Saga}},\ }\href@noop {} {\bibfield  {journal} {\bibinfo  {journal} {Journal
  of Cosmology and Astroparticle Physics}\ }\textbf {\bibinfo {volume}
  {2013}}\bibinfo  {number} { (11)},\ \bibinfo {pages} {051}}\BibitemShut
  {NoStop}%
\bibitem [{\citenamefont {Shiraishi}(2013)}]{shiraishi2013probing}%
  \BibitemOpen
\bibfield  {number} {  }\bibfield  {author} {\bibinfo {author} {\bibfnamefont
  {M.}~\bibnamefont {Shiraishi}},\ }\href@noop {} {}\ (\bibinfo  {publisher}
  {Springer Science \& Business Media},\ \bibinfo {year} {2013})\BibitemShut
  {NoStop}%
\bibitem [{\citenamefont {Cook}\ and\ \citenamefont
  {Sorbo}(2013)}]{cook2013inflationary}%
  \BibitemOpen
  \bibfield  {author} {\bibinfo {author} {\bibfnamefont {J.~L.}\ \bibnamefont
  {Cook}}\ and\ \bibinfo {author} {\bibfnamefont {L.}~\bibnamefont {Sorbo}},\
  }\href@noop {} {\bibfield  {journal} {\bibinfo  {journal} {Journal of
  Cosmology and Astroparticle Physics}\ }\textbf {\bibinfo {volume}
  {2013}}\bibinfo  {number} { (11)},\ \bibinfo {pages} {047}}\BibitemShut
  {NoStop}%
\bibitem [{\citenamefont {Shiraishi}\ \emph {et~al.}(2015)\citenamefont
  {Shiraishi}, \citenamefont {Liguori},\ and\ \citenamefont
  {Fergusson}}]{shiraishi2015observed}%
  \BibitemOpen
\bibfield  {number} {  }\bibfield  {author} {\bibinfo {author} {\bibfnamefont
  {M.}~\bibnamefont {Shiraishi}}, \bibinfo {author} {\bibfnamefont
  {M.}~\bibnamefont {Liguori}},\ and\ \bibinfo {author} {\bibfnamefont {J.~R.}\
  \bibnamefont {Fergusson}},\ }\href@noop {} {\bibfield  {journal} {\bibinfo
  {journal} {Journal of Cosmology and Astroparticle Physics}\ }\textbf
  {\bibinfo {volume} {2015}}\bibinfo  {number} { (01)},\ \bibinfo {pages}
  {007}}\BibitemShut {NoStop}%
\bibitem [{\citenamefont {Shiraishi}(2016)}]{shiraishi2016parity}%
  \BibitemOpen
\bibfield  {number} {  }\bibfield  {author} {\bibinfo {author} {\bibfnamefont
  {M.}~\bibnamefont {Shiraishi}},\ }\href@noop {} {\bibfield  {journal}
  {\bibinfo  {journal} {Physical Review D}\ }\textbf {\bibinfo {volume} {94}},\
  \bibinfo {pages} {083503} (\bibinfo {year} {2016})}\BibitemShut {NoStop}%
\bibitem [{\citenamefont {Meerburg}\ \emph {et~al.}(2016)\citenamefont
  {Meerburg}, \citenamefont {Meyers}, \citenamefont {Van~Engelen},\ and\
  \citenamefont {Ali-Ha{\"\i}moud}}]{meerburg2016cmb}%
  \BibitemOpen
  \bibfield  {author} {\bibinfo {author} {\bibfnamefont {P.~D.}\ \bibnamefont
  {Meerburg}}, \bibinfo {author} {\bibfnamefont {J.}~\bibnamefont {Meyers}},
  \bibinfo {author} {\bibfnamefont {A.}~\bibnamefont {Van~Engelen}},\ and\
  \bibinfo {author} {\bibfnamefont {Y.}~\bibnamefont {Ali-Ha{\"\i}moud}},\
  }\href@noop {} {\bibfield  {journal} {\bibinfo  {journal} {Physical Review
  D}\ }\textbf {\bibinfo {volume} {93}},\ \bibinfo {pages} {123511} (\bibinfo
  {year} {2016})}\BibitemShut {NoStop}%
\bibitem [{\citenamefont {Bartolo}\ and\ \citenamefont
  {Orlando}(2017)}]{bartolo2017parity}%
  \BibitemOpen
  \bibfield  {author} {\bibinfo {author} {\bibfnamefont {N.}~\bibnamefont
  {Bartolo}}\ and\ \bibinfo {author} {\bibfnamefont {G.}~\bibnamefont
  {Orlando}},\ }\href@noop {} {\bibfield  {journal} {\bibinfo  {journal}
  {Journal of Cosmology and Astroparticle Physics}\ }\textbf {\bibinfo {volume}
  {2017}}\bibinfo  {number} { (07)},\ \bibinfo {pages} {034}}\BibitemShut
  {NoStop}%
\bibitem [{\citenamefont {Bartolo}\ \emph {et~al.}(2019)\citenamefont
  {Bartolo}, \citenamefont {Orlando},\ and\ \citenamefont
  {Shiraishi}}]{bartolo2019measuring}%
  \BibitemOpen
\bibfield  {number} {  }\bibfield  {author} {\bibinfo {author} {\bibfnamefont
  {N.}~\bibnamefont {Bartolo}}, \bibinfo {author} {\bibfnamefont
  {G.}~\bibnamefont {Orlando}},\ and\ \bibinfo {author} {\bibfnamefont
  {M.}~\bibnamefont {Shiraishi}},\ }\href@noop {} {\bibfield  {journal}
  {\bibinfo  {journal} {Journal of Cosmology and Astroparticle Physics}\
  }\textbf {\bibinfo {volume} {2019}}\bibinfo  {number} { (01)},\ \bibinfo
  {pages} {050}}\BibitemShut {NoStop}%
\bibitem [{\citenamefont {Aghanim}\ \emph {et~al.}(2020)\citenamefont
  {Aghanim}, \citenamefont {Akrami}, \citenamefont {Ashdown}, \citenamefont
  {Aumont}, \citenamefont {Baccigalupi}, \citenamefont {Ballardini},
  \citenamefont {Banday}, \citenamefont {Barreiro}, \citenamefont {Bartolo},
  \citenamefont {Basak} \emph {et~al.}}]{aghanim2020planck}%
  \BibitemOpen
\bibfield  {number} {  }\bibfield  {author} {\bibinfo {author} {\bibfnamefont
  {N.}~\bibnamefont {Aghanim}}, \bibinfo {author} {\bibfnamefont
  {Y.}~\bibnamefont {Akrami}}, \bibinfo {author} {\bibfnamefont
  {M.}~\bibnamefont {Ashdown}}, \bibinfo {author} {\bibfnamefont
  {J.}~\bibnamefont {Aumont}}, \bibinfo {author} {\bibfnamefont
  {C.}~\bibnamefont {Baccigalupi}}, \bibinfo {author} {\bibfnamefont
  {M.}~\bibnamefont {Ballardini}}, \bibinfo {author} {\bibfnamefont
  {A.}~\bibnamefont {Banday}}, \bibinfo {author} {\bibfnamefont
  {R.}~\bibnamefont {Barreiro}}, \bibinfo {author} {\bibfnamefont
  {N.}~\bibnamefont {Bartolo}}, \bibinfo {author} {\bibfnamefont
  {S.}~\bibnamefont {Basak}}, \emph {et~al.},\ }\href@noop {} {\bibfield
  {journal} {\bibinfo  {journal} {Astronomy \& Astrophysics}\ }\textbf
  {\bibinfo {volume} {641}},\ \bibinfo {pages} {A6} (\bibinfo {year}
  {2020})}\BibitemShut {NoStop}%
\bibitem [{\citenamefont {Liu}\ \emph {et~al.}(2020)\citenamefont {Liu},
  \citenamefont {Tong}, \citenamefont {Wang},\ and\ \citenamefont
  {Xianyu}}]{liu2020probing}%
  \BibitemOpen
  \bibfield  {author} {\bibinfo {author} {\bibfnamefont {T.}~\bibnamefont
  {Liu}}, \bibinfo {author} {\bibfnamefont {X.}~\bibnamefont {Tong}}, \bibinfo
  {author} {\bibfnamefont {Y.}~\bibnamefont {Wang}},\ and\ \bibinfo {author}
  {\bibfnamefont {Z.-Z.}\ \bibnamefont {Xianyu}},\ }\href@noop {} {\bibfield
  {journal} {\bibinfo  {journal} {Journal of High Energy Physics}\ }\textbf
  {\bibinfo {volume} {2020}},\ \bibinfo {pages} {1} (\bibinfo {year}
  {2020})}\BibitemShut {NoStop}%
\bibitem [{\citenamefont {Duivenvoorden}\ \emph {et~al.}(2020)\citenamefont
  {Duivenvoorden}, \citenamefont {Meerburg},\ and\ \citenamefont
  {Freese}}]{duivenvoorden2020cmb}%
  \BibitemOpen
  \bibfield  {author} {\bibinfo {author} {\bibfnamefont {A.~J.}\ \bibnamefont
  {Duivenvoorden}}, \bibinfo {author} {\bibfnamefont {P.~D.}\ \bibnamefont
  {Meerburg}},\ and\ \bibinfo {author} {\bibfnamefont {K.}~\bibnamefont
  {Freese}},\ }\href@noop {} {\bibfield  {journal} {\bibinfo  {journal}
  {Physical Review D}\ }\textbf {\bibinfo {volume} {102}},\ \bibinfo {pages}
  {023521} (\bibinfo {year} {2020})}\BibitemShut {NoStop}%
\bibitem [{\citenamefont {Bartolo}\ \emph {et~al.}(2021)\citenamefont
  {Bartolo}, \citenamefont {Caloni}, \citenamefont {Orlando},\ and\
  \citenamefont {Ricciardone}}]{bartolo2021tensor}%
  \BibitemOpen
  \bibfield  {author} {\bibinfo {author} {\bibfnamefont {N.}~\bibnamefont
  {Bartolo}}, \bibinfo {author} {\bibfnamefont {L.}~\bibnamefont {Caloni}},
  \bibinfo {author} {\bibfnamefont {G.}~\bibnamefont {Orlando}},\ and\ \bibinfo
  {author} {\bibfnamefont {A.}~\bibnamefont {Ricciardone}},\ }\href@noop {}
  {\bibfield  {journal} {\bibinfo  {journal} {Journal of Cosmology and
  Astroparticle Physics}\ }\textbf {\bibinfo {volume} {2021}}\bibinfo  {number}
  { (03)},\ \bibinfo {pages} {073}}\BibitemShut {NoStop}%
\bibitem [{\citenamefont {Niu}\ \emph {et~al.}(2022{\natexlab{a}})\citenamefont
  {Niu}, \citenamefont {Rahat}, \citenamefont {Srinivasan},\ and\ \citenamefont
  {Xue}}]{niu2022parity}%
  \BibitemOpen
\bibfield  {number} {  }\bibfield  {author} {\bibinfo {author} {\bibfnamefont
  {X.}~\bibnamefont {Niu}}, \bibinfo {author} {\bibfnamefont {M.~H.}\
  \bibnamefont {Rahat}}, \bibinfo {author} {\bibfnamefont {K.}~\bibnamefont
  {Srinivasan}},\ and\ \bibinfo {author} {\bibfnamefont {W.}~\bibnamefont
  {Xue}},\ }\href@noop {} {\bibfield  {journal} {\bibinfo  {journal} {arXiv
  preprint arXiv:2211.14324}\ } (\bibinfo {year}
  {2022}{\natexlab{a}})}\BibitemShut {NoStop}%
\bibitem [{\citenamefont {Philcox}(2023)}]{philcox2023cmb}%
  \BibitemOpen
  \bibfield  {author} {\bibinfo {author} {\bibfnamefont {O.~H.~E.}\
  \bibnamefont {Philcox}},\ }\href@noop {} {} (\bibinfo {year} {2023}),\
  \Eprint {https://arxiv.org/abs/2303.12106} {arXiv:2303.12106 [astro-ph.CO]}
  \BibitemShut {NoStop}%
\bibitem [{\citenamefont {Philcox}\ and\ \citenamefont
  {Shiraishi}(2023{\natexlab{a}})}]{philcox2023testing}%
  \BibitemOpen
  \bibfield  {author} {\bibinfo {author} {\bibfnamefont {O.~H.~E.}\
  \bibnamefont {Philcox}}\ and\ \bibinfo {author} {\bibfnamefont
  {M.}~\bibnamefont {Shiraishi}},\ }\href@noop {} {} (\bibinfo {year}
  {2023}{\natexlab{a}}),\ \Eprint {https://arxiv.org/abs/2308.03831}
  {arXiv:2308.03831 [astro-ph.CO]} \BibitemShut {NoStop}%
\bibitem [{\citenamefont {Philcox}\ and\ \citenamefont
  {Shiraishi}(2023{\natexlab{b}})}]{philcox2023testing2}%
  \BibitemOpen
  \bibfield  {author} {\bibinfo {author} {\bibfnamefont {O.~H.~E.}\
  \bibnamefont {Philcox}}\ and\ \bibinfo {author} {\bibfnamefont
  {M.}~\bibnamefont {Shiraishi}},\ }\href@noop {} {} (\bibinfo {year}
  {2023}{\natexlab{b}}),\ \Eprint {https://arxiv.org/abs/2312.12498}
  {arXiv:2312.12498 [astro-ph.CO]} \BibitemShut {NoStop}%
\bibitem [{\citenamefont {Dai}\ \emph {et~al.}(2016)\citenamefont {Dai},
  \citenamefont {Kamionkowski}, \citenamefont {Kovetz}, \citenamefont
  {Raccanelli},\ and\ \citenamefont {Shiraishi}}]{dai2016antisymmetric}%
  \BibitemOpen
  \bibfield  {author} {\bibinfo {author} {\bibfnamefont {L.}~\bibnamefont
  {Dai}}, \bibinfo {author} {\bibfnamefont {M.}~\bibnamefont {Kamionkowski}},
  \bibinfo {author} {\bibfnamefont {E.~D.}\ \bibnamefont {Kovetz}}, \bibinfo
  {author} {\bibfnamefont {A.}~\bibnamefont {Raccanelli}},\ and\ \bibinfo
  {author} {\bibfnamefont {M.}~\bibnamefont {Shiraishi}},\ }\href@noop {}
  {\bibfield  {journal} {\bibinfo  {journal} {Physical Review D}\ }\textbf
  {\bibinfo {volume} {93}},\ \bibinfo {pages} {023507} (\bibinfo {year}
  {2016})}\BibitemShut {NoStop}%
\bibitem [{\citenamefont {Cahn}\ \emph {et~al.}(2021)\citenamefont {Cahn},
  \citenamefont {Slepian},\ and\ \citenamefont {Hou}}]{cahn2021test}%
  \BibitemOpen
  \bibfield  {author} {\bibinfo {author} {\bibfnamefont {R.~N.}\ \bibnamefont
  {Cahn}}, \bibinfo {author} {\bibfnamefont {Z.}~\bibnamefont {Slepian}},\ and\
  \bibinfo {author} {\bibfnamefont {J.}~\bibnamefont {Hou}},\ }\href@noop {}
  {\bibfield  {journal} {\bibinfo  {journal} {arXiv preprint arXiv:2110.12004}\
  } (\bibinfo {year} {2021})}\BibitemShut {NoStop}%
\bibitem [{\citenamefont {Hou}\ \emph {et~al.}(2022)\citenamefont {Hou},
  \citenamefont {Slepian},\ and\ \citenamefont {Cahn}}]{hou2022measurement}%
  \BibitemOpen
  \bibfield  {author} {\bibinfo {author} {\bibfnamefont {J.}~\bibnamefont
  {Hou}}, \bibinfo {author} {\bibfnamefont {Z.}~\bibnamefont {Slepian}},\ and\
  \bibinfo {author} {\bibfnamefont {R.~N.}\ \bibnamefont {Cahn}},\ }\href@noop
  {} {\bibfield  {journal} {\bibinfo  {journal} {arXiv preprint
  arXiv:2206.03625}\ } (\bibinfo {year} {2022})}\BibitemShut {NoStop}%
\bibitem [{\citenamefont {Philcox}(2022)}]{philcox2022probing}%
  \BibitemOpen
  \bibfield  {author} {\bibinfo {author} {\bibfnamefont {O.~H.}\ \bibnamefont
  {Philcox}},\ }\href@noop {} {\bibfield  {journal} {\bibinfo  {journal}
  {Physical Review D}\ }\textbf {\bibinfo {volume} {106}},\ \bibinfo {pages}
  {063501} (\bibinfo {year} {2022})}\BibitemShut {NoStop}%
\bibitem [{\citenamefont {Cabass}\ \emph
  {et~al.}(2023{\natexlab{a}})\citenamefont {Cabass}, \citenamefont {Ivanov},\
  and\ \citenamefont {Philcox}}]{cabass2023colliders}%
  \BibitemOpen
  \bibfield  {author} {\bibinfo {author} {\bibfnamefont {G.}~\bibnamefont
  {Cabass}}, \bibinfo {author} {\bibfnamefont {M.~M.}\ \bibnamefont {Ivanov}},\
  and\ \bibinfo {author} {\bibfnamefont {O.~H.}\ \bibnamefont {Philcox}},\
  }\href@noop {} {\bibfield  {journal} {\bibinfo  {journal} {Physical Review
  D}\ }\textbf {\bibinfo {volume} {107}},\ \bibinfo {pages} {023523} (\bibinfo
  {year} {2023}{\natexlab{a}})}\BibitemShut {NoStop}%
\bibitem [{\citenamefont {Cabass}\ \emph
  {et~al.}(2023{\natexlab{b}})\citenamefont {Cabass}, \citenamefont {Jazayeri},
  \citenamefont {Pajer},\ and\ \citenamefont {Stefanyszyn}}]{cabass2023parity}%
  \BibitemOpen
  \bibfield  {author} {\bibinfo {author} {\bibfnamefont {G.}~\bibnamefont
  {Cabass}}, \bibinfo {author} {\bibfnamefont {S.}~\bibnamefont {Jazayeri}},
  \bibinfo {author} {\bibfnamefont {E.}~\bibnamefont {Pajer}},\ and\ \bibinfo
  {author} {\bibfnamefont {D.}~\bibnamefont {Stefanyszyn}},\ }\href@noop {}
  {\bibfield  {journal} {\bibinfo  {journal} {Journal of High Energy Physics}\
  }\textbf {\bibinfo {volume} {2023}},\ \bibinfo {pages} {1} (\bibinfo {year}
  {2023}{\natexlab{b}})}\BibitemShut {NoStop}%
\bibitem [{\citenamefont {Creque-Sarbinowski}\ \emph
  {et~al.}(2022)\citenamefont {Creque-Sarbinowski}, \citenamefont {Alexander},
  \citenamefont {Kamionkowski},\ and\ \citenamefont
  {Philcox}}]{creque2023parity}%
  \BibitemOpen
  \bibfield  {author} {\bibinfo {author} {\bibfnamefont {C.}~\bibnamefont
  {Creque-Sarbinowski}}, \bibinfo {author} {\bibfnamefont {S.}~\bibnamefont
  {Alexander}}, \bibinfo {author} {\bibfnamefont {M.}~\bibnamefont
  {Kamionkowski}},\ and\ \bibinfo {author} {\bibfnamefont {O.}~\bibnamefont
  {Philcox}},\ }\href@noop {} {\bibfield  {journal} {\bibinfo  {journal} {arXiv
  preprint arXiv:2303.04815}\ } (\bibinfo {year} {2022})}\BibitemShut {NoStop}%
\bibitem [{\citenamefont {Coulton}\ \emph {et~al.}(2023)\citenamefont
  {Coulton}, \citenamefont {Philcox},\ and\ \citenamefont
  {Villaescusa-Navarro}}]{coulton2023signatures}%
  \BibitemOpen
  \bibfield  {author} {\bibinfo {author} {\bibfnamefont {W.~R.}\ \bibnamefont
  {Coulton}}, \bibinfo {author} {\bibfnamefont {O.~H.~E.}\ \bibnamefont
  {Philcox}},\ and\ \bibinfo {author} {\bibfnamefont {F.}~\bibnamefont
  {Villaescusa-Navarro}},\ }\href@noop {} {} (\bibinfo {year} {2023}),\ \Eprint
  {https://arxiv.org/abs/2306.11782} {arXiv:2306.11782 [astro-ph.CO]}
  \BibitemShut {NoStop}%
\bibitem [{\citenamefont {Philcox}\ \emph {et~al.}(2023)\citenamefont
  {Philcox}, \citenamefont {König}, \citenamefont {Alexander},\ and\
  \citenamefont {Spergel}}]{philcox2023galaxy}%
  \BibitemOpen
  \bibfield  {author} {\bibinfo {author} {\bibfnamefont {O.~H.~E.}\
  \bibnamefont {Philcox}}, \bibinfo {author} {\bibfnamefont {M.~J.}\
  \bibnamefont {König}}, \bibinfo {author} {\bibfnamefont {S.}~\bibnamefont
  {Alexander}},\ and\ \bibinfo {author} {\bibfnamefont {D.~N.}\ \bibnamefont
  {Spergel}},\ }\href@noop {} {} (\bibinfo {year} {2023}),\ \Eprint
  {https://arxiv.org/abs/2309.08653} {arXiv:2309.08653 [astro-ph.CO]}
  \BibitemShut {NoStop}%
\bibitem [{\citenamefont {Taylor}\ \emph {et~al.}(2023)\citenamefont {Taylor},
  \citenamefont {Craigie},\ and\ \citenamefont
  {Ting}}]{taylor2023unsupervised}%
  \BibitemOpen
  \bibfield  {author} {\bibinfo {author} {\bibfnamefont {P.~L.}\ \bibnamefont
  {Taylor}}, \bibinfo {author} {\bibfnamefont {M.}~\bibnamefont {Craigie}},\
  and\ \bibinfo {author} {\bibfnamefont {Y.-S.}\ \bibnamefont {Ting}},\
  }\href@noop {} {} (\bibinfo {year} {2023}),\ \Eprint
  {https://arxiv.org/abs/2312.09287} {arXiv:2312.09287 [astro-ph.CO]}
  \BibitemShut {NoStop}%
\bibitem [{\citenamefont {Yunes}\ \emph {et~al.}(2010)\citenamefont {Yunes},
  \citenamefont {O’Shaughnessy}, \citenamefont {Owen},\ and\ \citenamefont
  {Alexander}}]{yunes2010testing}%
  \BibitemOpen
  \bibfield  {author} {\bibinfo {author} {\bibfnamefont {N.}~\bibnamefont
  {Yunes}}, \bibinfo {author} {\bibfnamefont {R.}~\bibnamefont
  {O’Shaughnessy}}, \bibinfo {author} {\bibfnamefont {B.~J.}\ \bibnamefont
  {Owen}},\ and\ \bibinfo {author} {\bibfnamefont {S.}~\bibnamefont
  {Alexander}},\ }\href@noop {} {\bibfield  {journal} {\bibinfo  {journal}
  {Physical Review D}\ }\textbf {\bibinfo {volume} {82}},\ \bibinfo {pages}
  {064017} (\bibinfo {year} {2010})}\BibitemShut {NoStop}%
\bibitem [{\citenamefont {Crowder}\ \emph {et~al.}(2013)\citenamefont
  {Crowder}, \citenamefont {Namba}, \citenamefont {Mandic}, \citenamefont
  {Mukohyama},\ and\ \citenamefont {Peloso}}]{crowder2013measurement}%
  \BibitemOpen
  \bibfield  {author} {\bibinfo {author} {\bibfnamefont {S.}~\bibnamefont
  {Crowder}}, \bibinfo {author} {\bibfnamefont {R.}~\bibnamefont {Namba}},
  \bibinfo {author} {\bibfnamefont {V.}~\bibnamefont {Mandic}}, \bibinfo
  {author} {\bibfnamefont {S.}~\bibnamefont {Mukohyama}},\ and\ \bibinfo
  {author} {\bibfnamefont {M.}~\bibnamefont {Peloso}},\ }\href@noop {}
  {\bibfield  {journal} {\bibinfo  {journal} {Physics Letters B}\ }\textbf
  {\bibinfo {volume} {726}},\ \bibinfo {pages} {66} (\bibinfo {year}
  {2013})}\BibitemShut {NoStop}%
\bibitem [{\citenamefont {Kosteleck{\`y}}\ and\ \citenamefont
  {Mewes}(2016)}]{kostelecky2016testing}%
  \BibitemOpen
  \bibfield  {author} {\bibinfo {author} {\bibfnamefont {V.~A.}\ \bibnamefont
  {Kosteleck{\`y}}}\ and\ \bibinfo {author} {\bibfnamefont {M.}~\bibnamefont
  {Mewes}},\ }\href@noop {} {\bibfield  {journal} {\bibinfo  {journal} {Physics
  Letters B}\ }\textbf {\bibinfo {volume} {757}},\ \bibinfo {pages} {510}
  (\bibinfo {year} {2016})}\BibitemShut {NoStop}%
\bibitem [{\citenamefont {Alexander}\ and\ \citenamefont
  {Yunes}(2018)}]{alexander2018gravitational}%
  \BibitemOpen
  \bibfield  {author} {\bibinfo {author} {\bibfnamefont {S.~H.}\ \bibnamefont
  {Alexander}}\ and\ \bibinfo {author} {\bibfnamefont {N.}~\bibnamefont
  {Yunes}},\ }\href@noop {} {\bibfield  {journal} {\bibinfo  {journal}
  {Physical Review D}\ }\textbf {\bibinfo {volume} {97}},\ \bibinfo {pages}
  {064033} (\bibinfo {year} {2018})}\BibitemShut {NoStop}%
\bibitem [{\citenamefont {Yagi}\ and\ \citenamefont
  {Yang}(2018)}]{yagi2018probing}%
  \BibitemOpen
  \bibfield  {author} {\bibinfo {author} {\bibfnamefont {K.}~\bibnamefont
  {Yagi}}\ and\ \bibinfo {author} {\bibfnamefont {H.}~\bibnamefont {Yang}},\
  }\href@noop {} {\bibfield  {journal} {\bibinfo  {journal} {Physical Review
  D}\ }\textbf {\bibinfo {volume} {97}},\ \bibinfo {pages} {104018} (\bibinfo
  {year} {2018})}\BibitemShut {NoStop}%
\bibitem [{\citenamefont {Shao}(2020)}]{shao2020combined}%
  \BibitemOpen
  \bibfield  {author} {\bibinfo {author} {\bibfnamefont {L.}~\bibnamefont
  {Shao}},\ }\href@noop {} {\bibfield  {journal} {\bibinfo  {journal} {Physical
  Review D}\ }\textbf {\bibinfo {volume} {101}},\ \bibinfo {pages} {104019}
  (\bibinfo {year} {2020})}\BibitemShut {NoStop}%
\bibitem [{\citenamefont {Orlando}\ \emph {et~al.}(2021)\citenamefont
  {Orlando}, \citenamefont {Pieroni},\ and\ \citenamefont
  {Ricciardone}}]{orlando2021measuring}%
  \BibitemOpen
  \bibfield  {author} {\bibinfo {author} {\bibfnamefont {G.}~\bibnamefont
  {Orlando}}, \bibinfo {author} {\bibfnamefont {M.}~\bibnamefont {Pieroni}},\
  and\ \bibinfo {author} {\bibfnamefont {A.}~\bibnamefont {Ricciardone}},\
  }\href@noop {} {\bibfield  {journal} {\bibinfo  {journal} {Journal of
  Cosmology and Astroparticle Physics}\ }\textbf {\bibinfo {volume}
  {2021}}\bibinfo  {number} { (03)},\ \bibinfo {pages} {069}}\BibitemShut
  {NoStop}%
\bibitem [{\citenamefont {Wang}\ \emph {et~al.}(2021)\citenamefont {Wang},
  \citenamefont {Niu}, \citenamefont {Zhu},\ and\ \citenamefont
  {Zhao}}]{wang2021gravitational}%
  \BibitemOpen
\bibfield  {number} {  }\bibfield  {author} {\bibinfo {author} {\bibfnamefont
  {Y.-F.}\ \bibnamefont {Wang}}, \bibinfo {author} {\bibfnamefont
  {R.}~\bibnamefont {Niu}}, \bibinfo {author} {\bibfnamefont {T.}~\bibnamefont
  {Zhu}},\ and\ \bibinfo {author} {\bibfnamefont {W.}~\bibnamefont {Zhao}},\
  }\href@noop {} {\bibfield  {journal} {\bibinfo  {journal} {The Astrophysical
  Journal}\ }\textbf {\bibinfo {volume} {908}},\ \bibinfo {pages} {58}
  (\bibinfo {year} {2021})}\BibitemShut {NoStop}%
\bibitem [{\citenamefont {Martinovic}\ \emph {et~al.}(2021)\citenamefont
  {Martinovic}, \citenamefont {Badger}, \citenamefont {Sakellariadou},\ and\
  \citenamefont {Mandic}}]{martinovic2021searching}%
  \BibitemOpen
  \bibfield  {author} {\bibinfo {author} {\bibfnamefont {K.}~\bibnamefont
  {Martinovic}}, \bibinfo {author} {\bibfnamefont {C.}~\bibnamefont {Badger}},
  \bibinfo {author} {\bibfnamefont {M.}~\bibnamefont {Sakellariadou}},\ and\
  \bibinfo {author} {\bibfnamefont {V.}~\bibnamefont {Mandic}},\ }\href@noop {}
  {\bibfield  {journal} {\bibinfo  {journal} {Physical Review D}\ }\textbf
  {\bibinfo {volume} {104}},\ \bibinfo {pages} {L081101} (\bibinfo {year}
  {2021})}\BibitemShut {NoStop}%
\bibitem [{\citenamefont {Niu}\ \emph {et~al.}(2022{\natexlab{b}})\citenamefont
  {Niu}, \citenamefont {Zhu},\ and\ \citenamefont
  {Zhao}}]{niu2022constraining}%
  \BibitemOpen
  \bibfield  {author} {\bibinfo {author} {\bibfnamefont {R.}~\bibnamefont
  {Niu}}, \bibinfo {author} {\bibfnamefont {T.}~\bibnamefont {Zhu}},\ and\
  \bibinfo {author} {\bibfnamefont {W.}~\bibnamefont {Zhao}},\ }\href@noop {}
  {\bibfield  {journal} {\bibinfo  {journal} {arXiv preprint arXiv:2202.05092}\
  } (\bibinfo {year} {2022}{\natexlab{b}})}\BibitemShut {NoStop}%
\bibitem [{\citenamefont {Okounkova}\ \emph {et~al.}(2022)\citenamefont
  {Okounkova}, \citenamefont {Farr}, \citenamefont {Isi},\ and\ \citenamefont
  {Stein}}]{okounkova2022constraining}%
  \BibitemOpen
  \bibfield  {author} {\bibinfo {author} {\bibfnamefont {M.}~\bibnamefont
  {Okounkova}}, \bibinfo {author} {\bibfnamefont {W.~M.}\ \bibnamefont {Farr}},
  \bibinfo {author} {\bibfnamefont {M.}~\bibnamefont {Isi}},\ and\ \bibinfo
  {author} {\bibfnamefont {L.~C.}\ \bibnamefont {Stein}},\ }\href@noop {}
  {\bibfield  {journal} {\bibinfo  {journal} {Physical Review D}\ }\textbf
  {\bibinfo {volume} {106}},\ \bibinfo {pages} {044067} (\bibinfo {year}
  {2022})}\BibitemShut {NoStop}%
\bibitem [{\citenamefont {Califano}\ \emph {et~al.}(2023)\citenamefont
  {Califano}, \citenamefont {D'Agostino},\ and\ \citenamefont
  {Vernieri}}]{califano2023parity}%
  \BibitemOpen
  \bibfield  {author} {\bibinfo {author} {\bibfnamefont {M.}~\bibnamefont
  {Califano}}, \bibinfo {author} {\bibfnamefont {R.}~\bibnamefont
  {D'Agostino}},\ and\ \bibinfo {author} {\bibfnamefont {D.}~\bibnamefont
  {Vernieri}},\ }\href@noop {} {} (\bibinfo {year} {2023}),\ \Eprint
  {https://arxiv.org/abs/2311.02161} {arXiv:2311.02161 [gr-qc]} \BibitemShut
  {NoStop}%
\bibitem [{\citenamefont {Callister}\ \emph {et~al.}(2023)\citenamefont
  {Callister}, \citenamefont {Jenks}, \citenamefont {Holz},\ and\ \citenamefont
  {Yunes}}]{callister2023new}%
  \BibitemOpen
  \bibfield  {author} {\bibinfo {author} {\bibfnamefont {T.}~\bibnamefont
  {Callister}}, \bibinfo {author} {\bibfnamefont {L.}~\bibnamefont {Jenks}},
  \bibinfo {author} {\bibfnamefont {D.}~\bibnamefont {Holz}},\ and\ \bibinfo
  {author} {\bibfnamefont {N.}~\bibnamefont {Yunes}},\ }\href@noop {} {}
  (\bibinfo {year} {2023}),\ \Eprint {https://arxiv.org/abs/2312.12532}
  {arXiv:2312.12532 [gr-qc]} \BibitemShut {NoStop}%
\bibitem [{\citenamefont {Carroll}\ \emph {et~al.}(1990)\citenamefont
  {Carroll}, \citenamefont {Field},\ and\ \citenamefont
  {Jackiw}}]{carroll1990limits}%
  \BibitemOpen
  \bibfield  {author} {\bibinfo {author} {\bibfnamefont {S.~M.}\ \bibnamefont
  {Carroll}}, \bibinfo {author} {\bibfnamefont {G.~B.}\ \bibnamefont {Field}},\
  and\ \bibinfo {author} {\bibfnamefont {R.}~\bibnamefont {Jackiw}},\
  }\href@noop {} {\bibfield  {journal} {\bibinfo  {journal} {Physical Review
  D}\ }\textbf {\bibinfo {volume} {41}},\ \bibinfo {pages} {1231} (\bibinfo
  {year} {1990})}\BibitemShut {NoStop}%
\bibitem [{\citenamefont {Lee}\ \emph {et~al.}(2019)\citenamefont {Lee},
  \citenamefont {Liu},\ and\ \citenamefont {Ng}}]{lee2019dark}%
  \BibitemOpen
  \bibfield  {author} {\bibinfo {author} {\bibfnamefont {S.}~\bibnamefont
  {Lee}}, \bibinfo {author} {\bibfnamefont {G.-C.}\ \bibnamefont {Liu}},\ and\
  \bibinfo {author} {\bibfnamefont {K.-W.}\ \bibnamefont {Ng}},\ }\href@noop {}
  {} (\bibinfo {year} {2019}),\ \Eprint {https://arxiv.org/abs/1912.12903}
  {arXiv:1912.12903 [astro-ph.CO]} \BibitemShut {NoStop}%
\bibitem [{\citenamefont {Namikawa}(2021)}]{namikawa2021cmb}%
  \BibitemOpen
  \bibfield  {author} {\bibinfo {author} {\bibfnamefont {T.}~\bibnamefont
  {Namikawa}},\ }\href@noop {} {\bibfield  {journal} {\bibinfo  {journal}
  {Monthly Notices of the Royal Astronomical Society}\ }\textbf {\bibinfo
  {volume} {506}},\ \bibinfo {pages} {1250} (\bibinfo {year}
  {2021})}\BibitemShut {NoStop}%
\bibitem [{\citenamefont {Naokawa}\ and\ \citenamefont
  {Namikawa}(2023)}]{naokawa2023gravitational}%
  \BibitemOpen
  \bibfield  {author} {\bibinfo {author} {\bibfnamefont {F.}~\bibnamefont
  {Naokawa}}\ and\ \bibinfo {author} {\bibfnamefont {T.}~\bibnamefont
  {Namikawa}},\ }\href@noop {} {} (\bibinfo {year} {2023}),\ \Eprint
  {https://arxiv.org/abs/2305.13976} {arXiv:2305.13976 [astro-ph.CO]}
  \BibitemShut {NoStop}%
\bibitem [{\citenamefont {Poulin}\ \emph {et~al.}(2019)\citenamefont {Poulin},
  \citenamefont {Smith}, \citenamefont {Karwal},\ and\ \citenamefont
  {Kamionkowski}}]{poulin2019early}%
  \BibitemOpen
  \bibfield  {author} {\bibinfo {author} {\bibfnamefont {V.}~\bibnamefont
  {Poulin}}, \bibinfo {author} {\bibfnamefont {T.~L.}\ \bibnamefont {Smith}},
  \bibinfo {author} {\bibfnamefont {T.}~\bibnamefont {Karwal}},\ and\ \bibinfo
  {author} {\bibfnamefont {M.}~\bibnamefont {Kamionkowski}},\ }\href@noop {}
  {\bibfield  {journal} {\bibinfo  {journal} {Physical Review Letters}\
  }\textbf {\bibinfo {volume} {122}},\ \bibinfo {pages} {221301} (\bibinfo
  {year} {2019})}\BibitemShut {NoStop}%
\bibitem [{\citenamefont {Choi}\ \emph {et~al.}(2021)\citenamefont {Choi},
  \citenamefont {Lin}, \citenamefont {Visinelli},\ and\ \citenamefont
  {Yanagida}}]{choi2021cosmic}%
  \BibitemOpen
  \bibfield  {author} {\bibinfo {author} {\bibfnamefont {G.}~\bibnamefont
  {Choi}}, \bibinfo {author} {\bibfnamefont {W.}~\bibnamefont {Lin}}, \bibinfo
  {author} {\bibfnamefont {L.}~\bibnamefont {Visinelli}},\ and\ \bibinfo
  {author} {\bibfnamefont {T.~T.}\ \bibnamefont {Yanagida}},\ }\href@noop {}
  {\bibfield  {journal} {\bibinfo  {journal} {Physical Review D}\ }\textbf
  {\bibinfo {volume} {104}},\ \bibinfo {pages} {L101302} (\bibinfo {year}
  {2021})}\BibitemShut {NoStop}%
\bibitem [{\citenamefont {Fujita}\ \emph {et~al.}(2021)\citenamefont {Fujita},
  \citenamefont {Murai}, \citenamefont {Nakatsuka},\ and\ \citenamefont
  {Tsujikawa}}]{fujita2021detection}%
  \BibitemOpen
  \bibfield  {author} {\bibinfo {author} {\bibfnamefont {T.}~\bibnamefont
  {Fujita}}, \bibinfo {author} {\bibfnamefont {K.}~\bibnamefont {Murai}},
  \bibinfo {author} {\bibfnamefont {H.}~\bibnamefont {Nakatsuka}},\ and\
  \bibinfo {author} {\bibfnamefont {S.}~\bibnamefont {Tsujikawa}},\ }\href@noop
  {} {\bibfield  {journal} {\bibinfo  {journal} {Physical Review D}\ }\textbf
  {\bibinfo {volume} {103}},\ \bibinfo {pages} {043509} (\bibinfo {year}
  {2021})}\BibitemShut {NoStop}%
\bibitem [{\citenamefont {Gasparotto}\ and\ \citenamefont
  {Obata}(2022)}]{gasparotto2022cosmic}%
  \BibitemOpen
  \bibfield  {author} {\bibinfo {author} {\bibfnamefont {S.}~\bibnamefont
  {Gasparotto}}\ and\ \bibinfo {author} {\bibfnamefont {I.}~\bibnamefont
  {Obata}},\ }\href@noop {} {\bibfield  {journal} {\bibinfo  {journal} {arXiv
  preprint arXiv:2203.09409}\ } (\bibinfo {year} {2022})}\BibitemShut {NoStop}%
\bibitem [{\citenamefont {Murai}\ \emph {et~al.}(2022)\citenamefont {Murai},
  \citenamefont {Naokawa}, \citenamefont {Namikawa},\ and\ \citenamefont
  {Komatsu}}]{murai2022isotropic}%
  \BibitemOpen
  \bibfield  {author} {\bibinfo {author} {\bibfnamefont {K.}~\bibnamefont
  {Murai}}, \bibinfo {author} {\bibfnamefont {F.}~\bibnamefont {Naokawa}},
  \bibinfo {author} {\bibfnamefont {T.}~\bibnamefont {Namikawa}},\ and\
  \bibinfo {author} {\bibfnamefont {E.}~\bibnamefont {Komatsu}},\ }\href@noop
  {} {\bibfield  {journal} {\bibinfo  {journal} {arXiv preprint
  arXiv:2209.07804}\ } (\bibinfo {year} {2022})}\BibitemShut {NoStop}%
\bibitem [{\citenamefont {Kamionkowski}\ and\ \citenamefont
  {Riess}(2022)}]{kamionkowski2022the}%
  \BibitemOpen
  \bibfield  {author} {\bibinfo {author} {\bibfnamefont {M.}~\bibnamefont
  {Kamionkowski}}\ and\ \bibinfo {author} {\bibfnamefont {A.~G.}\ \bibnamefont
  {Riess}},\ }\href@noop {} {} (\bibinfo {year} {2022})\BibitemShut {NoStop}%
\bibitem [{\citenamefont {Rezazadeh}\ \emph {et~al.}(2022)\citenamefont
  {Rezazadeh}, \citenamefont {Ashoorioon},\ and\ \citenamefont
  {Grin}}]{rezazadeh2022cascading}%
  \BibitemOpen
  \bibfield  {author} {\bibinfo {author} {\bibfnamefont {K.}~\bibnamefont
  {Rezazadeh}}, \bibinfo {author} {\bibfnamefont {A.}~\bibnamefont
  {Ashoorioon}},\ and\ \bibinfo {author} {\bibfnamefont {D.}~\bibnamefont
  {Grin}},\ }\href@noop {} {\bibfield  {journal} {\bibinfo  {journal} {arXiv
  preprint arXiv:2208.07631}\ } (\bibinfo {year} {2022})}\BibitemShut {NoStop}%
\bibitem [{\citenamefont {Yin}\ \emph {et~al.}(2023{\natexlab{a}})\citenamefont
  {Yin}, \citenamefont {Kochappan}, \citenamefont {Ghosh},\ and\ \citenamefont
  {Lee}}]{yin2023cosmic}%
  \BibitemOpen
  \bibfield  {author} {\bibinfo {author} {\bibfnamefont {L.}~\bibnamefont
  {Yin}}, \bibinfo {author} {\bibfnamefont {J.}~\bibnamefont {Kochappan}},
  \bibinfo {author} {\bibfnamefont {T.}~\bibnamefont {Ghosh}},\ and\ \bibinfo
  {author} {\bibfnamefont {B.-H.}\ \bibnamefont {Lee}},\ }\href@noop {} {}
  (\bibinfo {year} {2023}{\natexlab{a}}),\ \Eprint
  {https://arxiv.org/abs/2305.07937} {arXiv:2305.07937 [astro-ph.CO]}
  \BibitemShut {NoStop}%
\bibitem [{\citenamefont {Preskill}\ \emph {et~al.}(1983)\citenamefont
  {Preskill}, \citenamefont {Wise},\ and\ \citenamefont
  {Wilczek}}]{preskill1983cosmology}%
  \BibitemOpen
  \bibfield  {author} {\bibinfo {author} {\bibfnamefont {J.}~\bibnamefont
  {Preskill}}, \bibinfo {author} {\bibfnamefont {M.~B.}\ \bibnamefont {Wise}},\
  and\ \bibinfo {author} {\bibfnamefont {F.}~\bibnamefont {Wilczek}},\
  }\href@noop {} {\bibfield  {journal} {\bibinfo  {journal} {Physics Letters
  B}\ }\textbf {\bibinfo {volume} {120}},\ \bibinfo {pages} {127} (\bibinfo
  {year} {1983})}\BibitemShut {NoStop}%
\bibitem [{\citenamefont {Abbott}\ and\ \citenamefont
  {Sikivie}(1983)}]{abbott1983cosmological}%
  \BibitemOpen
  \bibfield  {author} {\bibinfo {author} {\bibfnamefont {L.~F.}\ \bibnamefont
  {Abbott}}\ and\ \bibinfo {author} {\bibfnamefont {P.}~\bibnamefont
  {Sikivie}},\ }\href@noop {} {\bibfield  {journal} {\bibinfo  {journal}
  {Physics Letters B}\ }\textbf {\bibinfo {volume} {120}},\ \bibinfo {pages}
  {133} (\bibinfo {year} {1983})}\BibitemShut {NoStop}%
\bibitem [{\citenamefont {Dine}\ and\ \citenamefont
  {Fischler}(1983)}]{dine1983not}%
  \BibitemOpen
  \bibfield  {author} {\bibinfo {author} {\bibfnamefont {M.}~\bibnamefont
  {Dine}}\ and\ \bibinfo {author} {\bibfnamefont {W.}~\bibnamefont
  {Fischler}},\ }\href@noop {} {\bibfield  {journal} {\bibinfo  {journal}
  {Physics Letters B}\ }\textbf {\bibinfo {volume} {120}},\ \bibinfo {pages}
  {137} (\bibinfo {year} {1983})}\BibitemShut {NoStop}%
\bibitem [{\citenamefont {Liu}\ and\ \citenamefont {Ng}(2017)}]{liu2017axion}%
  \BibitemOpen
  \bibfield  {author} {\bibinfo {author} {\bibfnamefont {G.}~\bibnamefont
  {Liu}}\ and\ \bibinfo {author} {\bibfnamefont {K.}~\bibnamefont {Ng}},\
  }\href@noop {} {\bibfield  {journal} {\bibinfo  {journal} {Physics of the
  dark universe}\ }\textbf {\bibinfo {volume} {16}},\ \bibinfo {pages} {22}
  (\bibinfo {year} {2017})}\BibitemShut {NoStop}%
\bibitem [{\citenamefont {Nakagawa}\ \emph {et~al.}(2021)\citenamefont
  {Nakagawa}, \citenamefont {Takahashi},\ and\ \citenamefont
  {Yamada}}]{nakagawa2021cosmic}%
  \BibitemOpen
  \bibfield  {author} {\bibinfo {author} {\bibfnamefont {S.}~\bibnamefont
  {Nakagawa}}, \bibinfo {author} {\bibfnamefont {F.}~\bibnamefont
  {Takahashi}},\ and\ \bibinfo {author} {\bibfnamefont {M.}~\bibnamefont
  {Yamada}},\ }\href@noop {} {\bibfield  {journal} {\bibinfo  {journal}
  {Physical review letters}\ }\textbf {\bibinfo {volume} {127}},\ \bibinfo
  {pages} {181103} (\bibinfo {year} {2021})}\BibitemShut {NoStop}%
\bibitem [{\citenamefont {Obata}(2022)}]{obata2022implications}%
  \BibitemOpen
  \bibfield  {author} {\bibinfo {author} {\bibfnamefont {I.}~\bibnamefont
  {Obata}},\ }\href@noop {} {\bibfield  {journal} {\bibinfo  {journal} {Journal
  of Cosmology and Astroparticle Physics}\ }\textbf {\bibinfo {volume}
  {2022}}\bibinfo  {number} { (09)},\ \bibinfo {pages} {062}}\BibitemShut
  {NoStop}%
\bibitem [{\citenamefont {Zhou}\ \emph {et~al.}(2023)\citenamefont {Zhou},
  \citenamefont {Huang},\ and\ \citenamefont {Geng}}]{zhou2023cosmic}%
  \BibitemOpen
\bibfield  {number} {  }\bibfield  {author} {\bibinfo {author} {\bibfnamefont
  {R.-P.}\ \bibnamefont {Zhou}}, \bibinfo {author} {\bibfnamefont
  {D.}~\bibnamefont {Huang}},\ and\ \bibinfo {author} {\bibfnamefont {C.-Q.}\
  \bibnamefont {Geng}},\ }\href@noop {} {\bibfield  {journal} {\bibinfo
  {journal} {arXiv preprint arXiv:2302.11140}\ } (\bibinfo {year}
  {2023})}\BibitemShut {NoStop}%
\bibitem [{\citenamefont {Arvanitaki}\ \emph {et~al.}(2010)\citenamefont
  {Arvanitaki}, \citenamefont {Dimopoulos}, \citenamefont {Dubovsky},
  \citenamefont {Kaloper},\ and\ \citenamefont
  {March-Russell}}]{arvanitaki2010string}%
  \BibitemOpen
  \bibfield  {author} {\bibinfo {author} {\bibfnamefont {A.}~\bibnamefont
  {Arvanitaki}}, \bibinfo {author} {\bibfnamefont {S.}~\bibnamefont
  {Dimopoulos}}, \bibinfo {author} {\bibfnamefont {S.}~\bibnamefont
  {Dubovsky}}, \bibinfo {author} {\bibfnamefont {N.}~\bibnamefont {Kaloper}},\
  and\ \bibinfo {author} {\bibfnamefont {J.}~\bibnamefont {March-Russell}},\
  }\href@noop {} {\bibfield  {journal} {\bibinfo  {journal} {Physical Review
  D}\ }\textbf {\bibinfo {volume} {81}},\ \bibinfo {pages} {123530} (\bibinfo
  {year} {2010})}\BibitemShut {NoStop}%
\bibitem [{\citenamefont {Hlozek}\ \emph {et~al.}(2015)\citenamefont {Hlozek},
  \citenamefont {Grin}, \citenamefont {Marsh},\ and\ \citenamefont
  {Ferreira}}]{hlozek2015search}%
  \BibitemOpen
  \bibfield  {author} {\bibinfo {author} {\bibfnamefont {R.}~\bibnamefont
  {Hlozek}}, \bibinfo {author} {\bibfnamefont {D.}~\bibnamefont {Grin}},
  \bibinfo {author} {\bibfnamefont {D.~J.}\ \bibnamefont {Marsh}},\ and\
  \bibinfo {author} {\bibfnamefont {P.~G.}\ \bibnamefont {Ferreira}},\
  }\href@noop {} {\bibfield  {journal} {\bibinfo  {journal} {Physical Review
  D}\ }\textbf {\bibinfo {volume} {91}},\ \bibinfo {pages} {103512} (\bibinfo
  {year} {2015})}\BibitemShut {NoStop}%
\bibitem [{\citenamefont {Marsh}(2016)}]{marsh2016axion}%
  \BibitemOpen
  \bibfield  {author} {\bibinfo {author} {\bibfnamefont {D.~J.}\ \bibnamefont
  {Marsh}},\ }\href@noop {} {\bibfield  {journal} {\bibinfo  {journal} {Physics
  Reports}\ }\textbf {\bibinfo {volume} {643}},\ \bibinfo {pages} {1} (\bibinfo
  {year} {2016})}\BibitemShut {NoStop}%
\bibitem [{\citenamefont {Poulin}\ \emph {et~al.}(2018)\citenamefont {Poulin},
  \citenamefont {Smith}, \citenamefont {Grin}, \citenamefont {Karwal},\ and\
  \citenamefont {Kamionkowski}}]{poulin2018cosmological}%
  \BibitemOpen
  \bibfield  {author} {\bibinfo {author} {\bibfnamefont {V.}~\bibnamefont
  {Poulin}}, \bibinfo {author} {\bibfnamefont {T.~L.}\ \bibnamefont {Smith}},
  \bibinfo {author} {\bibfnamefont {D.}~\bibnamefont {Grin}}, \bibinfo {author}
  {\bibfnamefont {T.}~\bibnamefont {Karwal}},\ and\ \bibinfo {author}
  {\bibfnamefont {M.}~\bibnamefont {Kamionkowski}},\ }\href@noop {} {\bibfield
  {journal} {\bibinfo  {journal} {Physical Review D}\ }\textbf {\bibinfo
  {volume} {98}},\ \bibinfo {pages} {083525} (\bibinfo {year}
  {2018})}\BibitemShut {NoStop}%
\bibitem [{\citenamefont {Kim}\ \emph {et~al.}(2021)\citenamefont {Kim},
  \citenamefont {Kim}, \citenamefont {Semertzidis}, \citenamefont {Shin},\ and\
  \citenamefont {Yin}}]{kim2021cosmic}%
  \BibitemOpen
  \bibfield  {author} {\bibinfo {author} {\bibfnamefont {D.}~\bibnamefont
  {Kim}}, \bibinfo {author} {\bibfnamefont {Y.}~\bibnamefont {Kim}}, \bibinfo
  {author} {\bibfnamefont {Y.~K.}\ \bibnamefont {Semertzidis}}, \bibinfo
  {author} {\bibfnamefont {Y.~C.}\ \bibnamefont {Shin}},\ and\ \bibinfo
  {author} {\bibfnamefont {W.}~\bibnamefont {Yin}},\ }\href@noop {} {\bibfield
  {journal} {\bibinfo  {journal} {Physical Review D}\ }\textbf {\bibinfo
  {volume} {104}},\ \bibinfo {pages} {095010} (\bibinfo {year}
  {2021})}\BibitemShut {NoStop}%
\bibitem [{\citenamefont {Jain}\ \emph {et~al.}(2021)\citenamefont {Jain},
  \citenamefont {Long},\ and\ \citenamefont {Amin}}]{jain2021cmb}%
  \BibitemOpen
  \bibfield  {author} {\bibinfo {author} {\bibfnamefont {M.}~\bibnamefont
  {Jain}}, \bibinfo {author} {\bibfnamefont {A.~J.}\ \bibnamefont {Long}},\
  and\ \bibinfo {author} {\bibfnamefont {M.~A.}\ \bibnamefont {Amin}},\
  }\href@noop {} {\bibfield  {journal} {\bibinfo  {journal} {Journal of
  Cosmology and Astroparticle Physics}\ }\textbf {\bibinfo {volume}
  {2021}}\bibinfo  {number} { (05)},\ \bibinfo {pages} {055}}\BibitemShut
  {NoStop}%
\bibitem [{\citenamefont {Cao}\ and\ \citenamefont
  {Boyanovsky}(2022)}]{cao2022non}%
  \BibitemOpen
\bibfield  {number} {  }\bibfield  {author} {\bibinfo {author} {\bibfnamefont
  {S.}~\bibnamefont {Cao}}\ and\ \bibinfo {author} {\bibfnamefont
  {D.}~\bibnamefont {Boyanovsky}},\ }\href@noop {} {\bibfield  {journal}
  {\bibinfo  {journal} {arXiv preprint arXiv:2212.05161}\ } (\bibinfo {year}
  {2022})}\BibitemShut {NoStop}%
\bibitem [{\citenamefont {Lin}\ and\ \citenamefont
  {Yanagida}(2022)}]{lin2022consistency}%
  \BibitemOpen
  \bibfield  {author} {\bibinfo {author} {\bibfnamefont {W.}~\bibnamefont
  {Lin}}\ and\ \bibinfo {author} {\bibfnamefont {T.~T.}\ \bibnamefont
  {Yanagida}},\ }\href@noop {} {\bibfield  {journal} {\bibinfo  {journal}
  {arXiv preprint arXiv:2208.06843}\ } (\bibinfo {year} {2022})}\BibitemShut
  {NoStop}%
\bibitem [{\citenamefont {Gonzalez}\ \emph {et~al.}(2022)\citenamefont
  {Gonzalez}, \citenamefont {Kitajima}, \citenamefont {Takahashi},\ and\
  \citenamefont {Yin}}]{gonzalez2022stability}%
  \BibitemOpen
  \bibfield  {author} {\bibinfo {author} {\bibfnamefont {D.}~\bibnamefont
  {Gonzalez}}, \bibinfo {author} {\bibfnamefont {N.}~\bibnamefont {Kitajima}},
  \bibinfo {author} {\bibfnamefont {F.}~\bibnamefont {Takahashi}},\ and\
  \bibinfo {author} {\bibfnamefont {W.}~\bibnamefont {Yin}},\ }\href@noop {}
  {\bibfield  {journal} {\bibinfo  {journal} {arXiv preprint arXiv:2211.06849}\
  } (\bibinfo {year} {2022})}\BibitemShut {NoStop}%
\bibitem [{\citenamefont {Diego-Palazuelos}(2023)}]{diegopalazuelos2023search}%
  \BibitemOpen
  \bibfield  {author} {\bibinfo {author} {\bibfnamefont {P.}~\bibnamefont
  {Diego-Palazuelos}},\ }\href@noop {} {} (\bibinfo {year} {2023}),\ \Eprint
  {https://arxiv.org/abs/2304.03647} {arXiv:2304.03647 [astro-ph.CO]}
  \BibitemShut {NoStop}%
\bibitem [{\citenamefont {Yin}\ \emph {et~al.}(2023{\natexlab{b}})\citenamefont
  {Yin}, \citenamefont {Dai},\ and\ \citenamefont {Ferraro}}]{yin2023testing}%
  \BibitemOpen
  \bibfield  {author} {\bibinfo {author} {\bibfnamefont {W.~W.}\ \bibnamefont
  {Yin}}, \bibinfo {author} {\bibfnamefont {L.}~\bibnamefont {Dai}},\ and\
  \bibinfo {author} {\bibfnamefont {S.}~\bibnamefont {Ferraro}},\ }\href@noop
  {} {} (\bibinfo {year} {2023}{\natexlab{b}}),\ \Eprint
  {https://arxiv.org/abs/2305.02318} {arXiv:2305.02318 [astro-ph.CO]}
  \BibitemShut {NoStop}%
\bibitem [{\citenamefont {Gasparotto}\ and\ \citenamefont
  {Sfakianakis}(2023)}]{gasparotto2023axiverse}%
  \BibitemOpen
  \bibfield  {author} {\bibinfo {author} {\bibfnamefont {S.}~\bibnamefont
  {Gasparotto}}\ and\ \bibinfo {author} {\bibfnamefont {E.~I.}\ \bibnamefont
  {Sfakianakis}},\ }\href@noop {} {} (\bibinfo {year} {2023}),\ \Eprint
  {https://arxiv.org/abs/2306.16355} {arXiv:2306.16355 [astro-ph.CO]}
  \BibitemShut {NoStop}%
\bibitem [{\citenamefont {Ferreira}\ \emph {et~al.}(2023)\citenamefont
  {Ferreira}, \citenamefont {Gasparotto}, \citenamefont {Hiramatsu},
  \citenamefont {Obata},\ and\ \citenamefont {Pujolas}}]{ferreira2023axionic}%
  \BibitemOpen
  \bibfield  {author} {\bibinfo {author} {\bibfnamefont {R.~Z.}\ \bibnamefont
  {Ferreira}}, \bibinfo {author} {\bibfnamefont {S.}~\bibnamefont
  {Gasparotto}}, \bibinfo {author} {\bibfnamefont {T.}~\bibnamefont
  {Hiramatsu}}, \bibinfo {author} {\bibfnamefont {I.}~\bibnamefont {Obata}},\
  and\ \bibinfo {author} {\bibfnamefont {O.}~\bibnamefont {Pujolas}},\
  }\href@noop {} {\bibfield  {journal} {\bibinfo  {journal} {arXiv preprint
  arXiv:2312.14104}\ } (\bibinfo {year} {2023})}\BibitemShut {NoStop}%
\bibitem [{\citenamefont {Caloni}\ \emph {et~al.}(2022)\citenamefont {Caloni},
  \citenamefont {Giardiello}, \citenamefont {Lembo}, \citenamefont {Gerbino},
  \citenamefont {Gubitosi}, \citenamefont {Lattanzi},\ and\ \citenamefont
  {Pagano}}]{caloni2022probing}%
  \BibitemOpen
  \bibfield  {author} {\bibinfo {author} {\bibfnamefont {L.}~\bibnamefont
  {Caloni}}, \bibinfo {author} {\bibfnamefont {S.}~\bibnamefont {Giardiello}},
  \bibinfo {author} {\bibfnamefont {M.}~\bibnamefont {Lembo}}, \bibinfo
  {author} {\bibfnamefont {M.}~\bibnamefont {Gerbino}}, \bibinfo {author}
  {\bibfnamefont {G.}~\bibnamefont {Gubitosi}}, \bibinfo {author}
  {\bibfnamefont {M.}~\bibnamefont {Lattanzi}},\ and\ \bibinfo {author}
  {\bibfnamefont {L.}~\bibnamefont {Pagano}},\ }\href@noop {} {\bibfield
  {journal} {\bibinfo  {journal} {arXiv preprint arXiv:2212.04867}\ } (\bibinfo
  {year} {2022})}\BibitemShut {NoStop}%
\bibitem [{\citenamefont {Nakai}\ \emph {et~al.}(2023)\citenamefont {Nakai},
  \citenamefont {Namba}, \citenamefont {Obata}, \citenamefont {Qiu},\ and\
  \citenamefont {Saito}}]{nakai2023explain}%
  \BibitemOpen
  \bibfield  {author} {\bibinfo {author} {\bibfnamefont {Y.}~\bibnamefont
  {Nakai}}, \bibinfo {author} {\bibfnamefont {R.}~\bibnamefont {Namba}},
  \bibinfo {author} {\bibfnamefont {I.}~\bibnamefont {Obata}}, \bibinfo
  {author} {\bibfnamefont {Y.-C.}\ \bibnamefont {Qiu}},\ and\ \bibinfo {author}
  {\bibfnamefont {R.}~\bibnamefont {Saito}},\ }\href@noop {} {} (\bibinfo
  {year} {2023}),\ \Eprint {https://arxiv.org/abs/2310.09152} {arXiv:2310.09152
  [astro-ph.CO]} \BibitemShut {NoStop}%
\bibitem [{\citenamefont {Nilsson}\ and\ \citenamefont
  {Poncin-Lafitte}(2023)}]{nilsson2023reexamining}%
  \BibitemOpen
  \bibfield  {author} {\bibinfo {author} {\bibfnamefont {N.~A.}\ \bibnamefont
  {Nilsson}}\ and\ \bibinfo {author} {\bibfnamefont {C.~L.}\ \bibnamefont
  {Poncin-Lafitte}},\ }\href@noop {} {} (\bibinfo {year} {2023}),\ \Eprint
  {https://arxiv.org/abs/2311.16368} {arXiv:2311.16368 [astro-ph.CO]}
  \BibitemShut {NoStop}%
\bibitem [{\citenamefont {Sherwin}\ and\ \citenamefont
  {Namikawa}(2021)}]{sherwin2021cosmic}%
  \BibitemOpen
  \bibfield  {author} {\bibinfo {author} {\bibfnamefont {B.~D.}\ \bibnamefont
  {Sherwin}}\ and\ \bibinfo {author} {\bibfnamefont {T.}~\bibnamefont
  {Namikawa}},\ }\href@noop {} {\bibfield  {journal} {\bibinfo  {journal}
  {arXiv preprint arXiv:2108.09287}\ } (\bibinfo {year} {2021})}\BibitemShut
  {NoStop}%
\bibitem [{\citenamefont {Nakatsuka}\ \emph {et~al.}(2022)\citenamefont
  {Nakatsuka}, \citenamefont {Namikawa},\ and\ \citenamefont
  {Komatsu}}]{nakatsuka2022cosmic}%
  \BibitemOpen
  \bibfield  {author} {\bibinfo {author} {\bibfnamefont {H.}~\bibnamefont
  {Nakatsuka}}, \bibinfo {author} {\bibfnamefont {T.}~\bibnamefont
  {Namikawa}},\ and\ \bibinfo {author} {\bibfnamefont {E.}~\bibnamefont
  {Komatsu}},\ }\href@noop {} {\bibfield  {journal} {\bibinfo  {journal} {arXiv
  preprint arXiv:2203.08560}\ } (\bibinfo {year} {2022})}\BibitemShut {NoStop}%
\bibitem [{\citenamefont {Lee}\ \emph {et~al.}(2022)\citenamefont {Lee},
  \citenamefont {Hotinli},\ and\ \citenamefont
  {Kamionkowski}}]{lee2022probing}%
  \BibitemOpen
  \bibfield  {author} {\bibinfo {author} {\bibfnamefont {N.}~\bibnamefont
  {Lee}}, \bibinfo {author} {\bibfnamefont {S.~C.}\ \bibnamefont {Hotinli}},\
  and\ \bibinfo {author} {\bibfnamefont {M.}~\bibnamefont {Kamionkowski}},\
  }\href@noop {} {\bibfield  {journal} {\bibinfo  {journal} {arXiv preprint
  arXiv:2207.05687}\ } (\bibinfo {year} {2022})}\BibitemShut {NoStop}%
\bibitem [{\citenamefont {Greco}\ \emph {et~al.}(2023)\citenamefont {Greco},
  \citenamefont {Bartolo},\ and\ \citenamefont {Gruppuso}}]{greco2023probing}%
  \BibitemOpen
  \bibfield  {author} {\bibinfo {author} {\bibfnamefont {A.}~\bibnamefont
  {Greco}}, \bibinfo {author} {\bibfnamefont {N.}~\bibnamefont {Bartolo}},\
  and\ \bibinfo {author} {\bibfnamefont {A.}~\bibnamefont {Gruppuso}},\
  }\href@noop {} {\bibfield  {journal} {\bibinfo  {journal} {Journal of
  Cosmology and Astroparticle Physics}\ }\textbf {\bibinfo {volume}
  {2023}}\bibinfo  {number} { (05)},\ \bibinfo {pages} {026}}\BibitemShut
  {NoStop}%
\bibitem [{\citenamefont {Namikawa}\ and\ \citenamefont
  {Obata}(2023)}]{namikawa2023cosmic}%
  \BibitemOpen
\bibfield  {number} {  }\bibfield  {author} {\bibinfo {author} {\bibfnamefont
  {T.}~\bibnamefont {Namikawa}}\ and\ \bibinfo {author} {\bibfnamefont
  {I.}~\bibnamefont {Obata}},\ }\href@noop {} {} (\bibinfo {year} {2023}),\
  \Eprint {https://arxiv.org/abs/2306.08875} {arXiv:2306.08875 [astro-ph.CO]}
  \BibitemShut {NoStop}%
\bibitem [{\citenamefont {Li}\ and\ \citenamefont
  {Zhang}(2008)}]{li2008cosmological}%
  \BibitemOpen
  \bibfield  {author} {\bibinfo {author} {\bibfnamefont {M.}~\bibnamefont
  {Li}}\ and\ \bibinfo {author} {\bibfnamefont {X.}~\bibnamefont {Zhang}},\
  }\href@noop {} {\bibfield  {journal} {\bibinfo  {journal} {Physical Review
  D}\ }\textbf {\bibinfo {volume} {78}},\ \bibinfo {pages} {103516} (\bibinfo
  {year} {2008})}\BibitemShut {NoStop}%
\bibitem [{\citenamefont {Caldwell}\ \emph {et~al.}(2011)\citenamefont
  {Caldwell}, \citenamefont {Gluscevic},\ and\ \citenamefont
  {Kamionkowski}}]{caldwell2011cross}%
  \BibitemOpen
  \bibfield  {author} {\bibinfo {author} {\bibfnamefont {R.~R.}\ \bibnamefont
  {Caldwell}}, \bibinfo {author} {\bibfnamefont {V.}~\bibnamefont
  {Gluscevic}},\ and\ \bibinfo {author} {\bibfnamefont {M.}~\bibnamefont
  {Kamionkowski}},\ }\href@noop {} {\bibfield  {journal} {\bibinfo  {journal}
  {Physical Review D}\ }\textbf {\bibinfo {volume} {84}},\ \bibinfo {pages}
  {043504} (\bibinfo {year} {2011})}\BibitemShut {NoStop}%
\bibitem [{\citenamefont {Zhao}\ and\ \citenamefont
  {Li}(2014)}]{zhao2014fluctuations}%
  \BibitemOpen
  \bibfield  {author} {\bibinfo {author} {\bibfnamefont {W.}~\bibnamefont
  {Zhao}}\ and\ \bibinfo {author} {\bibfnamefont {M.}~\bibnamefont {Li}},\
  }\href@noop {} {\bibfield  {journal} {\bibinfo  {journal} {Physical Review
  D}\ }\textbf {\bibinfo {volume} {89}},\ \bibinfo {pages} {103518} (\bibinfo
  {year} {2014})}\BibitemShut {NoStop}%
\bibitem [{\citenamefont {Zhai}\ \emph {et~al.}(2020)\citenamefont {Zhai},
  \citenamefont {Li}, \citenamefont {Li}, \citenamefont {Li},\ and\
  \citenamefont {Zhang}}]{zhai2020effects}%
  \BibitemOpen
  \bibfield  {author} {\bibinfo {author} {\bibfnamefont {H.}~\bibnamefont
  {Zhai}}, \bibinfo {author} {\bibfnamefont {S.}~\bibnamefont {Li}}, \bibinfo
  {author} {\bibfnamefont {M.}~\bibnamefont {Li}}, \bibinfo {author}
  {\bibfnamefont {H.}~\bibnamefont {Li}},\ and\ \bibinfo {author}
  {\bibfnamefont {X.}~\bibnamefont {Zhang}},\ }\href@noop {} {\bibfield
  {journal} {\bibinfo  {journal} {Journal of Cosmology and Astroparticle
  Physics}\ }\textbf {\bibinfo {volume} {2020}}\bibinfo  {number} { (12)},\
  \bibinfo {pages} {051}}\BibitemShut {NoStop}%
\bibitem [{\citenamefont {Capparelli}\ \emph {et~al.}(2020)\citenamefont
  {Capparelli}, \citenamefont {Caldwell},\ and\ \citenamefont
  {Melchiorri}}]{capparelli2020cosmic}%
  \BibitemOpen
\bibfield  {number} {  }\bibfield  {author} {\bibinfo {author} {\bibfnamefont
  {L.~M.}\ \bibnamefont {Capparelli}}, \bibinfo {author} {\bibfnamefont
  {R.~R.}\ \bibnamefont {Caldwell}},\ and\ \bibinfo {author} {\bibfnamefont
  {A.}~\bibnamefont {Melchiorri}},\ }\href@noop {} {\bibfield  {journal}
  {\bibinfo  {journal} {Physical Review D}\ }\textbf {\bibinfo {volume}
  {101}},\ \bibinfo {pages} {123529} (\bibinfo {year} {2020})}\BibitemShut
  {NoStop}%
\bibitem [{\citenamefont {Takahashi}\ and\ \citenamefont
  {Yin}(2021)}]{takahashi2021kilobyte}%
  \BibitemOpen
  \bibfield  {author} {\bibinfo {author} {\bibfnamefont {F.}~\bibnamefont
  {Takahashi}}\ and\ \bibinfo {author} {\bibfnamefont {W.}~\bibnamefont
  {Yin}},\ }\href@noop {} {\bibfield  {journal} {\bibinfo  {journal} {Journal
  of Cosmology and Astroparticle Physics}\ }\textbf {\bibinfo {volume}
  {2021}}\bibinfo  {number} { (04)},\ \bibinfo {pages} {007}}\BibitemShut
  {NoStop}%
\bibitem [{\citenamefont {Greco}\ \emph {et~al.}(2022)\citenamefont {Greco},
  \citenamefont {Bartolo},\ and\ \citenamefont {Gruppuso}}]{greco2022cosmic}%
  \BibitemOpen
\bibfield  {number} {  }\bibfield  {author} {\bibinfo {author} {\bibfnamefont
  {A.}~\bibnamefont {Greco}}, \bibinfo {author} {\bibfnamefont
  {N.}~\bibnamefont {Bartolo}},\ and\ \bibinfo {author} {\bibfnamefont
  {A.}~\bibnamefont {Gruppuso}},\ }\href@noop {} {\bibfield  {journal}
  {\bibinfo  {journal} {Journal of Cosmology and Astroparticle Physics}\
  }\textbf {\bibinfo {volume} {2022}}\bibinfo  {number} { (03)},\ \bibinfo
  {pages} {050}}\BibitemShut {NoStop}%
\bibitem [{\citenamefont {Cai}\ \emph {et~al.}(2022)\citenamefont {Cai},
  \citenamefont {Guan}, \citenamefont {Namikawa},\ and\ \citenamefont
  {Kosowky}}]{cai2022impact}%
  \BibitemOpen
\bibfield  {number} {  }\bibfield  {author} {\bibinfo {author} {\bibfnamefont
  {H.}~\bibnamefont {Cai}}, \bibinfo {author} {\bibfnamefont {Y.}~\bibnamefont
  {Guan}}, \bibinfo {author} {\bibfnamefont {T.~N.}\ \bibnamefont {Namikawa}},\
  and\ \bibinfo {author} {\bibfnamefont {A.}~\bibnamefont {Kosowky}},\
  }\href@noop {} {\bibfield  {journal} {\bibinfo  {journal} {arXiv preprint
  arXiv:2209.08749}\ } (\bibinfo {year} {2022})}\BibitemShut {NoStop}%
\bibitem [{\citenamefont {Hagimoto}\ and\ \citenamefont
  {Long}(2023)}]{hagimoto2023measures}%
  \BibitemOpen
  \bibfield  {author} {\bibinfo {author} {\bibfnamefont {R.}~\bibnamefont
  {Hagimoto}}\ and\ \bibinfo {author} {\bibfnamefont {A.~J.}\ \bibnamefont
  {Long}},\ }\href@noop {} {} (\bibinfo {year} {2023}),\ \Eprint
  {https://arxiv.org/abs/2306.07351} {arXiv:2306.07351 [astro-ph.CO]}
  \BibitemShut {NoStop}%
\bibitem [{\citenamefont {Lee}\ \emph {et~al.}(2023)\citenamefont {Lee},
  \citenamefont {Kang}, \citenamefont {Gong}, \citenamefont {Jeong},
  \citenamefont {Jung},\ and\ \citenamefont {Park}}]{lee2023cosmological}%
  \BibitemOpen
  \bibfield  {author} {\bibinfo {author} {\bibfnamefont {S.~M.}\ \bibnamefont
  {Lee}}, \bibinfo {author} {\bibfnamefont {D.~W.}\ \bibnamefont {Kang}},
  \bibinfo {author} {\bibfnamefont {J.-O.}\ \bibnamefont {Gong}}, \bibinfo
  {author} {\bibfnamefont {D.}~\bibnamefont {Jeong}}, \bibinfo {author}
  {\bibfnamefont {D.-W.}\ \bibnamefont {Jung}},\ and\ \bibinfo {author}
  {\bibfnamefont {S.~C.}\ \bibnamefont {Park}},\ }\href@noop {} {} (\bibinfo
  {year} {2023}),\ \Eprint {https://arxiv.org/abs/2307.14798} {arXiv:2307.14798
  [hep-ph]} \BibitemShut {NoStop}%
\bibitem [{\citenamefont {Feng}\ \emph {et~al.}(2006)\citenamefont {Feng},
  \citenamefont {Li}, \citenamefont {Xia}, \citenamefont {Chen},\ and\
  \citenamefont {Zhang}}]{feng2006searching}%
  \BibitemOpen
  \bibfield  {author} {\bibinfo {author} {\bibfnamefont {B.}~\bibnamefont
  {Feng}}, \bibinfo {author} {\bibfnamefont {M.}~\bibnamefont {Li}}, \bibinfo
  {author} {\bibfnamefont {J.-Q.}\ \bibnamefont {Xia}}, \bibinfo {author}
  {\bibfnamefont {X.}~\bibnamefont {Chen}},\ and\ \bibinfo {author}
  {\bibfnamefont {X.}~\bibnamefont {Zhang}},\ }\href@noop {} {\bibfield
  {journal} {\bibinfo  {journal} {Physical review letters}\ }\textbf {\bibinfo
  {volume} {96}},\ \bibinfo {pages} {221302} (\bibinfo {year}
  {2006})}\BibitemShut {NoStop}%
\bibitem [{\citenamefont {Jarosik}\ \emph {et~al.}(2011)\citenamefont
  {Jarosik}, \citenamefont {Bennett}, \citenamefont {Dunkley}, \citenamefont
  {Gold}, \citenamefont {Greason}, \citenamefont {Halpern}, \citenamefont
  {Hill}, \citenamefont {Hinshaw}, \citenamefont {Kogut}, \citenamefont
  {Komatsu} \emph {et~al.}}]{jarosik2011seven}%
  \BibitemOpen
  \bibfield  {author} {\bibinfo {author} {\bibfnamefont {N.}~\bibnamefont
  {Jarosik}}, \bibinfo {author} {\bibfnamefont {C.}~\bibnamefont {Bennett}},
  \bibinfo {author} {\bibfnamefont {J.}~\bibnamefont {Dunkley}}, \bibinfo
  {author} {\bibfnamefont {B.}~\bibnamefont {Gold}}, \bibinfo {author}
  {\bibfnamefont {M.}~\bibnamefont {Greason}}, \bibinfo {author} {\bibfnamefont
  {M.}~\bibnamefont {Halpern}}, \bibinfo {author} {\bibfnamefont
  {R.}~\bibnamefont {Hill}}, \bibinfo {author} {\bibfnamefont {G.}~\bibnamefont
  {Hinshaw}}, \bibinfo {author} {\bibfnamefont {A.}~\bibnamefont {Kogut}},
  \bibinfo {author} {\bibfnamefont {E.}~\bibnamefont {Komatsu}}, \emph
  {et~al.},\ }\href@noop {} {\bibfield  {journal} {\bibinfo  {journal} {The
  Astrophysical Journal Supplement Series}\ }\textbf {\bibinfo {volume}
  {192}},\ \bibinfo {pages} {14} (\bibinfo {year} {2011})}\BibitemShut
  {NoStop}%
\bibitem [{\citenamefont {Gluscevic}\ \emph {et~al.}(2012)\citenamefont
  {Gluscevic}, \citenamefont {Hanson}, \citenamefont {Kamionkowski},\ and\
  \citenamefont {Hirata}}]{gluscevic2012first}%
  \BibitemOpen
  \bibfield  {author} {\bibinfo {author} {\bibfnamefont {V.}~\bibnamefont
  {Gluscevic}}, \bibinfo {author} {\bibfnamefont {D.}~\bibnamefont {Hanson}},
  \bibinfo {author} {\bibfnamefont {M.}~\bibnamefont {Kamionkowski}},\ and\
  \bibinfo {author} {\bibfnamefont {C.~M.}\ \bibnamefont {Hirata}},\
  }\href@noop {} {\bibfield  {journal} {\bibinfo  {journal} {Physical Review
  D}\ }\textbf {\bibinfo {volume} {86}},\ \bibinfo {pages} {103529} (\bibinfo
  {year} {2012})}\BibitemShut {NoStop}%
\bibitem [{\citenamefont {Eskilt}\ and\ \citenamefont
  {Komatsu}(2022)}]{eskilt2022improved}%
  \BibitemOpen
  \bibfield  {author} {\bibinfo {author} {\bibfnamefont {J.~R.}\ \bibnamefont
  {Eskilt}}\ and\ \bibinfo {author} {\bibfnamefont {E.}~\bibnamefont
  {Komatsu}},\ }\href@noop {} {\bibfield  {journal} {\bibinfo  {journal} {arXiv
  preprint arXiv:2205.13962}\ } (\bibinfo {year} {2022})}\BibitemShut {NoStop}%
\bibitem [{\citenamefont {Ade}\ \emph {et~al.}(2015)\citenamefont {Ade},
  \citenamefont {Arnold}, \citenamefont {Atlas}, \citenamefont {Baccigalupi},
  \citenamefont {Barron}, \citenamefont {Boettger}, \citenamefont {Borrill},
  \citenamefont {Chapman}, \citenamefont {Chinone}, \citenamefont {Cukierman}
  \emph {et~al.}}]{ade2015polarbear}%
  \BibitemOpen
  \bibfield  {author} {\bibinfo {author} {\bibfnamefont {P.~A.}\ \bibnamefont
  {Ade}}, \bibinfo {author} {\bibfnamefont {K.}~\bibnamefont {Arnold}},
  \bibinfo {author} {\bibfnamefont {M.}~\bibnamefont {Atlas}}, \bibinfo
  {author} {\bibfnamefont {C.}~\bibnamefont {Baccigalupi}}, \bibinfo {author}
  {\bibfnamefont {D.}~\bibnamefont {Barron}}, \bibinfo {author} {\bibfnamefont
  {D.}~\bibnamefont {Boettger}}, \bibinfo {author} {\bibfnamefont
  {J.}~\bibnamefont {Borrill}}, \bibinfo {author} {\bibfnamefont
  {S.}~\bibnamefont {Chapman}}, \bibinfo {author} {\bibfnamefont
  {Y.}~\bibnamefont {Chinone}}, \bibinfo {author} {\bibfnamefont
  {A.}~\bibnamefont {Cukierman}}, \emph {et~al.},\ }\href@noop {} {\bibfield
  {journal} {\bibinfo  {journal} {Physical Review D}\ }\textbf {\bibinfo
  {volume} {92}},\ \bibinfo {pages} {123509} (\bibinfo {year}
  {2015})}\BibitemShut {NoStop}%
\bibitem [{\citenamefont {{The POLARBEAR Collaboration}}\ \emph
  {et~al.}(2023)\citenamefont {{The POLARBEAR Collaboration}}, \citenamefont
  {Adachi}, \citenamefont {Adkins}, \citenamefont {Arnold}, \citenamefont
  {Baccigalupi}, \citenamefont {Barron}, \citenamefont {Cheung}, \citenamefont
  {Chinone}, \citenamefont {Crowley}, \citenamefont {Errard}, \citenamefont
  {Fabbian}, \citenamefont {Feng}, \citenamefont {Flauger}, \citenamefont
  {Fujino}, \citenamefont {Green}, \citenamefont {Hasegawa}, \citenamefont
  {Hazumi}, \citenamefont {Kaneko}, \citenamefont {Katayama}, \citenamefont
  {Keating}, \citenamefont {Kusaka}, \citenamefont {Lee}, \citenamefont
  {Minami}, \citenamefont {Nishino}, \citenamefont {Reichardt}, \citenamefont
  {Siritanasak}, \citenamefont {Spisak}, \citenamefont {Tajima}, \citenamefont
  {Takakura}, \citenamefont {Takatori}, \citenamefont {Teply},\ and\
  \citenamefont {Yamada}}]{polarbear2023constraints}%
  \BibitemOpen
  \bibfield  {author} {\bibinfo {author} {\bibnamefont {{The POLARBEAR
  Collaboration}}}, \bibinfo {author} {\bibfnamefont {S.}~\bibnamefont
  {Adachi}}, \bibinfo {author} {\bibfnamefont {T.}~\bibnamefont {Adkins}},
  \bibinfo {author} {\bibfnamefont {K.}~\bibnamefont {Arnold}}, \bibinfo
  {author} {\bibfnamefont {C.}~\bibnamefont {Baccigalupi}}, \bibinfo {author}
  {\bibfnamefont {D.}~\bibnamefont {Barron}}, \bibinfo {author} {\bibfnamefont
  {K.}~\bibnamefont {Cheung}}, \bibinfo {author} {\bibfnamefont
  {Y.}~\bibnamefont {Chinone}}, \bibinfo {author} {\bibfnamefont {K.~T.}\
  \bibnamefont {Crowley}}, \bibinfo {author} {\bibfnamefont {J.}~\bibnamefont
  {Errard}}, \bibinfo {author} {\bibfnamefont {G.}~\bibnamefont {Fabbian}},
  \bibinfo {author} {\bibfnamefont {C.}~\bibnamefont {Feng}}, \bibinfo {author}
  {\bibfnamefont {R.}~\bibnamefont {Flauger}}, \bibinfo {author} {\bibfnamefont
  {T.}~\bibnamefont {Fujino}}, \bibinfo {author} {\bibfnamefont
  {D.}~\bibnamefont {Green}}, \bibinfo {author} {\bibfnamefont
  {M.}~\bibnamefont {Hasegawa}}, \bibinfo {author} {\bibfnamefont
  {M.}~\bibnamefont {Hazumi}}, \bibinfo {author} {\bibfnamefont
  {D.}~\bibnamefont {Kaneko}}, \bibinfo {author} {\bibfnamefont
  {N.}~\bibnamefont {Katayama}}, \bibinfo {author} {\bibfnamefont
  {B.}~\bibnamefont {Keating}}, \bibinfo {author} {\bibfnamefont
  {A.}~\bibnamefont {Kusaka}}, \bibinfo {author} {\bibfnamefont {A.~T.}\
  \bibnamefont {Lee}}, \bibinfo {author} {\bibfnamefont {Y.}~\bibnamefont
  {Minami}}, \bibinfo {author} {\bibfnamefont {H.}~\bibnamefont {Nishino}},
  \bibinfo {author} {\bibfnamefont {C.~L.}\ \bibnamefont {Reichardt}}, \bibinfo
  {author} {\bibfnamefont {P.}~\bibnamefont {Siritanasak}}, \bibinfo {author}
  {\bibfnamefont {J.}~\bibnamefont {Spisak}}, \bibinfo {author} {\bibfnamefont
  {O.}~\bibnamefont {Tajima}}, \bibinfo {author} {\bibfnamefont
  {S.}~\bibnamefont {Takakura}}, \bibinfo {author} {\bibfnamefont
  {S.}~\bibnamefont {Takatori}}, \bibinfo {author} {\bibfnamefont {G.~P.}\
  \bibnamefont {Teply}},\ and\ \bibinfo {author} {\bibfnamefont
  {K.}~\bibnamefont {Yamada}},\ }\href@noop {} {} (\bibinfo {year}
  {2023})\BibitemShut {NoStop}%
\bibitem [{\citenamefont {Namikawa}\ \emph {et~al.}(2020)\citenamefont
  {Namikawa}, \citenamefont {Guan}, \citenamefont {Darwish}, \citenamefont
  {Sherwin}, \citenamefont {Aiola}, \citenamefont {Battaglia}, \citenamefont
  {Beall}, \citenamefont {Becker}, \citenamefont {Bond}, \citenamefont
  {Calabrese} \emph {et~al.}}]{namikawa2020atacama}%
  \BibitemOpen
  \bibfield  {author} {\bibinfo {author} {\bibfnamefont {T.}~\bibnamefont
  {Namikawa}}, \bibinfo {author} {\bibfnamefont {Y.}~\bibnamefont {Guan}},
  \bibinfo {author} {\bibfnamefont {O.}~\bibnamefont {Darwish}}, \bibinfo
  {author} {\bibfnamefont {B.~D.}\ \bibnamefont {Sherwin}}, \bibinfo {author}
  {\bibfnamefont {S.}~\bibnamefont {Aiola}}, \bibinfo {author} {\bibfnamefont
  {N.}~\bibnamefont {Battaglia}}, \bibinfo {author} {\bibfnamefont {J.~A.}\
  \bibnamefont {Beall}}, \bibinfo {author} {\bibfnamefont {D.~T.}\ \bibnamefont
  {Becker}}, \bibinfo {author} {\bibfnamefont {J.~R.}\ \bibnamefont {Bond}},
  \bibinfo {author} {\bibfnamefont {E.}~\bibnamefont {Calabrese}}, \emph
  {et~al.},\ }\href@noop {} {\bibfield  {journal} {\bibinfo  {journal}
  {Physical Review D}\ }\textbf {\bibinfo {volume} {101}},\ \bibinfo {pages}
  {083527} (\bibinfo {year} {2020})}\BibitemShut {NoStop}%
\bibitem [{\citenamefont {Bianchini}\ \emph {et~al.}(2020)\citenamefont
  {Bianchini}, \citenamefont {Wu}, \citenamefont {Ade}, \citenamefont
  {Anderson}, \citenamefont {Austermann}, \citenamefont {Avva}, \citenamefont
  {Balkenhol}, \citenamefont {Baxter}, \citenamefont {Beall}, \citenamefont
  {Bender} \emph {et~al.}}]{bianchini2020searching}%
  \BibitemOpen
  \bibfield  {author} {\bibinfo {author} {\bibfnamefont {F.}~\bibnamefont
  {Bianchini}}, \bibinfo {author} {\bibfnamefont {W.}~\bibnamefont {Wu}},
  \bibinfo {author} {\bibfnamefont {P.}~\bibnamefont {Ade}}, \bibinfo {author}
  {\bibfnamefont {A.}~\bibnamefont {Anderson}}, \bibinfo {author}
  {\bibfnamefont {J.}~\bibnamefont {Austermann}}, \bibinfo {author}
  {\bibfnamefont {J.}~\bibnamefont {Avva}}, \bibinfo {author} {\bibfnamefont
  {L.}~\bibnamefont {Balkenhol}}, \bibinfo {author} {\bibfnamefont
  {E.}~\bibnamefont {Baxter}}, \bibinfo {author} {\bibfnamefont
  {J.}~\bibnamefont {Beall}}, \bibinfo {author} {\bibfnamefont
  {A.}~\bibnamefont {Bender}}, \emph {et~al.},\ }\href@noop {} {\bibfield
  {journal} {\bibinfo  {journal} {Physical Review D}\ }\textbf {\bibinfo
  {volume} {102}},\ \bibinfo {pages} {083504} (\bibinfo {year}
  {2020})}\BibitemShut {NoStop}%
\bibitem [{\citenamefont {Ade}\ \emph {et~al.}(2017)\citenamefont {Ade},
  \citenamefont {Aikin}, \citenamefont {Bock}, \citenamefont {Brevik},
  \citenamefont {Filippini}, \citenamefont {Ghosh}, \citenamefont
  {Hildebrandt}, \citenamefont {Hui}, \citenamefont {Kefeli}, \citenamefont
  {Moncelsi} \emph {et~al.}}]{ade2017bicep2}%
  \BibitemOpen
  \bibfield  {author} {\bibinfo {author} {\bibfnamefont {P.}~\bibnamefont
  {Ade}}, \bibinfo {author} {\bibfnamefont {R.}~\bibnamefont {Aikin}}, \bibinfo
  {author} {\bibfnamefont {J.}~\bibnamefont {Bock}}, \bibinfo {author}
  {\bibfnamefont {J.}~\bibnamefont {Brevik}}, \bibinfo {author} {\bibfnamefont
  {J.}~\bibnamefont {Filippini}}, \bibinfo {author} {\bibfnamefont
  {T.}~\bibnamefont {Ghosh}}, \bibinfo {author} {\bibfnamefont
  {S.}~\bibnamefont {Hildebrandt}}, \bibinfo {author} {\bibfnamefont
  {H.}~\bibnamefont {Hui}}, \bibinfo {author} {\bibfnamefont {S.}~\bibnamefont
  {Kefeli}}, \bibinfo {author} {\bibfnamefont {L.}~\bibnamefont {Moncelsi}},
  \emph {et~al.},\ }\href@noop {} {\bibfield  {journal} {\bibinfo  {journal}
  {Physical Review D}\ }\textbf {\bibinfo {volume} {96}},\ \bibinfo {pages}
  {Art} (\bibinfo {year} {2017})}\BibitemShut {NoStop}%
\bibitem [{\citenamefont {{Keck Collaboration}}\ \emph
  {et~al.}(2022)\citenamefont {{Keck Collaboration}}, \citenamefont {Ade},
  \citenamefont {Ahmed}, \citenamefont {Amiri}, \citenamefont {Barkats},
  \citenamefont {Thakur}, \citenamefont {Beck}, \citenamefont {Bischoff},
  \citenamefont {Bock}, \citenamefont {Boenish}, \citenamefont {Bullock},
  \citenamefont {Buza}, \citenamefont {Cheshire}, \citenamefont {Connors},
  \citenamefont {Cornelison}, \citenamefont {Crumrine}, \citenamefont
  {Cukierman}, \citenamefont {Denison}, \citenamefont {Dierickx}, \citenamefont
  {Duband}, \citenamefont {Eiben}, \citenamefont {Fatigoni}, \citenamefont
  {Filippini}, \citenamefont {Fliescher}, \citenamefont {Giannakopoulos},
  \citenamefont {Goeckner-Wald}, \citenamefont {Goldfinger}, \citenamefont
  {Grayson}, \citenamefont {Grimes}, \citenamefont {Halal}, \citenamefont
  {Hall}, \citenamefont {Halpern}, \citenamefont {Hand}, \citenamefont
  {Harrison}, \citenamefont {Henderson}, \citenamefont {Hildebrandt},
  \citenamefont {Hubmayr}, \citenamefont {Hui}, \citenamefont {Irwin},
  \citenamefont {Kang}, \citenamefont {Karkare}, \citenamefont {Karpel},
  \citenamefont {Kefeli}, \citenamefont {Kernasovskiy}, \citenamefont {Kovac},
  \citenamefont {Kuo}, \citenamefont {Lau}, \citenamefont {Leitch},
  \citenamefont {Lennox}, \citenamefont {Megerian}, \citenamefont {Minutolo},
  \citenamefont {Moncelsi}, \citenamefont {Nakato}, \citenamefont {Namikawa},
  \citenamefont {Nguyen}, \citenamefont {O'Brient}, \citenamefont {Ogburn},
  \citenamefont {Palladino}, \citenamefont {Petroff}, \citenamefont {Prouve},
  \citenamefont {Pryke}, \citenamefont {Racine}, \citenamefont {Reintsema},
  \citenamefont {Richter}, \citenamefont {Schillaci}, \citenamefont {Schmitt},
  \citenamefont {Schwarz}, \citenamefont {Sheehy}, \citenamefont {Singari},
  \citenamefont {Soliman}, \citenamefont {Germaine}, \citenamefont {Steinbach},
  \citenamefont {Sudiwala}, \citenamefont {Teply}, \citenamefont {Thompson},
  \citenamefont {Tolan}, \citenamefont {Tucker}, \citenamefont {Turner},
  \citenamefont {Umilta}, \citenamefont {Verges}, \citenamefont {Vieregg},
  \citenamefont {Wandui}, \citenamefont {Weber}, \citenamefont {Wiebe},
  \citenamefont {Willmert}, \citenamefont {Wong}, \citenamefont {Wu},
  \citenamefont {Yang}, \citenamefont {Yoon}, \citenamefont {Young},
  \citenamefont {Yu}, \citenamefont {Zeng}, \citenamefont {Zhang},\ and\
  \citenamefont {Zhang}}]{keck2022line}%
  \BibitemOpen
  \bibfield  {author} {\bibinfo {author} {\bibnamefont {{Keck Collaboration}}},
  \bibinfo {author} {\bibfnamefont {P.~A.~R.}\ \bibnamefont {Ade}}, \bibinfo
  {author} {\bibfnamefont {Z.}~\bibnamefont {Ahmed}}, \bibinfo {author}
  {\bibfnamefont {M.}~\bibnamefont {Amiri}}, \bibinfo {author} {\bibfnamefont
  {D.}~\bibnamefont {Barkats}}, \bibinfo {author} {\bibfnamefont {R.~B.}\
  \bibnamefont {Thakur}}, \bibinfo {author} {\bibfnamefont {D.}~\bibnamefont
  {Beck}}, \bibinfo {author} {\bibfnamefont {C.~A.}\ \bibnamefont {Bischoff}},
  \bibinfo {author} {\bibfnamefont {J.~J.}\ \bibnamefont {Bock}}, \bibinfo
  {author} {\bibfnamefont {H.}~\bibnamefont {Boenish}}, \bibinfo {author}
  {\bibfnamefont {E.}~\bibnamefont {Bullock}}, \bibinfo {author} {\bibfnamefont
  {V.}~\bibnamefont {Buza}}, \bibinfo {author} {\bibfnamefont {J.~R.}\
  \bibnamefont {Cheshire}}, \bibinfo {author} {\bibfnamefont {J.}~\bibnamefont
  {Connors}}, \bibinfo {author} {\bibfnamefont {J.}~\bibnamefont {Cornelison}},
  \bibinfo {author} {\bibfnamefont {M.}~\bibnamefont {Crumrine}}, \bibinfo
  {author} {\bibfnamefont {A.}~\bibnamefont {Cukierman}}, \bibinfo {author}
  {\bibfnamefont {E.~V.}\ \bibnamefont {Denison}}, \bibinfo {author}
  {\bibfnamefont {M.}~\bibnamefont {Dierickx}}, \bibinfo {author}
  {\bibfnamefont {L.}~\bibnamefont {Duband}}, \bibinfo {author} {\bibfnamefont
  {M.}~\bibnamefont {Eiben}}, \bibinfo {author} {\bibfnamefont
  {S.}~\bibnamefont {Fatigoni}}, \bibinfo {author} {\bibfnamefont {J.~P.}\
  \bibnamefont {Filippini}}, \bibinfo {author} {\bibfnamefont {S.}~\bibnamefont
  {Fliescher}}, \bibinfo {author} {\bibfnamefont {C.}~\bibnamefont
  {Giannakopoulos}}, \bibinfo {author} {\bibfnamefont {N.}~\bibnamefont
  {Goeckner-Wald}}, \bibinfo {author} {\bibfnamefont {D.~C.}\ \bibnamefont
  {Goldfinger}}, \bibinfo {author} {\bibfnamefont {J.}~\bibnamefont {Grayson}},
  \bibinfo {author} {\bibfnamefont {P.}~\bibnamefont {Grimes}}, \bibinfo
  {author} {\bibfnamefont {G.}~\bibnamefont {Halal}}, \bibinfo {author}
  {\bibfnamefont {G.}~\bibnamefont {Hall}}, \bibinfo {author} {\bibfnamefont
  {M.}~\bibnamefont {Halpern}}, \bibinfo {author} {\bibfnamefont
  {E.}~\bibnamefont {Hand}}, \bibinfo {author} {\bibfnamefont {S.}~\bibnamefont
  {Harrison}}, \bibinfo {author} {\bibfnamefont {S.}~\bibnamefont {Henderson}},
  \bibinfo {author} {\bibfnamefont {S.~R.}\ \bibnamefont {Hildebrandt}},
  \bibinfo {author} {\bibfnamefont {J.}~\bibnamefont {Hubmayr}}, \bibinfo
  {author} {\bibfnamefont {H.}~\bibnamefont {Hui}}, \bibinfo {author}
  {\bibfnamefont {K.~D.}\ \bibnamefont {Irwin}}, \bibinfo {author}
  {\bibfnamefont {J.}~\bibnamefont {Kang}}, \bibinfo {author} {\bibfnamefont
  {K.~S.}\ \bibnamefont {Karkare}}, \bibinfo {author} {\bibfnamefont
  {E.}~\bibnamefont {Karpel}}, \bibinfo {author} {\bibfnamefont
  {S.}~\bibnamefont {Kefeli}}, \bibinfo {author} {\bibfnamefont {S.~A.}\
  \bibnamefont {Kernasovskiy}}, \bibinfo {author} {\bibfnamefont {J.~M.}\
  \bibnamefont {Kovac}}, \bibinfo {author} {\bibfnamefont {C.~L.}\ \bibnamefont
  {Kuo}}, \bibinfo {author} {\bibfnamefont {K.}~\bibnamefont {Lau}}, \bibinfo
  {author} {\bibfnamefont {E.~M.}\ \bibnamefont {Leitch}}, \bibinfo {author}
  {\bibfnamefont {A.}~\bibnamefont {Lennox}}, \bibinfo {author} {\bibfnamefont
  {K.~G.}\ \bibnamefont {Megerian}}, \bibinfo {author} {\bibfnamefont
  {L.}~\bibnamefont {Minutolo}}, \bibinfo {author} {\bibfnamefont
  {L.}~\bibnamefont {Moncelsi}}, \bibinfo {author} {\bibfnamefont
  {Y.}~\bibnamefont {Nakato}}, \bibinfo {author} {\bibfnamefont
  {T.}~\bibnamefont {Namikawa}}, \bibinfo {author} {\bibfnamefont {H.~T.}\
  \bibnamefont {Nguyen}}, \bibinfo {author} {\bibfnamefont {R.}~\bibnamefont
  {O'Brient}}, \bibinfo {author} {\bibfnamefont {R.~W.}\ \bibnamefont
  {Ogburn}}, \bibinfo {author} {\bibfnamefont {S.}~\bibnamefont {Palladino}},
  \bibinfo {author} {\bibfnamefont {M.~A.}\ \bibnamefont {Petroff}}, \bibinfo
  {author} {\bibfnamefont {T.}~\bibnamefont {Prouve}}, \bibinfo {author}
  {\bibfnamefont {C.}~\bibnamefont {Pryke}}, \bibinfo {author} {\bibfnamefont
  {B.}~\bibnamefont {Racine}}, \bibinfo {author} {\bibfnamefont {C.~D.}\
  \bibnamefont {Reintsema}}, \bibinfo {author} {\bibfnamefont {S.}~\bibnamefont
  {Richter}}, \bibinfo {author} {\bibfnamefont {A.}~\bibnamefont {Schillaci}},
  \bibinfo {author} {\bibfnamefont {B.~L.}\ \bibnamefont {Schmitt}}, \bibinfo
  {author} {\bibfnamefont {R.}~\bibnamefont {Schwarz}}, \bibinfo {author}
  {\bibfnamefont {C.~D.}\ \bibnamefont {Sheehy}}, \bibinfo {author}
  {\bibfnamefont {B.}~\bibnamefont {Singari}}, \bibinfo {author} {\bibfnamefont
  {A.}~\bibnamefont {Soliman}}, \bibinfo {author} {\bibfnamefont {T.~S.}\
  \bibnamefont {Germaine}}, \bibinfo {author} {\bibfnamefont {B.}~\bibnamefont
  {Steinbach}}, \bibinfo {author} {\bibfnamefont {R.~V.}\ \bibnamefont
  {Sudiwala}}, \bibinfo {author} {\bibfnamefont {G.~P.}\ \bibnamefont {Teply}},
  \bibinfo {author} {\bibfnamefont {K.~L.}\ \bibnamefont {Thompson}}, \bibinfo
  {author} {\bibfnamefont {J.~E.}\ \bibnamefont {Tolan}}, \bibinfo {author}
  {\bibfnamefont {C.}~\bibnamefont {Tucker}}, \bibinfo {author} {\bibfnamefont
  {A.~D.}\ \bibnamefont {Turner}}, \bibinfo {author} {\bibfnamefont
  {C.}~\bibnamefont {Umilta}}, \bibinfo {author} {\bibfnamefont
  {C.}~\bibnamefont {Verges}}, \bibinfo {author} {\bibfnamefont {A.~G.}\
  \bibnamefont {Vieregg}}, \bibinfo {author} {\bibfnamefont {A.}~\bibnamefont
  {Wandui}}, \bibinfo {author} {\bibfnamefont {A.~C.}\ \bibnamefont {Weber}},
  \bibinfo {author} {\bibfnamefont {D.~V.}\ \bibnamefont {Wiebe}}, \bibinfo
  {author} {\bibfnamefont {J.}~\bibnamefont {Willmert}}, \bibinfo {author}
  {\bibfnamefont {C.~L.}\ \bibnamefont {Wong}}, \bibinfo {author}
  {\bibfnamefont {W.~L.~K.}\ \bibnamefont {Wu}}, \bibinfo {author}
  {\bibfnamefont {H.}~\bibnamefont {Yang}}, \bibinfo {author} {\bibfnamefont
  {K.~W.}\ \bibnamefont {Yoon}}, \bibinfo {author} {\bibfnamefont
  {E.}~\bibnamefont {Young}}, \bibinfo {author} {\bibfnamefont
  {C.}~\bibnamefont {Yu}}, \bibinfo {author} {\bibfnamefont {L.}~\bibnamefont
  {Zeng}}, \bibinfo {author} {\bibfnamefont {C.}~\bibnamefont {Zhang}},\ and\
  \bibinfo {author} {\bibfnamefont {S.}~\bibnamefont {Zhang}},\ }\href@noop {}
  {\bibfield  {journal} {\bibinfo  {journal} {arXiv preprint arXiv:2210.08038}\
  } (\bibinfo {year} {2022})}\BibitemShut {NoStop}%
\bibitem [{\citenamefont {Aghanim}\ \emph {et~al.}(2016)\citenamefont
  {Aghanim}, \citenamefont {Ashdown}, \citenamefont {Aumont}, \citenamefont
  {Baccigalupi}, \citenamefont {Ballardini}, \citenamefont {Banday},
  \citenamefont {Barreiro}, \citenamefont {Bartolo}, \citenamefont {Basak},
  \citenamefont {Benabed} \emph {et~al.}}]{aghanim2016planck}%
  \BibitemOpen
  \bibfield  {author} {\bibinfo {author} {\bibfnamefont {N.}~\bibnamefont
  {Aghanim}}, \bibinfo {author} {\bibfnamefont {M.}~\bibnamefont {Ashdown}},
  \bibinfo {author} {\bibfnamefont {J.}~\bibnamefont {Aumont}}, \bibinfo
  {author} {\bibfnamefont {C.}~\bibnamefont {Baccigalupi}}, \bibinfo {author}
  {\bibfnamefont {M.}~\bibnamefont {Ballardini}}, \bibinfo {author}
  {\bibfnamefont {A.}~\bibnamefont {Banday}}, \bibinfo {author} {\bibfnamefont
  {R.}~\bibnamefont {Barreiro}}, \bibinfo {author} {\bibfnamefont
  {N.}~\bibnamefont {Bartolo}}, \bibinfo {author} {\bibfnamefont
  {S.}~\bibnamefont {Basak}}, \bibinfo {author} {\bibfnamefont
  {K.}~\bibnamefont {Benabed}}, \emph {et~al.},\ }\href@noop {} {\bibfield
  {journal} {\bibinfo  {journal} {Astronomy \& Astrophysics}\ }\textbf
  {\bibinfo {volume} {596}},\ \bibinfo {pages} {A110} (\bibinfo {year}
  {2016})}\BibitemShut {NoStop}%
\bibitem [{\citenamefont {Contreras}\ \emph {et~al.}(2017)\citenamefont
  {Contreras}, \citenamefont {Boubel},\ and\ \citenamefont
  {Scott}}]{contreras2017constraints}%
  \BibitemOpen
  \bibfield  {author} {\bibinfo {author} {\bibfnamefont {D.}~\bibnamefont
  {Contreras}}, \bibinfo {author} {\bibfnamefont {P.}~\bibnamefont {Boubel}},\
  and\ \bibinfo {author} {\bibfnamefont {D.}~\bibnamefont {Scott}},\
  }\href@noop {} {\bibfield  {journal} {\bibinfo  {journal} {Journal of
  Cosmology and Astroparticle Physics}\ }\textbf {\bibinfo {volume}
  {2017}}\bibinfo  {number} { (12)},\ \bibinfo {pages} {046}}\BibitemShut
  {NoStop}%
\bibitem [{\citenamefont {Minami}\ and\ \citenamefont
  {Komatsu}(2020)}]{minami2020new}%
  \BibitemOpen
\bibfield  {number} {  }\bibfield  {author} {\bibinfo {author} {\bibfnamefont
  {Y.}~\bibnamefont {Minami}}\ and\ \bibinfo {author} {\bibfnamefont
  {E.}~\bibnamefont {Komatsu}},\ }\href@noop {} {\bibfield  {journal} {\bibinfo
   {journal} {Physical Review Letters}\ }\textbf {\bibinfo {volume} {125}},\
  \bibinfo {pages} {221301} (\bibinfo {year} {2020})}\BibitemShut {NoStop}%
\bibitem [{\citenamefont {Gruppuso}\ \emph {et~al.}(2020)\citenamefont
  {Gruppuso}, \citenamefont {Molinari}, \citenamefont {Natoli},\ and\
  \citenamefont {Pagano}}]{gruppuso2020planck}%
  \BibitemOpen
  \bibfield  {author} {\bibinfo {author} {\bibfnamefont {A.}~\bibnamefont
  {Gruppuso}}, \bibinfo {author} {\bibfnamefont {D.}~\bibnamefont {Molinari}},
  \bibinfo {author} {\bibfnamefont {P.}~\bibnamefont {Natoli}},\ and\ \bibinfo
  {author} {\bibfnamefont {L.}~\bibnamefont {Pagano}},\ }\href@noop {}
  {\bibfield  {journal} {\bibinfo  {journal} {Journal of Cosmology and
  Astroparticle Physics}\ }\textbf {\bibinfo {volume} {2020}}\bibinfo  {number}
  { (11)},\ \bibinfo {pages} {066}}\BibitemShut {NoStop}%
\bibitem [{\citenamefont {Bortolami}\ \emph {et~al.}(2022)\citenamefont
  {Bortolami}, \citenamefont {Billi}, \citenamefont {Gruppuso}, \citenamefont
  {Natoli},\ and\ \citenamefont {Pagano}}]{bortolami2022planck}%
  \BibitemOpen
\bibfield  {number} {  }\bibfield  {author} {\bibinfo {author} {\bibfnamefont
  {M.}~\bibnamefont {Bortolami}}, \bibinfo {author} {\bibfnamefont
  {M.}~\bibnamefont {Billi}}, \bibinfo {author} {\bibfnamefont
  {A.}~\bibnamefont {Gruppuso}}, \bibinfo {author} {\bibfnamefont
  {P.}~\bibnamefont {Natoli}},\ and\ \bibinfo {author} {\bibfnamefont
  {L.}~\bibnamefont {Pagano}},\ }\href@noop {} {\bibfield  {journal} {\bibinfo
  {journal} {arXiv preprint arXiv:2206.01635}\ } (\bibinfo {year}
  {2022})}\BibitemShut {NoStop}%
\bibitem [{\citenamefont {Eskilt}(2022)}]{eskilt2022frequency}%
  \BibitemOpen
  \bibfield  {author} {\bibinfo {author} {\bibfnamefont {J.}~\bibnamefont
  {Eskilt}},\ }\href@noop {} {\bibfield  {journal} {\bibinfo  {journal}
  {Astronomy \& Astrophysics}\ }\textbf {\bibinfo {volume} {662}},\ \bibinfo
  {pages} {A10} (\bibinfo {year} {2022})}\BibitemShut {NoStop}%
\bibitem [{\citenamefont {Eskilt}\ \emph {et~al.}(2023)\citenamefont {Eskilt},
  \citenamefont {Herold}, \citenamefont {Komatsu}, \citenamefont {Murai},
  \citenamefont {Namikawa},\ and\ \citenamefont
  {Naokawa}}]{eskilt2023constraint}%
  \BibitemOpen
  \bibfield  {author} {\bibinfo {author} {\bibfnamefont {J.~R.}\ \bibnamefont
  {Eskilt}}, \bibinfo {author} {\bibfnamefont {L.}~\bibnamefont {Herold}},
  \bibinfo {author} {\bibfnamefont {E.}~\bibnamefont {Komatsu}}, \bibinfo
  {author} {\bibfnamefont {K.}~\bibnamefont {Murai}}, \bibinfo {author}
  {\bibfnamefont {T.}~\bibnamefont {Namikawa}},\ and\ \bibinfo {author}
  {\bibfnamefont {F.}~\bibnamefont {Naokawa}},\ }\href@noop {} {} (\bibinfo
  {year} {2023}),\ \Eprint {https://arxiv.org/abs/2303.15369} {arXiv:2303.15369
  [astro-ph.CO]} \BibitemShut {NoStop}%
\bibitem [{\citenamefont {Zagatti}\ \emph {et~al.}(2024)\citenamefont
  {Zagatti}, \citenamefont {Bortolami}, \citenamefont {Gruppuso}, \citenamefont
  {Natoli}, \citenamefont {Pagano},\ and\ \citenamefont
  {Fabbian}}]{zagatti2024planck}%
  \BibitemOpen
  \bibfield  {author} {\bibinfo {author} {\bibfnamefont {G.}~\bibnamefont
  {Zagatti}}, \bibinfo {author} {\bibfnamefont {M.}~\bibnamefont {Bortolami}},
  \bibinfo {author} {\bibfnamefont {A.}~\bibnamefont {Gruppuso}}, \bibinfo
  {author} {\bibfnamefont {P.}~\bibnamefont {Natoli}}, \bibinfo {author}
  {\bibfnamefont {L.}~\bibnamefont {Pagano}},\ and\ \bibinfo {author}
  {\bibfnamefont {G.}~\bibnamefont {Fabbian}},\ }\href@noop {} {} (\bibinfo
  {year} {2024}),\ \Eprint {https://arxiv.org/abs/2401.11973} {arXiv:2401.11973
  [astro-ph.CO]} \BibitemShut {NoStop}%
\bibitem [{\citenamefont {Diego-Palazuelos}\ \emph
  {et~al.}(2022{\natexlab{a}})\citenamefont {Diego-Palazuelos}, \citenamefont
  {Eskilt}, \citenamefont {Minami}, \citenamefont {Tristram}, \citenamefont
  {Sullivan}, \citenamefont {Banday}, \citenamefont {Barreiro}, \citenamefont
  {Eriksen}, \citenamefont {G{\'o}rski}, \citenamefont {Keskitalo} \emph
  {et~al.}}]{diego2022cosmic}%
  \BibitemOpen
  \bibfield  {author} {\bibinfo {author} {\bibfnamefont {P.}~\bibnamefont
  {Diego-Palazuelos}}, \bibinfo {author} {\bibfnamefont {J.~R.}\ \bibnamefont
  {Eskilt}}, \bibinfo {author} {\bibfnamefont {Y.}~\bibnamefont {Minami}},
  \bibinfo {author} {\bibfnamefont {M.}~\bibnamefont {Tristram}}, \bibinfo
  {author} {\bibfnamefont {R.}~\bibnamefont {Sullivan}}, \bibinfo {author}
  {\bibfnamefont {A.}~\bibnamefont {Banday}}, \bibinfo {author} {\bibfnamefont
  {R.}~\bibnamefont {Barreiro}}, \bibinfo {author} {\bibfnamefont
  {H.}~\bibnamefont {Eriksen}}, \bibinfo {author} {\bibfnamefont
  {K.}~\bibnamefont {G{\'o}rski}}, \bibinfo {author} {\bibfnamefont
  {R.}~\bibnamefont {Keskitalo}}, \emph {et~al.},\ }\href@noop {} {\bibfield
  {journal} {\bibinfo  {journal} {Physical Review Letters}\ }\textbf {\bibinfo
  {volume} {128}},\ \bibinfo {pages} {091302} (\bibinfo {year}
  {2022}{\natexlab{a}})}\BibitemShut {NoStop}%
\bibitem [{\citenamefont {Miller}\ \emph {et~al.}(2009)\citenamefont {Miller},
  \citenamefont {Shimon},\ and\ \citenamefont {Keating}}]{miller2009cmb}%
  \BibitemOpen
  \bibfield  {author} {\bibinfo {author} {\bibfnamefont {N.}~\bibnamefont
  {Miller}}, \bibinfo {author} {\bibfnamefont {M.}~\bibnamefont {Shimon}},\
  and\ \bibinfo {author} {\bibfnamefont {B.}~\bibnamefont {Keating}},\
  }\href@noop {} {\bibfield  {journal} {\bibinfo  {journal} {Physical Review
  D}\ }\textbf {\bibinfo {volume} {79}},\ \bibinfo {pages} {103002} (\bibinfo
  {year} {2009})}\BibitemShut {NoStop}%
\bibitem [{\citenamefont {Keating}\ \emph {et~al.}(2012)\citenamefont
  {Keating}, \citenamefont {Shimon},\ and\ \citenamefont
  {Yadav}}]{keating2012self}%
  \BibitemOpen
  \bibfield  {author} {\bibinfo {author} {\bibfnamefont {B.~G.}\ \bibnamefont
  {Keating}}, \bibinfo {author} {\bibfnamefont {M.}~\bibnamefont {Shimon}},\
  and\ \bibinfo {author} {\bibfnamefont {A.~P.}\ \bibnamefont {Yadav}},\
  }\href@noop {} {\bibfield  {journal} {\bibinfo  {journal} {The Astrophysical
  Journal Letters}\ }\textbf {\bibinfo {volume} {762}},\ \bibinfo {pages} {L23}
  (\bibinfo {year} {2012})}\BibitemShut {NoStop}%
\bibitem [{\citenamefont {Minami}\ \emph {et~al.}(2019)\citenamefont {Minami},
  \citenamefont {Ochi}, \citenamefont {Ichiki}, \citenamefont {Katayama},
  \citenamefont {Komatsu},\ and\ \citenamefont
  {Matsumura}}]{minami2019simultaneous}%
  \BibitemOpen
  \bibfield  {author} {\bibinfo {author} {\bibfnamefont {Y.}~\bibnamefont
  {Minami}}, \bibinfo {author} {\bibfnamefont {H.}~\bibnamefont {Ochi}},
  \bibinfo {author} {\bibfnamefont {K.}~\bibnamefont {Ichiki}}, \bibinfo
  {author} {\bibfnamefont {N.}~\bibnamefont {Katayama}}, \bibinfo {author}
  {\bibfnamefont {E.}~\bibnamefont {Komatsu}},\ and\ \bibinfo {author}
  {\bibfnamefont {T.}~\bibnamefont {Matsumura}},\ }\href@noop {} {\bibfield
  {journal} {\bibinfo  {journal} {Progress of Theoretical and Experimental
  Physics}\ }\textbf {\bibinfo {volume} {2019}},\ \bibinfo {pages} {083E02}
  (\bibinfo {year} {2019})}\BibitemShut {NoStop}%
\bibitem [{\citenamefont {Clark}\ \emph {et~al.}(2021)\citenamefont {Clark},
  \citenamefont {Kim}, \citenamefont {Hill},\ and\ \citenamefont
  {Hensley}}]{clark2021origin}%
  \BibitemOpen
  \bibfield  {author} {\bibinfo {author} {\bibfnamefont {S.}~\bibnamefont
  {Clark}}, \bibinfo {author} {\bibfnamefont {C.-G.}\ \bibnamefont {Kim}},
  \bibinfo {author} {\bibfnamefont {J.~C.}\ \bibnamefont {Hill}},\ and\
  \bibinfo {author} {\bibfnamefont {B.~S.}\ \bibnamefont {Hensley}},\
  }\href@noop {} {\bibfield  {journal} {\bibinfo  {journal} {The Astrophysical
  Journal}\ }\textbf {\bibinfo {volume} {919}},\ \bibinfo {pages} {53}
  (\bibinfo {year} {2021})}\BibitemShut {NoStop}%
\bibitem [{\citenamefont {de~la Hoz}\ \emph {et~al.}(2022)\citenamefont {de~la
  Hoz}, \citenamefont {Diego-Palazuelos}, \citenamefont
  {Mart{\'\i}nez-Gonz{\'a}lez}, \citenamefont {Vielva}, \citenamefont
  {Barreiro},\ and\ \citenamefont {Bilbao-Ahedo}}]{de2022determination}%
  \BibitemOpen
  \bibfield  {author} {\bibinfo {author} {\bibfnamefont {E.}~\bibnamefont
  {de~la Hoz}}, \bibinfo {author} {\bibfnamefont {P.}~\bibnamefont
  {Diego-Palazuelos}}, \bibinfo {author} {\bibfnamefont {E.}~\bibnamefont
  {Mart{\'\i}nez-Gonz{\'a}lez}}, \bibinfo {author} {\bibfnamefont
  {P.}~\bibnamefont {Vielva}}, \bibinfo {author} {\bibfnamefont
  {R.}~\bibnamefont {Barreiro}},\ and\ \bibinfo {author} {\bibfnamefont
  {J.}~\bibnamefont {Bilbao-Ahedo}},\ }\href@noop {} {\bibfield  {journal}
  {\bibinfo  {journal} {Journal of Cosmology and Astroparticle Physics}\
  }\textbf {\bibinfo {volume} {2022}}\bibinfo  {number} { (03)},\ \bibinfo
  {pages} {032}}\BibitemShut {NoStop}%
\bibitem [{\citenamefont {Cukierman}\ \emph {et~al.}(2022)\citenamefont
  {Cukierman}, \citenamefont {Clark},\ and\ \citenamefont
  {Halal}}]{cukierman2022magnetic}%
  \BibitemOpen
\bibfield  {number} {  }\bibfield  {author} {\bibinfo {author} {\bibfnamefont
  {A.~J.}\ \bibnamefont {Cukierman}}, \bibinfo {author} {\bibfnamefont
  {S.}~\bibnamefont {Clark}},\ and\ \bibinfo {author} {\bibfnamefont
  {G.}~\bibnamefont {Halal}},\ }\href@noop {} {\bibfield  {journal} {\bibinfo
  {journal} {arXiv preprint arXiv:2208.07382}\ } (\bibinfo {year}
  {2022})}\BibitemShut {NoStop}%
\bibitem [{\citenamefont {Vacher}\ \emph {et~al.}(2022)\citenamefont {Vacher},
  \citenamefont {Aumont}, \citenamefont {Boulanger}, \citenamefont {Montier},
  \citenamefont {Guillet}, \citenamefont {Ritacco},\ and\ \citenamefont
  {Chluba}}]{vacher2022frequency}%
  \BibitemOpen
  \bibfield  {author} {\bibinfo {author} {\bibfnamefont {L.}~\bibnamefont
  {Vacher}}, \bibinfo {author} {\bibfnamefont {J.}~\bibnamefont {Aumont}},
  \bibinfo {author} {\bibfnamefont {F.}~\bibnamefont {Boulanger}}, \bibinfo
  {author} {\bibfnamefont {L.}~\bibnamefont {Montier}}, \bibinfo {author}
  {\bibfnamefont {V.}~\bibnamefont {Guillet}}, \bibinfo {author} {\bibfnamefont
  {A.}~\bibnamefont {Ritacco}},\ and\ \bibinfo {author} {\bibfnamefont
  {J.}~\bibnamefont {Chluba}},\ }\href@noop {} {\bibfield  {journal} {\bibinfo
  {journal} {arXiv preprint arXiv:2210.14768}\ } (\bibinfo {year}
  {2022})}\BibitemShut {NoStop}%
\bibitem [{\citenamefont {Diego-Palazuelos}\ \emph
  {et~al.}(2022{\natexlab{b}})\citenamefont {Diego-Palazuelos}, \citenamefont
  {Martínez-González}, \citenamefont {Vielva}, \citenamefont {Barreiro},
  \citenamefont {Tristram}, \citenamefont {de~la Hoz}, \citenamefont {Eskilt},
  \citenamefont {Minami}, \citenamefont {Sullivan}, \citenamefont {Banday},
  \citenamefont {Górski}, \citenamefont {Keskitalo}, \citenamefont {Komatsu},\
  and\ \citenamefont {Scott}}]{diego2022robustness}%
  \BibitemOpen
  \bibfield  {author} {\bibinfo {author} {\bibfnamefont {P.}~\bibnamefont
  {Diego-Palazuelos}}, \bibinfo {author} {\bibfnamefont {E.}~\bibnamefont
  {Martínez-González}}, \bibinfo {author} {\bibfnamefont {P.}~\bibnamefont
  {Vielva}}, \bibinfo {author} {\bibfnamefont {R.~B.}\ \bibnamefont
  {Barreiro}}, \bibinfo {author} {\bibfnamefont {M.}~\bibnamefont {Tristram}},
  \bibinfo {author} {\bibfnamefont {E.}~\bibnamefont {de~la Hoz}}, \bibinfo
  {author} {\bibfnamefont {J.~R.}\ \bibnamefont {Eskilt}}, \bibinfo {author}
  {\bibfnamefont {Y.}~\bibnamefont {Minami}}, \bibinfo {author} {\bibfnamefont
  {R.~M.}\ \bibnamefont {Sullivan}}, \bibinfo {author} {\bibfnamefont {A.~J.}\
  \bibnamefont {Banday}}, \bibinfo {author} {\bibfnamefont {K.~M.}\
  \bibnamefont {Górski}}, \bibinfo {author} {\bibfnamefont {R.}~\bibnamefont
  {Keskitalo}}, \bibinfo {author} {\bibfnamefont {E.}~\bibnamefont {Komatsu}},\
  and\ \bibinfo {author} {\bibfnamefont {D.}~\bibnamefont {Scott}},\
  }\href@noop {} {\bibfield  {journal} {\bibinfo  {journal} {arXiv preprint
  arXiv:2210.07655}\ } (\bibinfo {year} {2022}{\natexlab{b}})}\BibitemShut
  {NoStop}%
\bibitem [{\citenamefont {Monelli}\ \emph {et~al.}(2022)\citenamefont
  {Monelli}, \citenamefont {Komatsu}, \citenamefont {Adler}, \citenamefont
  {Billi}, \citenamefont {Campeti}, \citenamefont {Dachlythra}, \citenamefont
  {Duivenvoorden}, \citenamefont {Gudmundsson},\ and\ \citenamefont
  {Reinecke}}]{monelli2022impact}%
  \BibitemOpen
  \bibfield  {author} {\bibinfo {author} {\bibfnamefont {M.}~\bibnamefont
  {Monelli}}, \bibinfo {author} {\bibfnamefont {E.}~\bibnamefont {Komatsu}},
  \bibinfo {author} {\bibfnamefont {A.~E.}\ \bibnamefont {Adler}}, \bibinfo
  {author} {\bibfnamefont {M.}~\bibnamefont {Billi}}, \bibinfo {author}
  {\bibfnamefont {P.}~\bibnamefont {Campeti}}, \bibinfo {author} {\bibfnamefont
  {N.}~\bibnamefont {Dachlythra}}, \bibinfo {author} {\bibfnamefont {A.~J.}\
  \bibnamefont {Duivenvoorden}}, \bibinfo {author} {\bibfnamefont {J.~E.}\
  \bibnamefont {Gudmundsson}},\ and\ \bibinfo {author} {\bibfnamefont
  {M.}~\bibnamefont {Reinecke}},\ }\href@noop {} {} (\bibinfo {year}
  {2022})\BibitemShut {NoStop}%
\bibitem [{\citenamefont {Jost}\ \emph {et~al.}(2022)\citenamefont {Jost},
  \citenamefont {Errard},\ and\ \citenamefont
  {Stompor}}]{jost2022characterising}%
  \BibitemOpen
  \bibfield  {author} {\bibinfo {author} {\bibfnamefont {B.}~\bibnamefont
  {Jost}}, \bibinfo {author} {\bibfnamefont {J.}~\bibnamefont {Errard}},\ and\
  \bibinfo {author} {\bibfnamefont {R.}~\bibnamefont {Stompor}},\ }\href@noop
  {} {\bibfield  {journal} {\bibinfo  {journal} {arXiv preprint
  arXiv:2212.08007}\ } (\bibinfo {year} {2022})}\BibitemShut {NoStop}%
\bibitem [{\citenamefont {Monelli}\ \emph {et~al.}(2023)\citenamefont
  {Monelli}, \citenamefont {Komatsu}, \citenamefont {Ghigna}, \citenamefont
  {Matsumura}, \citenamefont {Pisano},\ and\ \citenamefont
  {Takaku}}]{monelli2023impact}%
  \BibitemOpen
  \bibfield  {author} {\bibinfo {author} {\bibfnamefont {M.}~\bibnamefont
  {Monelli}}, \bibinfo {author} {\bibfnamefont {E.}~\bibnamefont {Komatsu}},
  \bibinfo {author} {\bibfnamefont {T.}~\bibnamefont {Ghigna}}, \bibinfo
  {author} {\bibfnamefont {T.}~\bibnamefont {Matsumura}}, \bibinfo {author}
  {\bibfnamefont {G.}~\bibnamefont {Pisano}},\ and\ \bibinfo {author}
  {\bibfnamefont {R.}~\bibnamefont {Takaku}},\ }\href@noop {} {} (\bibinfo
  {year} {2023}),\ \Eprint {https://arxiv.org/abs/2311.07999} {arXiv:2311.07999
  [astro-ph.CO]} \BibitemShut {NoStop}%
\bibitem [{\citenamefont {Ritacco}\ \emph {et~al.}(2023)\citenamefont
  {Ritacco}, \citenamefont {Bizzarri}, \citenamefont {Boulanger}, \citenamefont
  {Pérault}, \citenamefont {Aumont}, \citenamefont {Bouchet}, \citenamefont
  {Calvo}, \citenamefont {Catalano}, \citenamefont {Darson}, \citenamefont
  {Désert}, \citenamefont {Errard}, \citenamefont {Feret}, \citenamefont
  {Macías-Pérez}, \citenamefont {Maffei}, \citenamefont {Monfardini},
  \citenamefont {Montier}, \citenamefont {Murgia}, \citenamefont {Morfin},
  \citenamefont {Nati}, \citenamefont {Pisano}, \citenamefont {Ponthieu},
  \citenamefont {Puget}, \citenamefont {Savorgnano}, \citenamefont {Segret},
  \citenamefont {Schuster}, \citenamefont {Treuttel},\ and\ \citenamefont
  {Zannoni}}]{ritacco2023polarization}%
  \BibitemOpen
  \bibfield  {author} {\bibinfo {author} {\bibfnamefont {A.}~\bibnamefont
  {Ritacco}}, \bibinfo {author} {\bibfnamefont {L.}~\bibnamefont {Bizzarri}},
  \bibinfo {author} {\bibfnamefont {F.}~\bibnamefont {Boulanger}}, \bibinfo
  {author} {\bibfnamefont {M.}~\bibnamefont {Pérault}}, \bibinfo {author}
  {\bibfnamefont {J.}~\bibnamefont {Aumont}}, \bibinfo {author} {\bibfnamefont
  {F.}~\bibnamefont {Bouchet}}, \bibinfo {author} {\bibfnamefont
  {M.}~\bibnamefont {Calvo}}, \bibinfo {author} {\bibfnamefont
  {A.}~\bibnamefont {Catalano}}, \bibinfo {author} {\bibfnamefont
  {D.}~\bibnamefont {Darson}}, \bibinfo {author} {\bibfnamefont {F.~X.}\
  \bibnamefont {Désert}}, \bibinfo {author} {\bibfnamefont {J.}~\bibnamefont
  {Errard}}, \bibinfo {author} {\bibfnamefont {A.}~\bibnamefont {Feret}},
  \bibinfo {author} {\bibfnamefont {J.~F.}\ \bibnamefont {Macías-Pérez}},
  \bibinfo {author} {\bibfnamefont {B.}~\bibnamefont {Maffei}}, \bibinfo
  {author} {\bibfnamefont {A.}~\bibnamefont {Monfardini}}, \bibinfo {author}
  {\bibfnamefont {L.}~\bibnamefont {Montier}}, \bibinfo {author} {\bibfnamefont
  {M.}~\bibnamefont {Murgia}}, \bibinfo {author} {\bibfnamefont
  {P.}~\bibnamefont {Morfin}}, \bibinfo {author} {\bibfnamefont
  {F.}~\bibnamefont {Nati}}, \bibinfo {author} {\bibfnamefont {G.}~\bibnamefont
  {Pisano}}, \bibinfo {author} {\bibfnamefont {N.}~\bibnamefont {Ponthieu}},
  \bibinfo {author} {\bibfnamefont {J.~L.}\ \bibnamefont {Puget}}, \bibinfo
  {author} {\bibfnamefont {S.}~\bibnamefont {Savorgnano}}, \bibinfo {author}
  {\bibfnamefont {B.}~\bibnamefont {Segret}}, \bibinfo {author} {\bibfnamefont
  {K.}~\bibnamefont {Schuster}}, \bibinfo {author} {\bibfnamefont
  {J.}~\bibnamefont {Treuttel}},\ and\ \bibinfo {author} {\bibfnamefont
  {M.}~\bibnamefont {Zannoni}},\ }\href@noop {} {} (\bibinfo {year} {2023}),\
  \Eprint {https://arxiv.org/abs/2311.08307} {arXiv:2311.08307 [astro-ph.IM]}
  \BibitemShut {NoStop}%
\bibitem [{\citenamefont {Fujita}\ \emph {et~al.}(2022)\citenamefont {Fujita},
  \citenamefont {Minami}, \citenamefont {Shiraishi},\ and\ \citenamefont
  {Yokoyama}}]{fujita2022can}%
  \BibitemOpen
  \bibfield  {author} {\bibinfo {author} {\bibfnamefont {T.}~\bibnamefont
  {Fujita}}, \bibinfo {author} {\bibfnamefont {Y.}~\bibnamefont {Minami}},
  \bibinfo {author} {\bibfnamefont {M.}~\bibnamefont {Shiraishi}},\ and\
  \bibinfo {author} {\bibfnamefont {S.}~\bibnamefont {Yokoyama}},\ }\href@noop
  {} {\bibfield  {journal} {\bibinfo  {journal} {arXiv preprint
  arXiv:2208.08101}\ } (\bibinfo {year} {2022})}\BibitemShut {NoStop}%
\bibitem [{\citenamefont {Loeb}\ and\ \citenamefont
  {Kosowsky}(1996)}]{loeb1996faraday}%
  \BibitemOpen
  \bibfield  {author} {\bibinfo {author} {\bibfnamefont {A.}~\bibnamefont
  {Loeb}}\ and\ \bibinfo {author} {\bibfnamefont {A.}~\bibnamefont
  {Kosowsky}},\ }\href@noop {} {\bibfield  {journal} {\bibinfo  {journal}
  {arXiv preprint astro-ph/9601055}\ } (\bibinfo {year} {1996})}\BibitemShut
  {NoStop}%
\bibitem [{\citenamefont {Liu}\ \emph {et~al.}(2006)\citenamefont {Liu},
  \citenamefont {Lee},\ and\ \citenamefont {Ng}}]{liu2006effect}%
  \BibitemOpen
  \bibfield  {author} {\bibinfo {author} {\bibfnamefont {G.-C.}\ \bibnamefont
  {Liu}}, \bibinfo {author} {\bibfnamefont {S.}~\bibnamefont {Lee}},\ and\
  \bibinfo {author} {\bibfnamefont {K.-W.}\ \bibnamefont {Ng}},\ }\href@noop {}
  {\bibfield  {journal} {\bibinfo  {journal} {Physical Review Letters}\
  }\textbf {\bibinfo {volume} {97}},\ \bibinfo {pages} {161303} (\bibinfo
  {year} {2006})}\BibitemShut {NoStop}%
\bibitem [{\citenamefont {Finelli}\ and\ \citenamefont
  {Galaverni}(2009)}]{finelli2009rotation}%
  \BibitemOpen
  \bibfield  {author} {\bibinfo {author} {\bibfnamefont {F.}~\bibnamefont
  {Finelli}}\ and\ \bibinfo {author} {\bibfnamefont {M.}~\bibnamefont
  {Galaverni}},\ }\href@noop {} {\bibfield  {journal} {\bibinfo  {journal}
  {Physical Review D}\ }\textbf {\bibinfo {volume} {79}},\ \bibinfo {pages}
  {063002} (\bibinfo {year} {2009})}\BibitemShut {NoStop}%
\bibitem [{\citenamefont {Gubitosi}\ \emph {et~al.}(2014)\citenamefont
  {Gubitosi}, \citenamefont {Martinelli},\ and\ \citenamefont
  {Pagano}}]{gubitosi2014including}%
  \BibitemOpen
  \bibfield  {author} {\bibinfo {author} {\bibfnamefont {G.}~\bibnamefont
  {Gubitosi}}, \bibinfo {author} {\bibfnamefont {M.}~\bibnamefont
  {Martinelli}},\ and\ \bibinfo {author} {\bibfnamefont {L.}~\bibnamefont
  {Pagano}},\ }\href@noop {} {\bibfield  {journal} {\bibinfo  {journal}
  {Journal of Cosmology and Astroparticle Physics}\ }\textbf {\bibinfo {volume}
  {2014}}\bibinfo  {number} { (12)},\ \bibinfo {pages} {020}}\BibitemShut
  {NoStop}%
\bibitem [{\citenamefont {Galaverni}\ \emph {et~al.}(2023)\citenamefont
  {Galaverni}, \citenamefont {Finelli},\ and\ \citenamefont
  {Paoletti}}]{galaverni2023redshift}%
  \BibitemOpen
\bibfield  {number} {  }\bibfield  {author} {\bibinfo {author} {\bibfnamefont
  {M.}~\bibnamefont {Galaverni}}, \bibinfo {author} {\bibfnamefont
  {F.}~\bibnamefont {Finelli}},\ and\ \bibinfo {author} {\bibfnamefont
  {D.}~\bibnamefont {Paoletti}},\ }\href@noop {} {\bibfield  {journal}
  {\bibinfo  {journal} {arXiv preprint arXiv:2301.07971}\ } (\bibinfo {year}
  {2023})}\BibitemShut {NoStop}%
\bibitem [{\citenamefont {Blas}\ \emph {et~al.}(2011)\citenamefont {Blas},
  \citenamefont {Lesgourgues},\ and\ \citenamefont {Tram}}]{blas2011cosmic}%
  \BibitemOpen
  \bibfield  {author} {\bibinfo {author} {\bibfnamefont {D.}~\bibnamefont
  {Blas}}, \bibinfo {author} {\bibfnamefont {J.}~\bibnamefont {Lesgourgues}},\
  and\ \bibinfo {author} {\bibfnamefont {T.}~\bibnamefont {Tram}},\ }\href@noop
  {} {\bibfield  {journal} {\bibinfo  {journal} {Journal of Cosmology and
  Astroparticle Physics}\ }\textbf {\bibinfo {volume} {2011}}\bibinfo  {number}
  { (07)},\ \bibinfo {pages} {034}}\BibitemShut {NoStop}%
\bibitem [{\citenamefont {Ali-Haimoud}\ and\ \citenamefont
  {Hirata}(2010)}]{Ali-Haimoud:2010tlj}%
  \BibitemOpen
\bibfield  {number} {  }\bibfield  {author} {\bibinfo {author} {\bibfnamefont
  {Y.}~\bibnamefont {Ali-Haimoud}}\ and\ \bibinfo {author} {\bibfnamefont
  {C.~M.}\ \bibnamefont {Hirata}},\ }\href
  {https://doi.org/10.1103/PhysRevD.82.063521} {\bibfield  {journal} {\bibinfo
  {journal} {Phys. Rev. D}\ }\textbf {\bibinfo {volume} {82}},\ \bibinfo
  {pages} {063521} (\bibinfo {year} {2010})},\ \Eprint
  {https://arxiv.org/abs/1006.1355} {arXiv:1006.1355 [astro-ph.CO]}
  \BibitemShut {NoStop}%
\bibitem [{\citenamefont {Ali-Haimoud}\ and\ \citenamefont
  {Hirata}(2011)}]{Ali-Haimoud:2010hou}%
  \BibitemOpen
  \bibfield  {author} {\bibinfo {author} {\bibfnamefont {Y.}~\bibnamefont
  {Ali-Haimoud}}\ and\ \bibinfo {author} {\bibfnamefont {C.~M.}\ \bibnamefont
  {Hirata}},\ }\href {https://doi.org/10.1103/PhysRevD.83.043513} {\bibfield
  {journal} {\bibinfo  {journal} {Phys. Rev. D}\ }\textbf {\bibinfo {volume}
  {83}},\ \bibinfo {pages} {043513} (\bibinfo {year} {2011})},\ \Eprint
  {https://arxiv.org/abs/1011.3758} {arXiv:1011.3758 [astro-ph.CO]}
  \BibitemShut {NoStop}%
\bibitem [{\citenamefont {Lee}\ and\ \citenamefont
  {Ali-Ha\"\i{}moud}(2020)}]{Lee:2020obi}%
  \BibitemOpen
  \bibfield  {author} {\bibinfo {author} {\bibfnamefont {N.}~\bibnamefont
  {Lee}}\ and\ \bibinfo {author} {\bibfnamefont {Y.}~\bibnamefont
  {Ali-Ha\"\i{}moud}},\ }\href {https://doi.org/10.1103/PhysRevD.102.083517}
  {\bibfield  {journal} {\bibinfo  {journal} {Phys. Rev. D}\ }\textbf {\bibinfo
  {volume} {102}},\ \bibinfo {pages} {083517} (\bibinfo {year} {2020})},\
  \Eprint {https://arxiv.org/abs/2007.14114} {arXiv:2007.14114 [astro-ph.CO]}
  \BibitemShut {NoStop}%
\bibitem [{\citenamefont {Kosowsky}(1995)}]{kosowsky1995cosmic}%
  \BibitemOpen
  \bibfield  {author} {\bibinfo {author} {\bibfnamefont {A.}~\bibnamefont
  {Kosowsky}},\ }\href@noop {} {\bibfield  {journal} {\bibinfo  {journal}
  {arXiv preprint astro-ph/9501045}\ } (\bibinfo {year} {1995})}\BibitemShut
  {NoStop}%
\bibitem [{\citenamefont {Hu}\ and\ \citenamefont {White}(1997)}]{hu1997cmb}%
  \BibitemOpen
  \bibfield  {author} {\bibinfo {author} {\bibfnamefont {W.}~\bibnamefont
  {Hu}}\ and\ \bibinfo {author} {\bibfnamefont {M.}~\bibnamefont {White}},\
  }\href@noop {} {\bibfield  {journal} {\bibinfo  {journal} {Physical Review
  D}\ }\textbf {\bibinfo {volume} {56}},\ \bibinfo {pages} {596} (\bibinfo
  {year} {1997})}\BibitemShut {NoStop}%
\bibitem [{\citenamefont {Zaldarriaga}\ and\ \citenamefont
  {Seljak}(1997)}]{zaldarriaga1997all}%
  \BibitemOpen
  \bibfield  {author} {\bibinfo {author} {\bibfnamefont {M.}~\bibnamefont
  {Zaldarriaga}}\ and\ \bibinfo {author} {\bibfnamefont {U.}~\bibnamefont
  {Seljak}},\ }\href@noop {} {\bibfield  {journal} {\bibinfo  {journal}
  {Physical Review D}\ }\textbf {\bibinfo {volume} {55}},\ \bibinfo {pages}
  {1830} (\bibinfo {year} {1997})}\BibitemShut {NoStop}%
\bibitem [{\citenamefont {Liu}\ \emph {et~al.}(2002)\citenamefont {Liu},
  \citenamefont {Sugiyama}, \citenamefont {Benson}, \citenamefont {Lacey},\
  and\ \citenamefont {Nusser}}]{liu2002polarization}%
  \BibitemOpen
  \bibfield  {author} {\bibinfo {author} {\bibfnamefont {G.-C.}\ \bibnamefont
  {Liu}}, \bibinfo {author} {\bibfnamefont {N.}~\bibnamefont {Sugiyama}},
  \bibinfo {author} {\bibfnamefont {A.~J.}\ \bibnamefont {Benson}}, \bibinfo
  {author} {\bibfnamefont {C.}~\bibnamefont {Lacey}},\ and\ \bibinfo {author}
  {\bibfnamefont {A.}~\bibnamefont {Nusser}},\ }in\ \href@noop {} {\emph
  {\bibinfo {booktitle} {AIP Conference Proceedings}}},\ Vol.\ \bibinfo
  {volume} {609}\ (\bibinfo {organization} {American Institute of Physics},\
  \bibinfo {year} {2002})\ pp.\ \bibinfo {pages} {271--274}\BibitemShut
  {NoStop}%
\bibitem [{\citenamefont {Bartolo}\ \emph {et~al.}(2006)\citenamefont
  {Bartolo}, \citenamefont {Matarrese},\ and\ \citenamefont
  {Riotto}}]{bartolo2006cosmic}%
  \BibitemOpen
  \bibfield  {author} {\bibinfo {author} {\bibfnamefont {N.}~\bibnamefont
  {Bartolo}}, \bibinfo {author} {\bibfnamefont {S.}~\bibnamefont {Matarrese}},\
  and\ \bibinfo {author} {\bibfnamefont {A.}~\bibnamefont {Riotto}},\
  }\href@noop {} {\bibfield  {journal} {\bibinfo  {journal} {Journal of
  Cosmology and Astroparticle Physics}\ }\textbf {\bibinfo {volume}
  {2006}}\bibinfo  {number} { (06)},\ \bibinfo {pages} {024}}\BibitemShut
  {NoStop}%
\bibitem [{\citenamefont {Bartolo}\ \emph
  {et~al.}(2007{\natexlab{a}})\citenamefont {Bartolo}, \citenamefont
  {Matarrese},\ and\ \citenamefont {Riotto}}]{bartolo2007cmb}%
  \BibitemOpen
\bibfield  {number} {  }\bibfield  {author} {\bibinfo {author} {\bibfnamefont
  {N.}~\bibnamefont {Bartolo}}, \bibinfo {author} {\bibfnamefont
  {S.}~\bibnamefont {Matarrese}},\ and\ \bibinfo {author} {\bibfnamefont
  {A.}~\bibnamefont {Riotto}},\ }\href@noop {} {\bibfield  {journal} {\bibinfo
  {journal} {Journal of Cosmology and Astroparticle Physics}\ }\textbf
  {\bibinfo {volume} {2007}}\bibinfo  {number} { (01)},\ \bibinfo {pages}
  {019}}\BibitemShut {NoStop}%
\bibitem [{\citenamefont {Bartolo}\ \emph
  {et~al.}(2007{\natexlab{b}})\citenamefont {Bartolo}, \citenamefont
  {Matarrese},\ and\ \citenamefont {Riotto}}]{bartolo2007cosmic}%
  \BibitemOpen
\bibfield  {number} {  }\bibfield  {author} {\bibinfo {author} {\bibfnamefont
  {N.}~\bibnamefont {Bartolo}}, \bibinfo {author} {\bibfnamefont
  {S.}~\bibnamefont {Matarrese}},\ and\ \bibinfo {author} {\bibfnamefont
  {A.}~\bibnamefont {Riotto}},\ }in\ \href@noop {} {\emph {\bibinfo {booktitle}
  {Les Houches Summer School - Session 86: Particle Physics and Cosmology: The
  Fabric of Spacetime}}}\ (\bibinfo {year} {2007})\ \Eprint
  {https://arxiv.org/abs/astro-ph/0703496} {arXiv:astro-ph/0703496}
  \BibitemShut {NoStop}%
\bibitem [{\citenamefont {Naruko}\ \emph {et~al.}(2013)\citenamefont {Naruko},
  \citenamefont {Pitrou}, \citenamefont {Koyama},\ and\ \citenamefont
  {Sasaki}}]{naruko2013second}%
  \BibitemOpen
  \bibfield  {author} {\bibinfo {author} {\bibfnamefont {A.}~\bibnamefont
  {Naruko}}, \bibinfo {author} {\bibfnamefont {C.}~\bibnamefont {Pitrou}},
  \bibinfo {author} {\bibfnamefont {K.}~\bibnamefont {Koyama}},\ and\ \bibinfo
  {author} {\bibfnamefont {M.}~\bibnamefont {Sasaki}},\ }\href@noop {}
  {\bibfield  {journal} {\bibinfo  {journal} {Classical and Quantum Gravity}\
  }\textbf {\bibinfo {volume} {30}},\ \bibinfo {pages} {165008} (\bibinfo
  {year} {2013})}\BibitemShut {NoStop}%
\bibitem [{\citenamefont {Pettinari}\ \emph {et~al.}(2013)\citenamefont
  {Pettinari}, \citenamefont {Fidler}, \citenamefont {Crittenden},
  \citenamefont {Koyama},\ and\ \citenamefont
  {Wands}}]{pettinari2013intrinsic}%
  \BibitemOpen
  \bibfield  {author} {\bibinfo {author} {\bibfnamefont {G.~W.}\ \bibnamefont
  {Pettinari}}, \bibinfo {author} {\bibfnamefont {C.}~\bibnamefont {Fidler}},
  \bibinfo {author} {\bibfnamefont {R.}~\bibnamefont {Crittenden}}, \bibinfo
  {author} {\bibfnamefont {K.}~\bibnamefont {Koyama}},\ and\ \bibinfo {author}
  {\bibfnamefont {D.}~\bibnamefont {Wands}},\ }\href@noop {} {\bibfield
  {journal} {\bibinfo  {journal} {Journal of Cosmology and Astroparticle
  Physics}\ }\textbf {\bibinfo {volume} {2013}}\bibinfo  {number} { (04)},\
  \bibinfo {pages} {003}}\BibitemShut {NoStop}%
\bibitem [{\citenamefont {Saito}\ \emph {et~al.}(2014)\citenamefont {Saito},
  \citenamefont {Naruko}, \citenamefont {Hiramatsu},\ and\ \citenamefont
  {Sasaki}}]{saito2014geodesic}%
  \BibitemOpen
\bibfield  {number} {  }\bibfield  {author} {\bibinfo {author} {\bibfnamefont
  {R.}~\bibnamefont {Saito}}, \bibinfo {author} {\bibfnamefont
  {A.}~\bibnamefont {Naruko}}, \bibinfo {author} {\bibfnamefont
  {T.}~\bibnamefont {Hiramatsu}},\ and\ \bibinfo {author} {\bibfnamefont
  {M.}~\bibnamefont {Sasaki}},\ }\href@noop {} {\bibfield  {journal} {\bibinfo
  {journal} {Journal of Cosmology and Astroparticle Physics}\ }\textbf
  {\bibinfo {volume} {2014}}\bibinfo  {number} { (10)},\ \bibinfo {pages}
  {051}}\BibitemShut {NoStop}%
\bibitem [{\citenamefont {Su}\ and\ \citenamefont
  {Lim}(2014)}]{su2014formulating}%
  \BibitemOpen
\bibfield  {number} {  }\bibfield  {author} {\bibinfo {author} {\bibfnamefont
  {S.-C.}\ \bibnamefont {Su}}\ and\ \bibinfo {author} {\bibfnamefont {E.~A.}\
  \bibnamefont {Lim}},\ }\href@noop {} {\bibfield  {journal} {\bibinfo
  {journal} {Physical Review D}\ }\textbf {\bibinfo {volume} {89}},\ \bibinfo
  {pages} {123006} (\bibinfo {year} {2014})}\BibitemShut {NoStop}%
\bibitem [{\citenamefont {Mehrem}(2011)}]{mehrem2011plane}%
  \BibitemOpen
  \bibfield  {author} {\bibinfo {author} {\bibfnamefont {R.}~\bibnamefont
  {Mehrem}},\ }\href@noop {} {\bibfield  {journal} {\bibinfo  {journal}
  {Applied mathematics and computation}\ }\textbf {\bibinfo {volume} {217}},\
  \bibinfo {pages} {5360} (\bibinfo {year} {2011})}\BibitemShut {NoStop}%
\bibitem [{\citenamefont {Varshalovich}(1988)}]{varshalovich1988quantum}%
  \BibitemOpen
  \bibfield  {author} {\bibinfo {author} {\bibfnamefont {D.}~\bibnamefont
  {Varshalovich}},\ }\href@noop {} {}\ (\bibinfo  {publisher} {World
  Scientific},\ \bibinfo {year} {1988})\BibitemShut {NoStop}%
\bibitem [{\citenamefont {Landau}\ and\ \citenamefont
  {Lifshitz}(2013)}]{landau2013quantum}%
  \BibitemOpen
  \bibfield  {author} {\bibinfo {author} {\bibfnamefont {L.~D.}\ \bibnamefont
  {Landau}}\ and\ \bibinfo {author} {\bibfnamefont {E.~M.}\ \bibnamefont
  {Lifshitz}},\ }\href@noop {} {}Vol.~\bibinfo {volume} {3}\ (\bibinfo
  {publisher} {Elsevier},\ \bibinfo {year} {2013})\BibitemShut {NoStop}%
\bibitem [{\citenamefont {Goldberg}\ \emph {et~al.}(1967)\citenamefont
  {Goldberg}, \citenamefont {MacFarlane}, \citenamefont {Newman}, \citenamefont
  {Rohrlich},\ and\ \citenamefont {Sudarshan}}]{goldberg1967spin}%
  \BibitemOpen
  \bibfield  {author} {\bibinfo {author} {\bibfnamefont {J.~N.}\ \bibnamefont
  {Goldberg}}, \bibinfo {author} {\bibfnamefont {A.~J.}\ \bibnamefont
  {MacFarlane}}, \bibinfo {author} {\bibfnamefont {E.~T.}\ \bibnamefont
  {Newman}}, \bibinfo {author} {\bibfnamefont {F.}~\bibnamefont {Rohrlich}},\
  and\ \bibinfo {author} {\bibfnamefont {E.~G.}\ \bibnamefont {Sudarshan}},\
  }\href@noop {} {\bibfield  {journal} {\bibinfo  {journal} {Journal of
  Mathematical Physics}\ }\textbf {\bibinfo {volume} {8}},\ \bibinfo {pages}
  {2155} (\bibinfo {year} {1967})}\BibitemShut {NoStop}%
\end{thebibliography}%
	
\end{document}